\def\tr{{\rm tr}\,}
\def\Tr{{\rm Tr}\,}
\def\sgn{{\rm sgn\,}}
\def\b{\bibitem}
\documentstyle[aps,prb,eqsecnum,psfig,floats]{revtex}
\begin{document}
\def\SNG{{\em Physical Review Style and Notation Guide}}
\def\LUG {{\em \LaTeX{} User's Guide \& Reference Manual}}
\def\btt#1{{\tt$\backslash$\string#1}}%
\def\REVTeX{REV\TeX}
\def\AmS{{\protect\the\textfont2
        A\kern-.1667em\lower.5ex\hbox{M}\kern-.125emS}}
\def\AmSLaTeX{\AmS-\LaTeX}
\def\BibTeX{\rm B{\sc ib}\TeX}
\twocolumn[\hsize\textwidth\columnwidth\hsize\csname@twocolumnfalse%
\endcsname
 
\title{Theory of many-fermion systems\\
       \small{$[$ Phys. Rev. B {\bf 56}, 6513 (1997) $]$}}
\author{D.Belitz}
\address{Department of Physics and Materials Science Institute\\
University of Oregon,\\
Eugene, OR 97403}
\author{T.R.Kirkpatrick}
\address{Institute for Physical Science and Technology, and Department of Physics\\
University of Maryland,\\ 
College Park, MD 20742}

\date{\today}
\maketitle

\begin{abstract}
A general field--theoretical description of many--fermion systems, both with
and without quenched disorder, is developed. Starting from the basic
Grassmannian action for interacting fermions, we first bosonize the theory
by introducing composite matrix variables that correspond to two--fermion
excitations and integrating out the fermion degrees of freedom. The saddle
point solution of the resulting matrix field theory reproduces a disordered
Hartree--Fock approximation, and an expansion to Gaussian order about the
saddle point corresponds to a disordered RPA--like theory. In the clean limit
they reduce to the ordinary Hartree--Fock and random--phase approximations. 
We first concentrate 
on disordered systems, and perform a symmetry analysis that allows for a 
systematic separation of the massless modes from the massive ones. By treating 
the massive modes in a simple approximation, one obtains a technically 
satisfactory derivation of the generalized nonlinear sigma--model that
has been used in the theory of metal--insulator transitions. The theory
also allows for the treatment of other phase transitions in the disordered
Fermi liquid. We further use renormalization group techniques to establish the
existence of a disordered Fermi--liquid fixed point, and show that it is
stable for all dimensions $d>2$. The so--called weak--localization effects can 
be understood as corrections to scaling near this fixed point. The general 
theory also allows for studying the clean limit. For that case we develop a 
loop expansion that corresponds to an expansion in powers of the screened 
Coulomb interaction, and that represents a systematic improvement over RPA.
We corroborate the existence of a Fermi--liquid fixed point that is stable for 
all dimensions $d>1$, in agreement with other recent field--theoretical
treatments of clean Fermi liquids.
\end{abstract}
\pacs{PACS numbers: }
]
\tableofcontents

\section{Introduction}
\label{sec:I}

The many--fermion problem has a long history, due to its importance with
respect to electrons in condensed matter
systems.\cite{PinesNozieres,FetterWalecka,AGD}
In recent years there has been a renewed interest in fundamental aspects
of this problem, especially for electrons
at low or zero temperature,
both with and without quenched disorder. The important
physical systems for which these studies are relevant include,
superconductors, both high--T$_{\rm c}$ and conventional ones,
doped semiconductors in both their
metallic and insulating phases, amorphous alloys, and Quantum Hall systems.

Historically, there have been two important techniques to tackle the
many--fermion problem: Landau's phenomenological Fermi--liquid
theory,\cite{BaymPethick} and the microscopic many--body perturbation theory
or Feynman diagram approach,\cite{FetterWalecka,AGD}
which starts from a canonically quantized
Hamiltonian of the problem. While both techniques are very useful, and
have yielded many important results, either one of them has serious
limitations. Fermi--liquid theory has been extended to include
disorder,\cite{CastellaniDiCastro,CastellaniKotliarLee} but often a microscopic
approach is preferred.
For the microscopic many--body perturbation theory, the inclusion of disorder
is extremely awkward, mostly because of a large number
of diagrammatic contributions that are individually
finite, but that ultimately cancel each other. More recent approaches have
used functional or field--theoretical
methods,\cite{NegeleOrland} which allow for the application of renormalization
group (RG) techniques. A significant recent advance in this area has been
Shankar's RG technique\cite{Shankar}
for clean fermion systems that is applied directly to the
Grassmannian field theory for fermions. This approach, which is still in
its infancy, is very promising. For instance, it has already led
to a derivation of Fermi--liquid theory from a microscopic starting
point.\cite{Shankar,Dupuis} It also has been used to provide RG 
derivations of the Cooper instability problem, and
of RPA--like screening.\cite{Shankar} Other applications, for instance a
discussion of corrections to scaling
near the RG fixed point (FP) that describes the clean Fermi liquid,
should be possible. However, for reasons that will be
discussed in detail in the present paper, it is not easy to include
the effects of quenched disorder into this approach either.

An alternative approach for clean electronic systems has been pursued by
Houghton, Marston, and others,\cite{HoughtonMarston} and by Fr{\"o}hlich and
G{\"o}tschmann.\cite{FrohlichGotschmann}
These authors have generalized the bosonization techniques that have been used 
successfully in $d=1$ to higher
dimensions. In Ref.\ \onlinecite{FrohlichGotschmann} the bosonization was
also combined with RG techniques.
These theories have been used, for example, to rederive certain
nonanalyticities that occur in Fermi--liquid theory, and
to study the possibility of marginal Fermi liquids in dimensions $d\geq 2$.
Finally, clean electronic systems have been bosonized by means of a
Hubbard--Stratonovich transformation and the use of the classical 
Hubbard--Stratonovich field as the fundamental field of the 
theory.\cite{KopietzSchonhammer} Within this latter approach, disorder
has been included in the single-particle Green function at the level of
the lowest order Born approximation.\cite{Kopietz} More sophisticated
effects of quenched disorder have so far not been included into any
of these approaches.

For systems of disordered electrons, a completely different theory has been
in use for some time, that describes composite fermionic, i.e. effectively
bosonic, degrees of freedom. For noninteracting electrons, this effective 
field theory takes the form of a nonlinear sigma--model,\cite{Wegner}
which was generalized to the case of interacting electrons by 
Finkel'stein.\cite{Finkelstein,R} It is custom tailored for the 
description of the metal--insulator transition near $d=2$, and it does
not allow for the clean limit to be taken. No technically
satisfactory derivation of the interacting theory has ever been given,
but rather its structure has been guessed, based on more rigorous
derivations of Wegner's effective model for the metal--insulator transition
in noninteracting electronic systems.\cite{SchaferWegner,PruiskenSchafer} 
The key idea underlying these
effective field theories is to keep explicitly only those degrees of
freedom that are likely to be relevant for the problem under consideration,
and to integrate out all others in some simple approximation.

It is clearly desirable to develop a more flexible field--theoretical approach
for the many--fermion problem. Some of the desired features of such a
theory are as follows. (1) Both clean and disordered systems should be 
describable within a single framework. (2) The theory should allow for a RG 
description of both clean and disordered Fermi liquids, of the metal--insulator
transition, as well as of other quantum phase transitions, like e.g.
the recently investigated magnetic ones.\cite{fm_dirty,fm_clean,afm} 
(3) It should offer flexibility with respect to how many, and which, degrees
of freedom one keeps explicitly, depending on the problem under consideration.
(4) It should provide a satisfactory derivation of the nonlinear sigma--model 
used to describe the metal--insulator transition in
interacting disordered electronic systems. (5) It should allow for explicit
calculations of physically relevant observables, such as thermodynamic and 
transport properties. While all of the existing theories fulfill some of
these criteria, none of them meet all of them simultaneously.

This is the first of two papers where we develop such a theory. In the present
paper we discuss our general physical ideas, and mainly focus on the
disordered interacting fermion problem, although we will discuss some aspects
of the clean limit. In a future, second paper we plan to use the same 
approach to thoroughly
discuss clean interacting fermion systems, as well as the connection 
between our approach and others. The outline of this paper is as follows. 
In Sec.\ \ref{sec:II} we introduce our starting point, 
a microscopic model for an electron fluid, in general in the
presence of quenched disorder. We then show how to 
transform this Grassmannian field theory into one for classical or bosonic 
fields, and discuss the physical motivation for this
transformation. In this section we also construct a saddle--point solution
for the resulting composite--fermion field theory, and expand to Gaussian
order about the saddle point. The result is equivalent to what one obtains
within many--body diagrammatic theory from the random phase approximation (RPA),
modified by disorder.
In Sec.\ \ref{sec:III} we perform a symmetry analysis of our field theory.
The results, together with some conclusions drawn from Sec.\ \ref{sec:II}, 
suggest
slightly separate ways to proceed from here for clean and disordered fermions,
respectively. Focusing on the latter, we use the symmetry analysis to
identify and classify all of the slow or soft modes of the system. This is
crucial for a RG description of the problem. Using
these results, we then derive an effective field theory for the slow modes in
a disordered fermion system. We identify a FP that describes a
disordered Fermi liquid,
and show that it is stable for $d>2$. We then show that the so--called
weak--localization effects can be interpreted as corrections to scaling near
this FP. We derive Finkel'stein's generalization of Wegner's nonlinear 
sigma--model that
has been used to describe metal--insulator transitions near $d=2$,
and discuss the critical FP that corresponds to this quantum
phase transition. The section is concluded by a brief discussion of a
magnetic phase transition that is described by the theory, and of the
relation between the theory presented here and
earlier theories of magnetic transitions in disordered systems. In Sec.\ 
\ref{sec:IV} we consider the clean limit of our theory and discuss the clean 
Fermi--liquid FP. There we show that certain nonanalyticities that appear in
Fermi--liquid theory can be interpreted as corrections to scaling near this FP,
and we point out far--reaching analogies between clean and disordered Fermi
systems. In Sec.\ \ref{sec:V} we present our conclusions. Appendix\ 
\ref{app:A} spells out a technical point that one encounters in 
Sec.\ \ref{sec:II}, and Appendix\ \ref{app:B} contains a pedagogical
discussion of an $O(N)$ symmetric $\phi^4$ theory.

\section{Matrix field theory}
\label{sec:II}

\subsection{Grassmannian field theory}
\label{subsec:II.A}

Let us start with a field--theoretical description of a system of interacting,
disordered fermions in terms of Grassmann variables.\cite{NegeleOrland} 
The partition function of the system is
\begin{equation}
Z=\int D[\bar{\psi},\psi]\ e^{S[\bar{\psi},\psi]}\quad. 
\label{eq:2.1}
\end{equation}
Here the functional integration is with respect to Grassmann valued fields,
$\bar\psi$ and $\psi$, and the action, $S$, is given by
\begin{mathletters}
\label{eqs:2.2}
\begin{equation}
S = -\int dx\sum_{\sigma}\ \bar{\psi}_{\sigma}(x)\,\partial_{\tau}\,
\psi_{\sigma}(x) + S_0 + S_{dis} + S_{int}\quad. 
\label{eq:2.2a}
\end{equation}
We use a ($d+1$)-vector notation, with $x=({\bf x},\tau)$, and
$\int dx=\int_V d{\bf x}\int_{0}^{\beta} d\tau$. ${\bf x}$ denotes position,
$\tau$ imaginary time, $V$ is the system volume, $\beta =1/T$ is the inverse 
temperature, $\sigma $ is the spin label, and we use units such that 
$\hbar = k_B = 1$. $S_0$ describes free fermions with chemical potential $\mu$,
\begin{equation}
S_0 = \int dx\sum_{\sigma}\ \bar{\psi}_\sigma (x)\,\left(\frac{\nabla^2}{2m}
       + \mu\right)\,\psi_{\sigma}(x)\quad, 
\label{eq:2.2b}
\end{equation}
with $m$ the fermion mass. The Laplacian will be denoted by $\nabla^2$
throughout. We will mostly be concerned with systems at $T=0$,
where $\mu = \epsilon_F$, with $\epsilon_F = k_F^2/2m$ the Fermi energy,
and $k_F$ the Fermi momentum.
$S_{dis}$ describes a static random potential, 
$u({\bf x})$, coupling to the fermionic number density,
\begin{equation}
S_{dis} = -\int dx\sum_{\sigma}\ u({\bf x})\,\bar{\psi}_\sigma (x)\,
                                             \psi_{\sigma}(x)\quad, 
\label{eq:2.2c}
\end{equation}
and S$_{int}$ describes a spin--independent two--particle interaction,
\begin{eqnarray}
S_{int} &=&-\frac{1}{2}\int dx_1\,dx_2\ \sum_{\sigma_1,\sigma_2}\ 
            v({\bf x}_1 - {\bf x}_2)
\nonumber\\
          &&\times\bar{\psi}_{\sigma _1}(x_1)\,\bar{\psi}_{\sigma _2}(x_2)\,
             \psi _{\sigma_2}(x_2)\,\psi _{\sigma _1}(x_1)\quad. 
\label{eq:2.2d}
\end{eqnarray}
\end{mathletters}%
The interaction potential $v({\bf x})$ will be specified below.

For simplicity, the random potential in Eq.\ (\ref{eq:2.2c}) is taken to 
be Gaussian distributed with a second moment that is given by a function
$U({\bf x})$ with the dimension of an energy density,
\begin{equation}
\left\{ u({\bf x})\,u({\bf y})\right\}_{dis} = \frac{1}{\pi N_F}\ 
           U({\bf x}-{\bf y})\quad,
\label{eq:2.3}
\end{equation}
where $\left\{\ldots\right\}_{dis}$ denotes the disorder average.
$N_F$ is a normalization factor with the dimension of a density of states.
One could use the free electron density of states at the Fermi level, but it
will turn out to be more convenient to include some trivial disorder and
interaction renormalizations. We will explicitly define $N_F$ in
Eq.\ (\ref{eq:2.42'}) below. The full physical density of states at the
Fermi level will also be 
encountered later, and will be denoted by $N(\epsilon_F)$.
More general disorder potentials can be considered, but as long as they couple 
only to the electron number density, the difference between the more general
potentials and Eq.\ (\ref{eq:2.3}) can be shown to be RG irrelevant at the 
disordered Fermi--liquid FP. For other purposes, e.g. a description of the
Anderson--Mott metal--insulator transition in the system, the difference is 
likewise expected to be RG irrelevant.

Since the system contains quenched disorder, it is necessary to average 
the free energy or $\ln Z$.
This is accomplished by means of the replica trick,\cite{ReplicaTrick} 
which is based on the identity
\begin{equation}
\ln Z = \lim_{n\rightarrow 0}\ (Z^n - 1)/n\quad.
\label{eq:2.4}
\end{equation}
Introducing $n$ identical replicas of the system (with $n$ an
integer), labeled by the index $\alpha $, and carrying out the disorder
average, we obtain
\begin{equation}
{\tilde Z}\equiv\{Z^n\}_{dis} 
      = \int\prod_{\alpha =1}^{n} D\left[\bar{\psi}^{\alpha},
                                                         \psi^{\alpha}\right]
\exp [\tilde S\,]\quad.
\label{eq:2.5}
\end{equation}
We again separate $\tilde S$ into free, disordered and
interaction parts as follows,
\begin{mathletters}
\label{eqs:2.6}
\begin{equation}
\tilde{S} = \sum_{\alpha =1}^{n}\ \left(\tilde{S}_{0}^{\,\alpha} 
                 +\tilde{S}_{dis}^{\,\alpha} + \tilde{S}_{int}^{\,\alpha}\right)
                         \quad.
\label{eq:2.6a}
\end{equation}
It is useful to go to a Fourier representation with wave vectors
{\bf k} and fermionic Matsubara frequencies $\omega _n=2\pi T(n+1/2)$, and
a $(d+1)$-vector notation, $k = ({\bf k},\omega_n)$. Then
\begin{equation}
\tilde{S}_{0}^{\,\alpha} = \sum_{k,\sigma}\,\bar{\psi}_\sigma^{\alpha}(k)\,
   \left[i\omega_n - {\bf k}^2/2m + \mu\right]\,\psi_\sigma^{\alpha}(k)\quad, 
\label{eq:2.6b}
\end{equation}
and,
\begin{eqnarray}
{\tilde S}_{dis}^{\,\alpha}&=&\frac{1}{2\pi N_F}\ \sum_{\beta =1}^{n}
         \sum_{\{{\bf k}_i\}}\sum_{n,m}\sum_{\sigma,\sigma^{\prime}}\ 
         \delta_{{\bf k}_1+{\bf k}_3,{\bf k}_2+{\bf k}_4} 
\nonumber\\
&&\times U({\bf k}_1 - {\bf k}_2)\ \bar{\psi}_{n\sigma}^{\alpha}({\bf k}_1)\,
\psi_{n\sigma }^{\alpha}
({\bf k}_2)\,\bar{\psi}_{m\sigma^{\prime}}^{\beta}({\bf k}_3)
\nonumber\\
&&\qquad\qquad\qquad\qquad \times\psi_{m\sigma^{\prime}}^\beta ({\bf k}_4)\quad,
\label{eq:2.6c}
\end{eqnarray}
\begin{eqnarray}
\tilde{S}_{int}^{\,\alpha}&=&-\frac T2\sum_{\sigma _1,\sigma
_2}\sum_{\{k_i\}}\ \delta _{k_1+k_2,k_3+k_4}\ v({\bf k}_2-{\bf k}_3)  
\nonumber\\
&&\times\bar{\psi}_{\sigma _1^{}}^\alpha (k_1)\,\bar{\psi}_{\sigma_2}^\alpha
(k_2)\,\psi_{\sigma_2}^\alpha (k_3)\,\psi_{\sigma_1}^\alpha (k_4)\quad. 
\label{eq:2.6d}
\end{eqnarray}
\end{mathletters}%
Here we have defined
\begin{mathletters}
\label{eqs:2.6'}
\begin{equation}
\psi _{n\sigma }({\bf k})\equiv\psi _\sigma (k) = \sqrt{T/V}\int dx\,
   e^{-i({\bf kx} - \omega_n\tau)}\,\psi_{\sigma}(x)\quad,\
\label{eq:2.6'a}
\end{equation}
\begin{equation}
{\bar\psi}_{n\sigma }({\bf k})\equiv{\bar\psi}_\sigma (k) = \sqrt{T/V}\int dx\
   e^{i({\bf kx} - \omega_n\tau)}{\bar\psi}_{\sigma}(x)\quad.
\label{eq:2.6'b}
\end{equation}
\end{mathletters}%
Ultimately, we will be interested in long--wavelength, low--frequency processes.
For clean electrons, this in general means that only the scattering of
electrons and holes close to the Fermi surface is important. In a clean
Fermi liquid this is true because the lifetimes of the single--particle 
excitations become infinitely long as the Fermi surface is approached. For the
disordered electron problem the situation is different. All single--particle
momentum and frequency excitations are damped by the single--particle
relaxation rate, $1/\tau_{rel}$, which in general is not small. For this case, 
the only slow or soft excitations are two--particle excitations. This implies
that the soft excitations in the two cases are fundamentally
different, and that clean systems have in general more soft modes than
disordered ones. This observation will be of great importance below, and
will suggest somewhat different effective field theories for dealing with
clean and disordered systems, respectively. In fact, this difference is
the reason why the approaches of Shankar\cite{Shankar} and
Houghton and Marston\cite{HoughtonMarston} cannot be simply extended to the 
dirty case.

\begin{figure}[t]
\centerline{\psfig{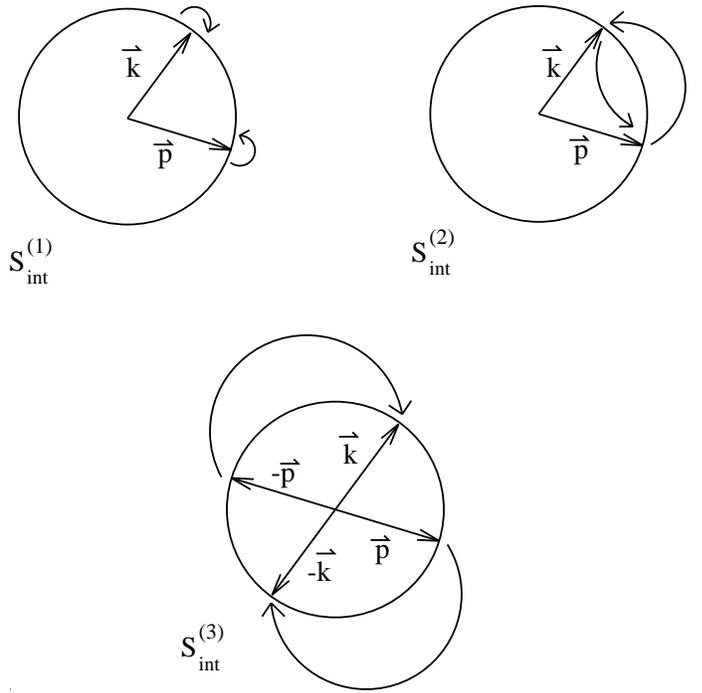}\vspace*{5mm}}
\caption{Typical small--angle (1), large--angle (2), and $2k_F$-scattering
 processes (3) near the Fermi surface in $d=2$.}
\label{fig:1}
\end{figure}

To examine this important point in more detail, we first consider
the clean case. It is customary and convenient
to divide the possible scattering processes into three classes: (1)
small--angle scattering, (2) large--angle scattering, and 
(3) $2k_F$-scattering. These classes are also referred to as 
the particle--hole channel
for classes (1) and (2), and the particle--particle
or Cooper channel for class (3),
respectively. The corresponding scattering processes are schematically
depicted in Fig.\ \ref{fig:1}. We next make explicit the phase space 
decomposition that is
%
%
inherent to this classification by writing $\tilde{S}_{int}^{\,\alpha}$,
Eq.\ (\ref{eq:2.6d}), as
\begin{mathletters}
\label{eqs:2.7}
\begin{equation}
\tilde{S}_{int}^{\,\alpha} = \tilde{S}_{int}^{\,\alpha\,(1)} + 
   \tilde{S}_{int}^{\,\alpha\,(2)} + \tilde{S}_{int}^{\,\alpha\,(3)}\quad , 
\label{eq:2.7a}
\end{equation}
where,
\begin{eqnarray}
\tilde{S}_{int}^{\,\alpha\,(1)}&=&-\frac T2\sum_{\sigma _1\sigma _2}\sum_{k,p}
   {\sum_q}^{\,\prime}
v({\bf q})\ \bar{\psi}_{\sigma _1}^\alpha (k)\,\bar{\psi}_{\sigma _2}^\alpha
(p+q)\,
\nonumber\\
      &&\times\psi _{\sigma _2}^\alpha (p)\psi _{\sigma _1}^\alpha (k+q)\quad, 
\label{eq:2.7b}
\end{eqnarray}
\begin{eqnarray}
\tilde{S}_{int}^{\,\alpha\,(2)}&=&-\frac T2\sum_{\sigma _1\sigma
_2}\sum\limits_{k,p}{\sum_q}^{\,\prime}
v({\bf p}-{\bf k})\ \bar{\psi}_{\sigma _1}^\alpha (k)\,\bar{\psi}_{\sigma
_2}^\alpha (p+q)\,
\nonumber\\
&&\times \psi _{\sigma _2}^\alpha (k+q)\psi _{\sigma _1}^\alpha (p)\quad,
\label{eq:2.7c}
\end{eqnarray}
\begin{eqnarray}
\tilde{S}_{int}^{\,\alpha\,(3)}&=&-\frac T2\sum_{\sigma _1\neq \sigma
_2}\sum_{k,p}{\sum_q}^{\,\prime}
v({\bf k}+{\bf p})\ \bar{\psi}_{\sigma _1}^\alpha (-k)\,\bar{\psi}_{\sigma
_2}^\alpha (k+q)\,
\nonumber\\ 
&&\times\psi _{\sigma _2}^\alpha (-p+q)\,\psi _{\sigma _1}^\alpha (p)\quad.
\label{eq:2.7d}
\end{eqnarray}
\end{mathletters}%
Here the prime on the $q$-summation indicates that only momenta up
to some momentum cutoff, $\lambda $, are integrated over. This restriction
is necessary to avoid double counting, since each of the three expressions,
Eqs.\ (\ref{eq:2.7b}) - (\ref{eq:2.7d}), 
represent all of $\tilde{S}_{int}^{\,\alpha}$ if all
wave vectors are summed over. In general the long--wavelength physics we are
interested in will not depend on $\lambda $. Finally, one often writes the
Eqs.\ (\ref{eqs:2.7}) in terms of singlet ($s$) and triplet ($t$) contributions.
We introduce spinors
\begin{mathletters}
\begin{equation}
\psi_{n}^{\alpha}({\bf x}) = \left( \begin{array}{c}
           \psi_{n\uparrow }^{\alpha}({\bf x})\\
           \psi_{n\downarrow}^{\alpha}({\bf x})\end{array}\right)\quad,
\label{eq:2.8a}
\end{equation}
and their Fourier transforms
\begin{equation}
\psi^{\alpha}(k)\equiv\psi_{n}^{\alpha}({\bf k}) = \left(\begin{array}{c} 
                       \psi_{n\uparrow}^{\alpha}({\bf k})\\
                       \psi_{n\downarrow}^{\alpha}({\bf k})\end{array}\right)
                                                                         \quad,
\label{eq:2.8b}
\end{equation}
\end{mathletters}
as well as their adjoints, $\bar\psi^{\alpha}(k)$,
and a scalar product in spinor space, $(\psi ,\psi )=\bar{\psi}\cdot\psi$, 
where the dot denotes the matrix product. Then we can
write the interaction term as
\begin{mathletters}
\label{eqs:2.9}
\begin{equation}
\tilde{S}_{int}^{\,\alpha} = \tilde{S}_{int}^{\,\alpha\,(s)} + 
   \tilde{S}_{int}^{\,\alpha\,(t)} + \tilde{S}_{int}^{\,\alpha\,(3)}\quad, 
\label{eq:2.9a}
\end{equation}
with,
\begin{eqnarray}
\tilde{S}_{int}^{\,\alpha\,(s)}&=&-\frac T2\sum_{k,p}{\sum_q}^{\,\prime}
\Gamma_{k,p}^{(s)}(q)\ \bigl(\psi ^\alpha (k),s_0\psi ^\alpha (k+q)\bigr)
\nonumber\\
&&\times \bigl(\psi ^\alpha (p+q),s_0\psi ^\alpha (p)\bigr)\quad, 
\label{eq:2.9b}
\end{eqnarray}
\begin{eqnarray}
\tilde{S}_{int}^{\,\alpha\,(t)}&=&-\frac T2\sum_{k,p}{\sum_q}^{\,\prime}
\Gamma _{k,p}^{(t)}(q)\sum_{i=1}^3 
\bigl(\psi ^\alpha (k),s_i\psi ^\alpha (k+q)\bigr)
\nonumber\\
&&\times \bigl(\psi ^\alpha (p+q),s_i\psi ^\alpha (p)\bigr)\quad. 
\label{eq:2.9c}
\end{eqnarray}
\end{mathletters}%
Here $s_j=i\sigma_j$, with $\sigma_{1,2,3}$ the Pauli matrices, and
$s_0=\sigma_0$ is the $2\times 2$ identity matrix. We have also defined
the singlet ($s$) and triplet ($t$) interaction amplitudes
\begin{mathletters}
\label{eqs:2.10}
\begin{equation}
\Gamma _{k,p}^{(t)}(q)=\frac{1}{2}\,v({\bf p}-{\bf k}) \quad,
\label{eq:2.10a}
\end{equation}
and
\begin{equation}
\Gamma_{k,p}^{(s)}(q) = v({\bf q})-\frac{1}{2}\,v({\bf p}-{\bf k}) \quad.
\label{eq:2.10b}
\end{equation}
In addition we define the Cooper channel or $2k_F$-scattering amplitude
\begin{equation}
\Gamma_{k,p}^{(c)}(q) = v({\bf k+p})\quad.
\label{eq:2.10c}
\end{equation}
\end{mathletters}%

The effective interaction potentials that appear in Eqs.\ (\ref{eqs:2.9}) 
and (\ref{eqs:2.10}) are all given by the basic interaction potential 
$v$, taken at different momenta. $\tilde{S}_{int}^{\,\alpha\,(1)}$, 
and $\tilde{S}_{int}^{\,\alpha\,(s)}$ as well, contains the direct
scattering contribution, or $v({\bf q})$, with ${\bf q}$ the restricted 
($\vert {\bf q}\vert < \lambda << k_F$)
momentum. If $v$ is chosen to be a bare Coulomb interaction, then the
direct scattering contribution leads to
singularities in perturbation theory that indicate the need for
infinite resummations to incorporate screening. For simplicity, we assume
that this procedure has already been carried out, and take $v$ to be a
statically screened Coulomb interaction. For small $\vert {\bf q}\vert$, 
it is then sufficient
to replace $v({\bf q})$ in Eq.\ (\ref{eq:2.10b}) by a number. 
The moduli of the other momenta,
{\bf k} and {\bf p} in Eqs. (2.10), are equal to $k_F$ for the most important
scattering processes since, as mentioned above, the soft or slow excitations 
in the clean Fermi system involve particles and holes close to the Fermi
surface. The angular dependences of these coupling constants is usually
taken into account by expanding them in Legendre polynomials on the Fermi 
surface.\cite{BaymPethick} For future reference we
note that if only the zero angular momentum channel, $\ell =0$, is retained,
then Eqs.\ (\ref{eqs:2.9}) and (\ref{eqs:2.10}) are valid with 
$\Gamma_{k,p}^{(t)}(0)$, $\Gamma_{k,p}^{(s)}(0)$, and $\Gamma_{k,p}^{(c)}(0)$ 
replaced by numbers, $\Gamma^{(s)}$, $\Gamma^{(t)}$, and $\Gamma^{(c)}$, 
which are the Fermi surface averages of the respective interaction amplitudes.
Shankar,\cite{Shankar} in his RG approach, has explicitly derived these
results, and has shown that all corrections to these approximations are
RG irrelevant with respect to a RG fixed point that describes a Fermi liquid.

For disordered Fermi systems the situation is different, since single--particle
excitations are massive modes with a mass that is proportional to the
inverse elastic mean--free time.
In Section III we will give a detailed discussion of
the soft or slow modes in a disordered Fermi system. The conclusion will be
that, of the modes that appear in ${\tilde S}_{int}$,
the dominant soft modes are those that involve fluctuations of either the 
particle number density, $n_n$, or the spin density, ${\bf n}_s$,
in the particle--hole channel, or density fluctuations, $n_c$, in the 
particle--particle channel.\cite{ConservationLawFootnote} In terms of our
spinors, the latter are given as (cf. Eqs.\ (\ref{eqs:2.9}))
\begin{mathletters}
\label{eqs:2.11}
\begin{equation}
n_n^{\alpha}(q) = \sqrt{T/V} \sum_k \bigl(\psi^{\alpha}(k),s_0\psi^{\alpha}
                                     (k+q)\bigr)\quad, 
\label{eq:2.11a}
\end{equation}
\begin{equation}
{\bf n}_s^{\alpha}(q) = \sqrt{T/V} \sum_k\bigl(\psi^{\alpha}(k),{\bf s}
                         \psi^{\alpha}(k+q)\bigr)\quad,
\label{eq:2.11b}
\end{equation}
\begin{equation}
n_c^{\alpha}(q) = \sqrt{T/V} \sum_k\bar{\psi}_{\uparrow}^{\alpha}(k)\,
                        \bar{\psi}_{\downarrow}^{\alpha}(-k+q)\quad.  
\label{eq:2.11c}
\end{equation}
\end{mathletters}%
In Eqs.\ (\ref{eqs:2.9}) this implies that we should project onto these density
fluctuations. This projection is performed in Appendix \ref{app:A}. 
The net result is that the interaction amplitudes can be replaced
by constants, namely their angular averages with a distribution of momenta near
the Fermi surface. We thus obtain the interaction part of the action in the
form,
\begin{mathletters}
\label{eqs:2.12}
\begin{eqnarray}
\tilde{S}_{int}^{\,\alpha\,(s)}&=&-\frac T2\,\Gamma^{(s)}
   \sum_{k,p}{\sum_q}^{\,\prime}\,\bigl(\psi^{\alpha}(k),s_0\psi^{\alpha}(k+q)
                                                                        \bigr)
\nonumber\\
&&\times \bigl(\psi ^\alpha (p+q),s_0\psi ^\alpha (p)\bigr)\quad,
\label{eq:2.12a}
\end{eqnarray}
\begin{eqnarray}
\tilde{S}_{int}^{\,\alpha\,(t)}&=&-\frac T2\,\Gamma^{(t)}
   \sum_{k,p}{\sum_q}^{\,\prime}\sum_{i=1}^3
             \bigl(\psi^\alpha (k),s_i\psi^{\alpha}(k+q)\bigr)
\nonumber\\
&&\times \bigl(\psi ^\alpha (p+q),s_i\psi ^\alpha (p)\bigr)\quad.
\label{eq:2.12b}
\end{eqnarray}
\begin{eqnarray}
\tilde{S}_{int}^{\,\alpha\,(3)}&=&-\frac T2\,\Gamma^{(c)}
   \sum_{\sigma_1\neq \sigma_2}\sum_{k,p}{\sum_q}^{\,\prime}
     \,\bar{\psi}_{\sigma _1}^\alpha (k)\,\bar{\psi}_{\sigma_2}^\alpha (-k+q)\,
\nonumber\\
&&\times\psi _{\sigma _2}^\alpha (p+q)\,\psi _{\sigma _1}^\alpha (-p)\quad.
\label{eq:2.12c}
\end{eqnarray}
\end{mathletters}%
The disordered Fermi system is thus
described by only three interaction parameters, $\Gamma^{(s)}$,
$\Gamma^{(t)}$, and $\Gamma^{(c)}$, which are given by Eq.\ (\ref{eq:A.4b})
and analogous expressions. As long as we are only interested in physics
that is dominated by soft modes, this restriction is exact.
Clean Fermi systems, on the other hand,
in general require an infinite number of Fermi--liquid parameters, and the
restriction to the Eqs.\ (\ref{eqs:2.12}), with the three interaction
constants related to the Fermi--liquid parameters in the $\ell = 0$ channel,
constitutes an approximation.
Physically this difference is due to the fact that there are more soft modes
in the clean case than in the dirty case.

\subsection{Composite variables: Matrix field theory}
\label{subsec:II.B}

\subsubsection{The action in terms of $Q$-matrices}
\label{subsubsec:II.B.1}

As noted above, the slow modes for the disordered fermion problem are
fluctuations of products of
fermion operators. It is therefore convenient to transform to a field
theory in terms of these composite variables. For both technical and
physical reasons, which will become clear in Sec.\ \ref{sec:III}, 
it is convenient to go to a bispinor and,
eventually, to a spin--quaternion representation. We define a 
bispinor\cite{EfetovLarkinKhmelnitskii}
\begin{mathletters}
\label{eqs:2.13}
\begin{equation}
\eta_{n}^{\alpha}({\bf x}) = \frac{1}{\sqrt{2}}\left(\begin{array}{c}
   \bar{\psi}_n^\alpha ({\bf x})\\
    s_2\,\psi _n^\alpha ({\bf x})\end{array}\right) = 
     \frac{1}{\sqrt{2}}\left(
    \begin{array}{c}\bar{\psi}_{n\uparrow}^{\alpha}({\bf x}) \\
                    \bar{\psi}_{n\downarrow}^{\alpha}({\bf x})\\
                    \psi_{n\downarrow}^{\alpha}({\bf x})\\ 
                    -\psi_{n\uparrow}^{\alpha}({\bf x})\end{array}
                                         \right)\quad,
\label{eq:2.13a}
\end{equation}
and an adjoint bispinor\cite{AdjointFootnote}
\begin{equation}
(\eta^+)_n^\alpha ({\bf x}) = ic\eta_n^\alpha ({\bf x})
     = \frac{i}{\sqrt{2}}\left(
    \begin{array}{c}-{\psi}_{n\uparrow}^{\alpha}({\bf x}) \\
                    -{\psi}_{n\downarrow}^{\alpha}({\bf x})\\
                    \bar{\psi}_{n\downarrow}^{\alpha}({\bf x})\\
                    -\bar{\psi}_{n\uparrow}^{\alpha}({\bf x})\end{array}
                                         \right)\quad,
\label{eq:2.13b}
\end{equation}
with $c$ the charge--conjugation matrix
\begin{equation}
c = \left(\begin{array}{cc}0&s_2\\
                           s_2&0\end{array}\right)
  = i\tau_1\otimes s_2\quad. 
\label{eq:2.13c}
\end{equation}
\end{mathletters}%
The four degrees of freedom represented by the bispinor are the 
particle--hole or number density degrees of freedom, and the two spin 
degrees of freedom. We have also defined a basis in spin--quaternion space
as $\tau_r\otimes s_i$ $(r,i=0,1,2,3)$, with $\tau_0=s_0$ the $2\times 2$
identity matrix, and $\tau_j=-s_j=-i\sigma_j$ $(j=1,2,3)$, with $\sigma_j$ the
Pauli matrices. In this basis, the channels $r=0,3$ and $r=1,2$ describe the
particle--hole and particle--particle degrees of freedom, and the
channels $i=0$ and $i=1,2,3$ describe the spin--singlet and spin--triplet,
respectively.

It is convenient to define a scalar product in bispinor space as
\begin{equation}
\bigl(\eta_{n}^{\alpha},\eta_{m}^{\beta}\bigr) 
     = \sum_{j=1}^{4} {_j\eta}_{n}^{\alpha}\,{^j\eta}_{m}^{\beta}\quad,
\label{eq:2.14}
\end{equation}
where $^j\eta$ denotes the elements of $\eta$, and $_j\eta$ denotes the
elements of $\eta^+$.
The adjoint of any operator $A$ in bispinor space with respect to this scalar
product is given by $c^T A^T c$.
In terms of the bispinors, the terms on the right--hand side of 
Eq.\ (\ref{eq:2.6a}) can be written
\begin{mathletters}
\label{eqs:2.15}
\begin{equation}
\tilde{S}_0 = -i\sum_{\alpha,k}\left(\eta^{\alpha}(k),
          [i\omega_n - {\bf k}^2/2m+\mu]\,\eta^{\alpha}(k)\right)\quad,
\label{eq:2.15a}
\end{equation}

\begin{eqnarray}
\tilde{S}_{dis} &=&\frac{-1}{2\pi N_F}\sum_{\alpha ,\beta
}\sum_{n,m}\sum_{\{{\bf k}_i\}}\,\delta _{{\bf k}_1+{\bf k}_3,{\bf k}_2+{\bf k}
_4}\,U({\bf k}_1 - {\bf k}_2)
\nonumber\\
&&\times \bigl(\eta _n^\alpha ({\bf k}_1),\eta _n^\alpha ({\bf k}_2)\bigr)
         \bigl(\eta_m^\beta ({\bf k}_3),\eta _m^\beta ({\bf k}_4)\bigr)\quad,
\label{eq:2.15b}
\end{eqnarray}
\end{mathletters}%
and
\begin{mathletters}
\label{eqs:2.16}
\begin{equation}
\tilde{S}_{int} = \tilde{S}_{int}^{\,(s)} + \tilde{S}_{int}^{\,(t)}
                  + \tilde{S}_{int}^{\,(c)}\quad, 
\label{eq:2.16a}
\end{equation}

\begin{eqnarray}
\tilde{S}_{int}^{\,(s)}&=&\frac{T\Gamma^{(s)}}2\sum_\alpha \sum_{k,p}
   {\sum_q}^{\,\prime}\sum_{r=0,3}(-1)^r
\nonumber\\
&&\times \bigl(\eta^\alpha (k),(\tau_r\otimes s_0)\eta^\alpha (k+q)\bigr) 
\nonumber \\
&&\times \bigl(\eta^\alpha (p+q),(\tau_r\otimes s_0)\eta^\alpha (p)\bigr)\quad,
\label{eq:2.16b}
\end{eqnarray}

\begin{eqnarray}
\tilde{S}_{int}^{\,(t)}&=&\frac{T\Gamma^{(t)}}2\sum_\alpha \sum_{k,p}
   {\sum_q}^{\,\prime}\sum_{r=0,3}(-1)^r\sum_{i=1}^3  
\nonumber \\
&&\times \bigl(\eta^\alpha (k),(\tau_r\otimes s_i)\eta^\alpha (k+q)\bigr)
\nonumber\\
&&\times \bigl((\eta^\alpha (p+q),(\tau_r\otimes s_i)\eta^\alpha (p)\bigr)
                \quad,
\label{eq:2.16c}
\end{eqnarray}

\begin{eqnarray}
\tilde{S}_{int}^{\,(c)}&=&\frac{T\Gamma^{(c)}}{2}\sum_{\alpha}\sum_{{\bf k,p}}
{\sum_{\bf q}}^{\,\prime}
\sum_{n_1,n_2,m}\sum_{r=1,2}  
\nonumber\\
&&\times \left(\eta_{n_1}^\alpha (-{\bf k}),\left(\tau_r\otimes s_0\right)
  \eta _{-n_1+m}^\alpha (- {\bf k}+{\bf q})\right)
\nonumber\\
&&\times \left(\eta_{-n_2}^\alpha (-{\bf p}),\left(\tau_r\otimes s_0\right)
  \eta_{n_2+m}^\alpha (-{\bf p}-{\bf q})\right)\quad.
\nonumber\\
\label{eq:2.16d}
\end{eqnarray}
\end{mathletters}%
The same phase space decomposition used to break up $S_{int}$
can also be used to rewrite ${\tilde S}_{dis}$ as
\begin{mathletters}
\label{eqs:2.17}
\begin{equation}
{\tilde S}_{dis} = {\tilde S}_{dis}^{\,(1)} + {\tilde S}_{dis}^{\,(2)}\quad,
\label{eq:2.17a}
\end{equation}
with,
\begin{eqnarray}
\tilde{S}_{dis}^{\,(1)} &=&\frac{-1}{2\pi N_F\tau_{1}}
                                                  \sum_{\alpha,\beta}
       \sum_{n,m}\sum_{{\bf k,p}}{\sum_{\bf q}}^{\,\prime}
\,\bigl(\eta_n^\alpha ({\bf k}),\eta_n^\alpha ({\bf k}+{\bf q})\bigr)
\nonumber\\
&&\times\bigl(\eta_m^{\beta}({\bf p}),\eta_m^{\beta}({\bf p}-{\bf q})\bigr)
                                   \quad,
\label{eq:2.17b}
\end{eqnarray}

\begin{eqnarray}
\tilde{S}_{dis}^{\,(2)} &=&\frac{-1}{\pi N_F\tau_{rel}}
                           \sum_{\alpha ,\beta}
\sum_{n,m}\sum_{{\bf k,p}}{\sum_{\bf q}}^{\,\prime}
\,\bigl(\eta_n^\alpha ({\bf k}),\eta_n^\alpha ({\bf p})\bigr)
\nonumber \\
&&\times\bigl(\eta_m^{\beta}({\bf p}+{\bf q}),\eta_m^{\beta}({\bf k}+{\bf q})
                                                          \bigr)\quad.
\label{eq:2.17c}
\end{eqnarray}
\end{mathletters}%
Here we have introduced two different scattering times, $\tau_1$ and 
$\tau_{rel}$, that will in general
renormalize differently. The reason for their appearance is related to
the potential correlation function $U({\bf k}_1 - {\bf k}_2)$ in
Eq.\ (\ref{eq:2.15b}), which
leads to a $U({\bf q})$ in ${\tilde S}_{dis}^{\,(1)}$, and a
$U({\bf k}-{\bf p})$ in ${\tilde S}_{dis}^{\,(2)}$. Projection onto the density,
as in the interaction terms, then leads to Eqs.\ (\ref{eqs:2.17}), with
$1/\tau_1 = U({\bf q}=0)$, and $1/\tau_{rel}$
a weighted average over $U({\bf k}-{\bf p})$. $\tau_{rel}$ is the 
single--particle relaxation or scattering time.

Next we define a Grassmannian matrix field $B_{12}$, with 
$1\equiv (n_1,\alpha_1)$, etc, by
\begin{mathletters}
\label{eqs:2.18}
\begin{equation}
B_{12}({\bf x}) = \eta^{+}_{1}({\bf x})\otimes\eta_{2}({\bf x})\quad,
\label{eq:2.18a}
\end{equation}
or, in Fourier space and with all components written out,
\begin{equation}
{_i{^jB}}_{nm}^{\alpha\beta}({\bf q}) = \sum_{{\bf k}}\,{_i\eta}_{n}^{\alpha}
({\bf k})\ {^j\eta}_{m}^{\beta}({\bf k}+{\bf q})\quad,
\label{eq:2.18b}
\end{equation}
\end{mathletters}%
The subset of the Grassmann algebra
that is given by the bilinear fields $B$ is isomorphic to a set of classical,
i.e. c-number valued, matrix fields. We exploit this isomorphism, which in 
particular maps the adjoint operation in $\psi$-space that we denoted by an 
overbar onto the complex conjugation operation, by
introducing a classical matrix field $Q$, and constraining $B$ to $Q$ by means
of a functional delta function.\cite{ConstraintFootnote}
We then use a functional integral
representation of the latter, with an auxiliary or ghost field $\tilde\Lambda$
that plays the role of a Lagrange multiplier, and integrate out the
fermions. This way we obtain,
\begin{eqnarray}
\tilde{Z} &=&\int D[\bar{\psi},\psi]\ e^{{\tilde S}[\bar\psi,\psi]}
                                                 \int D[Q]\ \delta [Q-B] 
\nonumber \\
&=&\int D[\eta]\ e^{{\tilde S}[\eta]}\int D[Q]\,D[\tilde\Lambda]\ 
   \exp \left(\tr\left[\tilde\Lambda (Q-B)\right]\right) 
\nonumber \\
&=&\int D[Q]\,D[\tilde\Lambda]\ e^{{\cal A}[Q,\tilde\Lambda]}
   \quad.
\label{eq:2.19}
\end{eqnarray}
In Eq.\ (\ref{eq:2.19}) we have defined an effective action,
\begin{eqnarray}
{\cal A}[Q,\tilde\Lambda] &=&{\cal A}_{dis}[Q] + {\cal A}_{int}[Q]+\frac{1}{2}\,
   \Tr\ln\left(G_0^{-1}-i\tilde\Lambda\right)
\nonumber\\
&&+\int d{\bf x}\ \tr\left(\tilde\Lambda({\bf x})\,Q({\bf x})\right)\quad.
\label{eq:2.20}
\end{eqnarray}
Here and in what follows, $\tr$ is a trace over all discrete degrees of 
freedom that are not shown explicitly, while $\Tr$
is a trace over all degrees of freedom, including an integral over ${\bf x}$.
\begin{equation}
G_0^{\,-1} = -\partial_{\tau} +\nabla^2 /2m+\mu \quad,
\label{eq:2.21}
\end{equation}
is the inverse of the free electron Green operator, and it is clear
from the structure of the $\Tr\ln$-term in Eq.\ (\ref{eq:2.20}) that the
physical interpretation of the ghost field is that of a self--energy. In
writing Eq.\ (\ref{eq:2.20}), we have replaced the dummy integration
variables $\tilde\Lambda$ and $Q$ by their transposeds, which carries a
Jacobian of unity, and have anticipated the fact that 
${\cal A}_{dis}[Q^T] = {\cal A}_{dis}[Q]$, and the same for ${\cal A}_{int}$. 
The latter two contributions to the action read
\begin{mathletters}
\label{eqs:2.22}
\begin{equation}
{\cal A}_{dis}[Q] = {\cal A}_{dis}^{\,(1)}[Q] + {\cal A}_{dis}^{\,(2)}[Q]\quad, 
\label{eq:2.22a}
\end{equation}
\begin{equation}
{\cal A}_{dis}^{\,(1)}[Q] = \frac{-1}{2\pi N_F\tau_{1}}\int d{\bf x}\ 
                       \Bigl(\tr Q({\bf x})\Bigr)^2\quad, 
\label{eq:2.22b}
\end{equation}
\begin{equation}
{\cal A}_{dis}^{\,(2)}[Q] = \frac{1}{\pi N_F\tau_{rel}}\int d{\bf x}\ 
                       \tr \bigl(Q({\bf x})\bigr)^2\quad, 
\label{eq:2.22c}
\end{equation}
\end{mathletters}%
and
\begin{mathletters}
\label{eqs:2.23}
\begin{equation}
{\cal A}_{int}[Q] = {\cal A}_{int}^{\,(s)} + {\cal A}_{int}^{\,(t)} 
                                           + {\cal A}_{int}^{\,(c)}\quad, 
\label{eq:2.23a}
\end{equation}
\begin{eqnarray}
{\cal A}_{int}^{\,(s)}&=&\frac{T\Gamma^{(s)}}{2}\int d{\bf x}\sum_{r=0,3}(-1)^r
\sum_{n_1,n_2,m}\sum_\alpha  
\nonumber\\
&&\times\left[\tr \left((\tau_r\otimes s_0)\,Q_{n_1,n_1+m}^{\alpha\alpha}
({\bf x})\right)\right]
\nonumber\\
&&\times\left[\tr \left((\tau_r\otimes s_0)\,Q_{n_2+m,n_2}^{\alpha\alpha}
({\bf x})\right)\right]\quad,
\label{eq:2.23b}
\end{eqnarray}

\begin{eqnarray}
{\cal A}_{int}^{\,(t)}&=&\frac{T\Gamma^{(t)}}{2}\int d{\bf x}\sum_{r=0,3}(-1)^r
\sum_{n_1,n_2,m}\sum_\alpha\sum_{i=1}^3
\nonumber\\
&&\times\left[\tr\left((\tau_r\otimes s_i)\,Q_{n_1,n_1+m}^{\alpha\alpha}
({\bf x})\right)\right]
\nonumber\\
&&\times\left[\tr\left((\tau_r\otimes s_i)\,Q_{n_2+m,n_2}^{\alpha\alpha}
({\bf x})\right)\right]\quad,
\label{eq:2.23c}
\end{eqnarray}

\begin{eqnarray}
{\cal A}_{int}^{\,(c)} &=&\frac{T\Gamma^{(c)}}{2}\int d{\bf x}\sum_{r=1,2}
                    \sum_{n_1,n_2,m}\sum_{\alpha}
\nonumber\\
&&\times\left[\tr\left((\tau_r\otimes s_0)\,Q_{n_1,-n_1+m}^{\alpha\alpha}
({\bf x})\right)\right]
\nonumber\\
&&\times\left[\tr\left((\tau_r\otimes s_0)\,Q_{-n_2,n_2+m}^{\alpha\alpha}
({\bf x})\right)\right]\quad.
\label{eq:2.23d}
\end{eqnarray}
\end{mathletters}%

\subsubsection{Properties of the $Q$-matrices}
\label{subsubsec:II.B.2}

We now derive some useful properties of the matrix field $Q$. Since $B$,
Eqs.\ (\ref{eqs:2.18}), is self-adjoint under the adjoint operation defined in
Eq.\ (\ref{eq:2.13b}) and denoted by a superscript `$+$', so is $Q$, $Q^+ = Q$.
Notice that $Q^+$ is distinct from the hermitian conjugate of $Q$, which we
write as $Q^{\dagger}\equiv (Q^T)^*$, with the star denoting complex
conjugation. The definition of the former adjoint implies $Q^+ = C^T Q^T C$, 
with
\begin{equation}
^{ij}C_{nm}^{\alpha\beta} = \delta_{nm}\,\delta_{\alpha\beta}\,c_{ij}\quad,
\label{eq:2.24}
\end{equation}
and $c$ defined in Eq.\ (\ref{eq:2.13c}). We thus have
\begin{mathletters}
\label{eqs:2.24'}
\begin{equation}
Q = C^T\,Q^T\,C\quad.
\label{eq:2.24'a}
\end{equation}
In addition, a direct calculation of the hermitian conjugate of $B$,
$B^{\dagger} \equiv \overline{B^T}$, reveals another constraint that is 
inherited by the $Q$, namely
\begin{equation}
Q^{\dagger} = -\Gamma\,Q\,\Gamma^{-1}\quad,
\label{eq:2.24'b}
\end{equation}
where the similarity transformation denoted by $\Gamma$ has the property
\begin{equation}
(\Gamma\,Q\,\Gamma^{-1})_{nm} = Q_{-n-1,-m-1}\quad.
\label{eq:2.24'c}
\end{equation}
\end{mathletters}%
We now expand our matrix fields in the spin--quaternion basis defined after
Eqs.\ (\ref{eqs:2.13}),
\begin{mathletters}
\label{eqs:2.25}
\begin{eqnarray}
Q_{12}({\bf x})&=&\sum_{r,i=0}^{3}\,\left(\tau_r\otimes s_i\right)
   \,{_r^iQ}_{12}({\bf x})\quad,
\label{eq:2.25a}\\
\tilde\Lambda_{12}({\bf x})&=&\sum_{r,i=0}^{3}\,\left(\tau_r\otimes s_i
   \right)\,{_r^i\tilde\Lambda}_{12}({\bf x})\quad.
\label{eq:2.25b}
\end{eqnarray}
\end{mathletters}%
where again $1\equiv (n_1,\alpha_1)$, etc. 
In this basis, the relations expressed by Eq.\ (\ref{eq:2.24'a})
imply the following symmetry properties,
\begin{mathletters}
\label{eqs:2.26}
\begin{eqnarray}
{^0_r Q}_{12}&=&(-)^r\,{^0_r Q}_{21}\quad,\quad (r=0,3)\quad,
\label{eq:2.26a}\\
{^i_r Q}_{12}&=&(-)^{r+1}\,{^i_r Q}_{21}\ ,\ (r=0,3;\ i=1,2,3)\quad,
\label{eq:2.26b}\\
{^0_r Q}_{12}&=&{^0_r Q}_{21}\quad,\quad (r=1,2)\quad,
\label{eq:2.26c}\\
{^i_r Q}_{12}&=&-{^i_r Q}_{21}\quad,\quad (r=1,2;\ i=1,2,3)\quad.
\label{eq:2.26d}
\end{eqnarray}
\end{mathletters}%
Together with the behavior under hermitian conjugation, Eq.\ (\ref{eq:2.24'b}),
this further implies
\begin{mathletters}
\label{eqs:2.27}
\begin{eqnarray}
{^i_r Q}_{12}^*&=&- {^i_r Q}_{-n_1-1,-n_2-1}^{\alpha_1\alpha_2}\quad,\quad 
                                                               (r=0,3)\quad,
\label{eq:2.27a}\\
{^i_r Q}_{12}^*&=& {^i_r Q}_{-n_1-1,-n_2-1}^{\alpha_1\alpha_2}\quad,\quad 
                                                               (r=1,2)\quad,
\label{eq:2.27b}
\end{eqnarray}
\end{mathletters}%
Analogous relations hold for $\tilde\Lambda$ by virtue of the linear
coupling between $Q$ and $\tilde\Lambda$.

This completes the derivation of the action in terms of classical matrix
fields. We reiterate that for disordered fermions, this action is
adequate as long as one is interested in physics dominated by soft modes, 
while for clean systems, a complete theory would
have to keep an infinite set of interaction constants. We also recall
that, by means of the $Q$-matrices, we keep explicitly only density
modes in both the particle--hole and the particle--particle channels.
While this is again physically well motivated in the disordered case,
it is more problematic in the clean case. Due to the larger number
of soft modes in the latter, our procedure means that in certain clean
system soft modes have been integrated out. While this integrating--out
procedure is exact, it leads to undesirable properties of the effective
field theory as we will discuss below. In the next subsection we will
find further indications for it being advantageous to derive somewhat
different effective theories for clean and disordered fermions,
respectively.

\subsection{Saddle--point solutions and Gaussian approximation}
\label{subsec:II.C}

The physical degrees of freedom are now given by the matrix elements of
the $Q$-matrices, and the physical correlation functions for number and 
spin density fluctuations can be expressed
in terms of $Q$-correlation functions. This implies that ultimately
we will want to integrate out the ghost field $\tilde\Lambda$ in 
Eq.\ (\ref{eq:2.20}). Before doing so, however,
we examine a saddle--point solution to the field theory defined by 
Eqs.\ (\ref{eq:2.20}) - (\ref{eqs:2.23}),
and the Gaussian fluctuations about this saddle point.

\subsubsection{Fermi--liquid saddle--point solution}
\label{subsubsec:II.C.1}

It is possible to develop a systematic theory for the expectation values of
$Q$ and $\tilde\Lambda$, rather than simply a saddle--point 
approximation for these quantities. We will come back to this point in
Sec.\ \ref{subsubsec:III.B.1} below.
We further note that there are many saddle--point solutions, 
and that in general 
they have different symmetry properties. Physically these different solutions 
correspond to states with different broken symmetries such as Fermi--liquid 
states, magnetically ordered states, or superconducting states. For now
we concentrate on a Fermi--liquid state.
In Sec.\ \ref{subsubsec:III.B.4} below, and 
elsewhere,\cite{ustbp} we will consider more exotic states.

The saddle--point condition is
\begin{equation}
\frac{\delta {\cal A}}{\delta Q}\bigg\vert_{Q_{sp},\tilde\Lambda _{sp}} 
= \frac{\delta {\cal A}}{\delta\tilde\Lambda}\bigg\vert_{Q_{sp},
                      \tilde\Lambda _{sp}} = 0\quad.
\label{eq:2.28}
\end{equation}
The Fermi--liquid saddle--point solution is spatially uniform, diagonal in
frequency and replica space, and there are no nonzero expectation values 
that would describe a spontaneous magnetization, or a BCS-like 
order parameter. In Fermi liquid--like phases, the saddle point values of both 
$Q$ and $\tilde\Lambda$ have the structures,
\begin{mathletters}
\label{eqs:2.29}
\begin{eqnarray}
{_r^iQ}_{12}({\bf x})\Bigl\vert_{sp}&=&\delta_{12}
   \,\delta_{r0}\,\delta_{i0}\,Q_n\quad,
\label{eq:2.29a}\\
{_r^i\tilde\Lambda}_{12}({\bf x})\Bigl\vert_{sp}&=&\delta_{12}\,\delta_{r0}\,
                                                \delta_{i0}\,\Lambda_n\quad,
\label{eq:2.29b}
\end{eqnarray}
\end{mathletters}%
and the average values of $Q$ and $\tilde\Lambda$ have the same properties. 
Using Eqs.\ (\ref{eq:2.20}) - (\ref{eqs:2.29}), we obtain the saddle
point equations
\begin{mathletters}
\label{eqs:2.30}
\begin{equation}
Q_n=\frac{i}{2V}\,\sum_{\bf p}\,\left[ i\omega_n - {\bf p}^2/2m 
                                 + \mu -i\Lambda_n\right]^{-1}\quad, 
\label{eq:2.30a}
\end{equation}
\begin{equation}
\Lambda_n = \frac{-2}{\pi N_F\tau_{rel}}\,Q_n 
           - 4T\Gamma^{(s)}\sum_{m}\,e^{i\omega_m 0}\,Q_{m}\quad.
\label{eq:2.30b}
\end{equation}
\end{mathletters}%
Here $e^{i\omega_m 0}$ is the usual convergence factor that resolves the
ambiguity presented by the equal-time Green function.\cite{FetterWalecka}
Notice that this saddle--point solution obeys all of the symmetry properties
of the $Q$ that were derived in Sec.\ \ref{subsubsec:II.B.2}. For a physical
interpretation of Eqs.\ (\ref{eqs:2.30}), we define
a self-energy, $\Sigma_{n} = i\Lambda_{n}$. In terms of $\Sigma$, the 
Eqs. (\ref{eqs:2.30}) can be rewritten,
\begin{eqnarray}
\Sigma_n&=&\frac{1}{\pi N_F\tau_{rel}}\,\frac{1}{V}\,
      \sum_{\bf p}\,\left[ i\omega_n 
    - {\bf p}^2/2m + \mu - \Sigma_{n}\right]^{-1} 
\nonumber\\
        &&+2\Gamma^{(s)}T\sum_{m}\,e^{i\omega_m 0}\,\frac{1}{V}\,
          \sum_{\bf p}\,\bigl[i\omega_{m}
          - {\bf p}^2/2m
\nonumber\\
        &&\qquad\qquad\qquad\qquad\qquad + \mu - \Sigma_{m}\bigr]^{-1}\quad.
\label{eq:2.31}
\end{eqnarray}
This integral equation for the frequency dependent self--energy has a familiar
structure. For vanishing interaction, $\Gamma^{(s)}=0$, Eq.\ ({\ref{eq:2.31}) 
reduces to the self--consistent Born approximation for the self--energy, or
the inverse mean--free time, in
a disordered fermion system. At zero disorder, on the other hand, 
Eq.\ (\ref{eq:2.31}) represents the
Hartree--Fock approximation for the self--energy. The full Eq.\ (\ref{eq:2.31})
we will refer to as the disordered Hartree--Fock approximation for the
self--energy. It interpolates smoothly between the clean and dirty cases.

In this context, a few remarks on ultraviolet divergencies and 
cutoffs are in order. 
In general dimensions, the sums or integrals in Eqs. (\ref{eqs:2.30}) and
(\ref{eq:2.31}) do not exist due to
ultraviolet divergencies. Clearly, this is due to the long--wavelength and
low--frequency approximations we have made by, for example, projecting onto
density modes only. Any ultraviolet problems encountered in the theory are
therefore artifacts, and should be cut off at some momentum scale $k_0$. In
principle there are a number of `large' momentum scales that are
candidates for $k_0$, e.g. the Fermi wave number $k_F$, the Thomas--Fermi 
wave number
$k_{TF}$, and the inverse elastic mean--free path $1/\ell$.
For metallic densities, $k_F\approx k_{TF}$, while $k_F \ell >> 1$
for weakly disordered systems, and $k_F\ell\approx 1$ for strongly disordered
ones. A popular choice is to use $k_0 = 2\pi/\ell$ in any disordered system, 
regardless of the strength of the disorder.\cite{Mott}
However, for weakly disordered systems this choice leads to qualitatively
wrong answers for various nonuniversal results, while using $k_0 = k_F$
is qualitatively correct.\cite{Cutoff} For the universal
phenomena that will be considered in the present paper, on the other hand, 
the precise value of $k_0$ is not important, and any of the above choices
are acceptable.

\subsubsection{Gaussian approximation}
\label{subsubsec:II.C.2}

We now examine the Gaussian fluctuations about the saddle point
discussed above. To this end, we write $Q$ and $\tilde\Lambda$ in 
Eqs.\ (\ref{eq:2.20}) - (\ref{eqs:2.23}) as,
\begin{mathletters}
\label{eqs:2.32}
\begin{eqnarray}
Q = Q_{sp} + \delta Q\quad,
\label{eq:2.32a}\\
\tilde\Lambda = \tilde\Lambda_{sp} + \delta\tilde\Lambda\quad,
\label{eq:2.32b}
\end{eqnarray}
\end{mathletters}%
and then expand to second or Gaussian order in the fluctuations $\delta Q$ and
$\delta\tilde\Lambda$. Denoting the constant saddle point contribution to the
action by ${\cal A}_{sp}$, and the Gaussian action by ${\cal A}_G$, we have
\begin{equation}
{\cal A}[Q,{\tilde\Lambda}] = {\cal A}_{sp} + {\cal A}_G[Q,{\tilde\Lambda}] 
                         + \Delta {\cal A}\quad,
\label{eq:2.32'}
\end{equation}
where $\Delta {\cal A}$ contains all terms of higher than quadratic order in the
small quantities $\delta Q$ and $\delta\tilde\Lambda$. The Gaussian action
reads,
\begin{mathletters}
\label{eqs:2.33}
\begin{eqnarray}
{\cal A}_G[Q,\tilde\Lambda]&=&{\cal A}_{dis}[\delta Q]
                             + {\cal A}_{int}[\delta Q]
  +\frac{1}{4}\,\Tr\left(G_{sp}\delta{\tilde\Lambda}G_{sp}\,
                                               \delta{\tilde\Lambda}\right)
\nonumber\\
&&+\int d{\bf x}\ \tr\left(\delta{\tilde\Lambda}({\bf x})\,\delta Q({\bf x})
                                                             \right)\quad,
\label{eq:2.33a}
\end{eqnarray}
with $G_{sp}$ the single--particle Green function in saddle--point
approximation,
\begin{equation}
G_{sp}({\bf p},\omega_n) = \left[i\omega_n - {\bf p}^2/2m + \mu -
   \Sigma_n\right]^{-1}\quad, 
\label{eq:2.33b}
\end{equation}
\end{mathletters}%
where $\Sigma_n$ is given by the solution of Eq.\ (\ref{eq:2.31}). The spectrum
of this saddle--point Green function determines the quantity $N_F$, see
Eq.\ (\ref{eq:2.42'}) below.

At this point it is important to notice that in the full action, 
Eq.\ (\ref{eq:2.32'}), the symmetry properties expressed by 
Eqs.\ (\ref{eqs:2.26}) are enforced by means of delta-function constraints.
Truncating the theory at the Gaussian level spoils this property.
Consequently, by using the constraints in different ways {\em before} the
truncation one obtains different Gaussian theories. Here we opt to use
the constraints to rewrite the theory entirely in terms of matrix
elements $\delta Q_{12}$ with $n_1 \geq n_2$. Correlation functions of
matrix elements that do not obey these frequency restrictions are related
to Eqs.\ (\ref{eqs:2.34}) below by means of Eqs.\ (\ref{eqs:2.26}). As we
will see, this
choice results in a Gaussian theory that reproduces results that are
well-known from many-body perturbation theory.
With this procedure, Eq.\ (\ref{eq:2.33a}) is a quadratic form that can be 
diagonalized using the
techniques discussed in Ref.\ \onlinecite{R}. One obtains
\begin{mathletters}
\label{eqs:2.34}
\begin{eqnarray}
\Bigl\langle{_r^i(\delta Q)}_{12}({\bf p}_1)\,{_s^j(\delta Q)}_{34}
   ({\bf p}_2)\Bigr\rangle_G &=&\frac{V}{16}\,
     \delta_{{\bf p}_1,-{\bf p}_2}\,\delta_{rs}\,\delta_{ij}
\nonumber\\
 &&\times\,{_r^iM}_{12,34}^{-1}({\bf p}_1)\ ,
\label{eq:2.34a}
\end{eqnarray}
\begin{eqnarray}
\Bigl\langle{_r^i{(\delta{\bar\Lambda})}}_{12}({\bf p}_1)\,
   {_s^j{(\delta{\bar\Lambda})}}_{34}({\bf p}_2)\Bigr\rangle_G
    &=&\frac{V}{16}\,\delta_{{\bf p}_1,-{\bf p}_2}\,\delta_{rs}\,\delta_{ij}
\nonumber\\
 &&\times\,{_r^iN}_{12,34}^{-1}(p_1)\ .
\label{eq:2.34b}
\end{eqnarray}
\end{mathletters}%
Here $\left\langle\ldots\right\rangle_G$ denotes an average
with a weight ${\cal A}_G$.
In Eq.\ (\ref{eq:2.34b}), we have defined a correlation
function for the fluctuations of a field, $\bar\Lambda$, 
that is decoupled from the
$Q$-correlation function. $\bar\Lambda$ is defined by
\begin{mathletters}
\label{eqs:2.35}
\begin{equation}
\bar\Lambda_{12} = \frac{1}{2}\,\varphi_{n_1n_2}\,{\tilde\Lambda}_{12} 
                   + Q_{12}\quad,
\label{eq:2.35a}
\end{equation}
with
\begin{equation}
\varphi_{nm}({\bf k}) = \frac{1}{V}\sum_{\bf p}\,
      G_{sp}({\bf p},\omega_{n})\,G_{sp}({\bf p}+{\bf k,}\omega_{m})\quad.
\label{eq:2.35b}
\end{equation}
\end{mathletters}%
The matrix inverse $M^{-1}$ for the particle-hole channel is given by,
\begin{mathletters}
\label{eqs:2.36}
\begin{eqnarray}
{_{0,3}^iM}_{12,34}^{-1}({\bf p})&=&\delta_{1-2,3-4}\ {^i_{0,3}I}_{12}\,
    \Bigl[\delta_{13}\,{\cal D}_{n_1n_2}({\bf p})
\nonumber\\
&& - \delta_{\alpha_1\alpha_2}\,\delta_{\alpha_3\alpha_4}\,2T\Gamma^{(i)}\,
   {\cal D}_{n_1n_2}^{(i)}({\bf p})\,{\cal D}_{n_3n_4}({\bf p})\Bigr]
\nonumber\\
&&-\frac{4\delta_{r0}\delta_{i0}}{\pi N_F\tau_{1}}\ 
                                                     \delta_{12}\delta_{34}
  {\cal D}_{n_1n_1}^{(s)}({\bf p}){\cal D}_{n_3n_3}^{(s)}({\bf p})\quad,
\nonumber\\
\label{eq:2.36a}
\end{eqnarray}
with
\begin{equation}
{^i_{0,3}I}_{12} = 1 + \delta_{n_1n_2}\bigl(1-2\delta_{r3}\bigr)\,
                       \Bigl(\delta_{i0} - \sum_{j=1}^{3}\delta_{ij}\Bigr)\quad,
\label{eq:2.36b}
\end{equation}
and $\Gamma^{(0)}=-\Gamma^{(s)}$, and $\Gamma^{(1,2,3)}=\Gamma^{(t)}$.
For the particle-particle channel, one finds,
\begin{eqnarray}
_{1,2}^iM_{12,34}^{-1}&(&{\bf p}) = -\delta _{1+2,3+4}\Bigl[\delta_{13}\ 
   {^i_{1,2}I}_{12}\,{\cal D}_{n_1n_2}({\bf p})
\nonumber\\
&-&\delta_{i0}\,\delta_{\alpha_1\alpha_2}\,\delta_{\alpha_3\alpha_4}\,
   4T\Gamma^{(c)}{\cal D}_{n_1n_2}^{(c)}
   ({\bf p})\,{\cal D}_{n_3n_4}({\bf p})\Bigr]\ ,
\nonumber\\
\label{eq:2.36c}
\end{eqnarray}
where
\begin{equation}
{^i_{1,2}I}_{12} = 1 + \delta_{n_1n_2}\Bigl(\delta_{i0} 
                           - \sum_{j=1}^{3}\delta_{ij}\Bigr)\quad.
\label{eq:2.36d}
\end{equation}
Similarly, one finds for the matrix inverse $N^{-1}$ in the particle-hole
channel,
\begin{equation}
{_{0,3}^iN}_{12,34}^{-1}({\bf p}) = -\delta_{13}\,\delta_{24}\ 
       {^i_{0,3}I}_{12}\ \varphi_{n_1n_2}({\bf p})\quad,
\label{eq:2.36e}
\end{equation}
and in the particle-particle channel,
\begin{equation}
{_{1,2}^iN}_{12,34}^{-1}({\bf p}) = \delta_{13}\,\delta_{24}\ 
        {^i_{1,2}I}_{12}\ \varphi_{n_1n_2}({\bf p})\quad,
\label{eq:2.36f}
\end{equation}
\end{mathletters}%
In the preceding equations,
\begin{mathletters}
\label{eqs:2.37}
\begin{equation}
{\cal D}_{n_1n_3}({\bf p}) = \varphi_{n_1n_2}({\bf p})\,
   \left[1 - \frac{1}{\pi N_F\tau_{rel}}\,\varphi_{n_1n_2}({\bf p})\right]^{-1}
                                                                    \quad,
\label{eq:2.37a}
\end{equation}
and,
\begin{eqnarray}
{\cal D}_{n_1n_2}^{(i)}({\bf p})&=&{\cal D}_{n_1n_2}({\bf p})\,
   \biggl[1+2T\Gamma^{(i)}
\nonumber\\
&&\qquad\quad\times\sum_{n}{\cal D}_{n,n+n_2-n_1}({\bf p})\biggr]^{-1}\quad,
\label{eq:2.37b}
\end{eqnarray}
with ${\cal D}^{(0)} = {\cal D}^{(s)}$, and 
${\cal D}^{(1)} = {\cal D}^{(2)} = {\cal D}^{(3)} = {\cal D}^{(t)}$, and
\begin{eqnarray}
{\cal D}_{n_1n_2}^{(c)}({\bf p})&=&{\cal D}_{n_1n_2}({\bf p})\,
   \biggl[1+2T\Gamma^{(c)}
\nonumber\\
&&\qquad\quad\times\sum_{n} {\cal D}_{n,n_1+n_2-n}({\bf p})\biggr]^{-1}\,.
\nonumber\\
\label{eq:2.37c}
\end{eqnarray}
\end{mathletters}%

\subsection{Physical correlation functions}
\label{subsec:II.D}

We conclude this section by interpreting the results obtained in the preceding
subsection, and expressing some physical quantities of interest in terms
of $Q$-correlations. 

First of all, $\varphi_{nm}({\bf k})$, Eq.\ (\ref{eq:2.35b}), 
is related to a disordered Lindhard function. We write
\begin{equation}
\varphi_{nm}({\bf k}) = \Theta (nm)\Phi + \Theta (-nm)
                                 \varphi ({\bf k},\Omega_{n-m})\quad,
\label{eq:2.38}
\end{equation}
with $\Omega_n = 2\pi Tn$ a bosonic Matsubara frequency.
For both clean and disordered systems, $\Phi$ approaches a finite constant
in the limit
of vanishing wave number and frequency. On the other hand, the behavior of
$\varphi$ in the same limit depends qualitatively
on the presence or absence of disorder.
For dirty systems, to leading order as $\tau_{rel}\rightarrow\infty$,
\begin{equation}
\varphi (\vert{\bf k}\vert\rightarrow 0,\Omega_n\rightarrow 0) =
                                                 \pi N_F\tau_{rel}\quad,
\label{eq:2.39}
\end{equation}
while for clean systems, $\varphi$ is singular in the long--wavelength,
low--frequency limit. For example, in $d=3$,
\begin{equation}
\varphi (\vert{\bf k}\vert\rightarrow 0,\Omega_n\rightarrow 0) =
     \frac{i\pi N_F}{2\vert{\bf k}\vert v_F}\,
        \sgn (\Omega_n)\ln\frac{i\Omega_n + \vert{\bf k}\vert v_F}
                               {i\Omega_n - \vert{\bf k}\vert v_F}\ .
\label{eq:2.40}
\end{equation}
Comparing Eqs.\ (\ref{eq:2.39}) and (\ref{eq:2.40}), we see that the
$1/\vert{\bf k}\vert$ and $1/\Omega_n$ singularities that occur
in the clean case
are cut off by a mean--free path, and a relaxation time, respectively, in
the disordered case. This implies that the $\bar\Lambda$-$\bar\Lambda$
correlation function is soft for clean systems, and massive for dirty
ones. This will turn out to be a fundamental difference between
clean and dirty systems, which
reflects the fact that single particle excitations are
soft in a clean system. These observations indicate that for the further
development of the theory it will be useful
to treat clean and disordered systems in somewhat different ways.

$\varphi$ determines the propagator $\cal D$, Eq.\ (\ref{eq:2.37a}). For
finite disorder, $n_1n_2\leq 0$, and in the long--wavelength, low--frequency
limit one finds
\begin{equation}
{\cal D}_{n_1n_2}({\bf p}) = \frac{\pi N_F}{D{\bf p}^2 + \vert\Omega_{n_1-n_2}
                                                          \vert}\quad, 
\label{eq:2.41}
\end{equation}
with $D=v_F^2\tau_{rel}/d$ a bare or Boltzmann diffusion 
coefficient.\cite{TransportTimeFootnote}
The propagators given by Eqs.\ (\ref{eq:2.34a}) and 
(\ref{eq:2.36a}) for $n_1n_2<0$ are
then the soft or massless propagators of the nonlinear sigma--model
that is used to
describe the metal--insulator transition in disordered interacting electronic
systems.\cite{Finkelstein,R}

Second, the physical correlation functions of the single--particle spectral
density, the particle number density, and the spin density, can be easily
expressed in terms
of the $Q$-correlation functions. By keeping an appropriate source
in the action while performing the transformation to
$Q$-variables, we obtain an expression for the density of states, $N$, as a
function of energy or frequency $\omega$,
\begin{equation}
N(\epsilon_F + \omega) = \frac{4}{\pi}\,{\rm Re}\,\Bigl\langle{^0_0 Q}_{nn}
        ({\bf x})\Bigr\rangle\Bigr\vert_{i\omega_n\rightarrow\omega + i0}
                                                                \quad.
\label{eq:2.42}
\end{equation}
In saddle point approximation, we have for the density of states at the
Fermi level,
\begin{equation}
N_F = \frac{-2}{\pi}\,\frac{1}{V}\,\sum_{\bf p} {\rm Im}\,
    G_{sp}({\bf p},i\omega_n\rightarrow i0)\quad.
\label{eq:2.42'}
\end{equation}
This defines the density of states that has been used throughout as an
energy normalization factor.
Similarly, one finds that the number density susceptibility, 
$\chi_n$, and the spin density susceptibility, $\chi_s$, are given by,
\begin{eqnarray}
\chi^{(i)}({\bf q},\Omega_n) = 16T\sum_{m_1,m_2}\sum_{r=0,3}\left\langle
{_r^i(\Delta Q)}_{m_1-n,m_1}^{\alpha\alpha}({\bf q})\right.
\nonumber\\
\times\left. {_r^i(\Delta Q)}_{m_2,m_2+n}^{\alpha\alpha}(-{\bf q})
      \right\rangle\quad , 
\label{eq:2.43}
\end{eqnarray}
with $\Delta Q = Q - \langle Q \rangle$, and $\chi^{(0)} = \chi_n$ and
$\chi^{(1,2,3)} = \chi_s$. By substituting the Gaussian
propagator we see that the Gaussian theory yields $\chi_n$ and $\chi_s$ 
in a disordered RPA-like approximation. Notice that, for weak disorder, the 
last term on the right--hand side of Eq.\ (\ref{eq:2.36a}) 
is smaller than the first
two terms by at least a factor of $1/\tau_1$, and therefore does not appear
in the usual resummation schemes of many--body perturbation theory. Neglecting
it, one finds, for small $\vert {\bf q}\vert$ and $\Omega_n$,
\begin{mathletters}
\label{eqs:2.43'}
\begin{equation}
\chi^{(i)}({\bf q},\Omega_n) = \chi^{(i)}_{st}\ \frac{-D^{(i)}{\bf q}^2}
   {\vert\Omega_n\vert + D^{(i)}{\bf q}^2}\quad,
\label{eq:2.43'a}
\end{equation}
where
\begin{equation}
\chi^{(i)}_{st} = N_F/\bigl(1 - N_F\Gamma^{(i)}\bigr)\quad,
\label{eq:2.43'b}
\end{equation}
is the static susceptibility in the long--wavelength limit, renormalized by
the interaction, and
\begin{equation}
D^{(i)} = D\,\bigl(1 - N_F\Gamma^{(i)}\bigr)\quad,
\label{eq:2.43'c}
\end{equation}
\end{mathletters}%
is the renormalized diffusion coefficient. 
Here we have used the identity $T\sum_n {\cal D}_{nn}({\bf q}\rightarrow 0)
= -N_F/2$, which expresses particle number conservation.
Notice that for $i=1,2,3$,
Eqs.\ (\ref{eqs:2.43'}) describe a disordered Stoner criterion:
For $N_F\Gamma^{(t)}=1$, the spin diffusion coefficient vanishes, and the
static spin susceptibility diverges.
In the clean limit one recovers RPA proper,
\begin{equation}
\chi^{(i)}({\bf q},\Omega_n) = 2\chi_{0}({\bf q},\Omega_n)\ 
   \left[1+\Gamma^{(i)}2\chi_{0}({\bf q},\Omega_n)\right]^{-1}\quad,
\label{eq:2.44}
\end{equation}
where $\chi_{0}({\bf q},\Omega_n) = T\sum_m \varphi_{m-n,m}({\bf q})$ is
the free electron density susceptibility per spin.

\section{The disordered fermion system}
\label{sec:III}

In this section we concentrate on the disordered system, and in particular
perform
a symmetry analysis in order to identify the soft modes in our system. This
will allow us to explicitly separate the soft modes from the massive ones,
and to formulate an effective field theory for which the soft modes remain
manifestly soft to all orders in perturbation theory.
In Appendix \ref{app:B} we point out that all of the considerations and
results of this section have precise analogies within the 
$\phi^4$--representation of the $O(N)$ Heisenberg model. 
These analogies are often helpful,
since $\phi^4$--theory is technically much simpler than the matrix field
theory under consideration here, and hence it displays the basic structure
of the symmetry analysis and its consequences more clearly.

\subsection{Symmetry analysis}
\label{subsec:III.A}

\subsubsection{Basic transformation properties}
\label{subsubsec:III.A.1}

Let us perform a symmetry analysis of our field theory. To this end, we
start with the action in fermionic form, formulated in terms of
bispinors, Eqs.\ (\ref{eqs:2.15}) - (\ref{eqs:2.17}). 
Our plan of attack is to consider the
various terms in the action separately. The crucial symmetries will involve
the free electron part, ${\tilde S}_0$, and the disorder part, 
${\tilde S}_{dis}$. The interaction part, ${\tilde S}_{int}$, 
Eqs.\ (\ref{eqs:2.16}), will turn out to always be effectively proportional 
to a frequency, at least near the disordered Fermi--liquid FP that we will
identify below. Since we are interested in low--frequency effects, this will 
ultimately imply that ${\tilde S}_{int}$ does not change the conclusions 
with respect to which two--particle excitations are soft or massless that 
are reached by considering the noninteracting action.

For zero external frequency, $i\omega_n =0$ in Eq.\ (\ref{eq:2.15a}), 
the free fermion action is invariant under transformations
that leave invariant the expression
\begin{equation}
\sum_n \sum_{\alpha} \tr\,\Bigl( (\eta^+)_n^{\alpha} \otimes \eta_n^{\alpha}
       \Bigr)\equiv \left(\eta,\eta\right)\quad,
\label{eq:3.1}
\end{equation}
with $(\eta,\eta)$ a generalization of
the scalar product defined in Eq.\ (\ref{eq:2.14}) that includes summation
over the frequency and replica indices. From the structure of the
disorder part of the action, Eq.\ (\ref{eq:2.15b}), it is clear that 
${\tilde S}_{dis}$ is also invariant under transformations that leave 
invariant $(\eta,\eta)$.

Now let $\eta$ be transformed by means of an operator $\hat T$:
$\eta \rightarrow {\hat T}\eta$. Remembering that $\eta^+$ is related to
$\eta$ by means of the charge conjugation matrix $C$, Eq.\ (\ref{eq:2.24}),
and using $C^{\,T} = C^{-1}$, we find that in order
to leave $(\eta,\eta)$ invariant, ${\hat T}$ must obey
\begin{equation}
{\hat T}^{\,T}\,C\,{\hat T} = C\quad.
\label{eq:3.2}
\end{equation}
For a system with $2N$ frequency labels ($N$ positive ones, including $0$, and
$N$ negative ones), and $n$ replicas, Eq.\ (\ref{eq:3.2}) defines a
representation of the
symplectic group ${\rm Sp}(8Nn,{\cal C})$ over the complex
numbers $\cal C$.\cite{Wybourne} Throughout this section we will denote the
real, complex, and quaternion number fields by $\cal R$, $\cal C$, and
$\cal Q$, respectively. In what follows we will need to consider only
small subgroups of the $8Nn(8Nn+1)$-parameter group ${\rm Sp}(8Nn,{\cal C})$.

Now we ask what transformations of our composite variables $Q$ correspond
to the above transformations of the bispinors. Since the $Q$-matrix is
constrained to the bilinear fermion field $B = \eta^+\otimes\eta$,
see Eq.\ (\ref{eq:2.19}),
$\eta \rightarrow {\hat T}\eta$ implies $Q\rightarrow TQT^{-1}$, with
$T = C{\hat T}C^{\,T}$. The ghost field $\tilde\Lambda$ 
couples linearly to $Q$ and $B$, and therefore transforms like $Q$, 
i.e. by means of the matrices $T$.
Together with the ${\hat T}$, the transformations $T$ also obey 
Eq.\ (\ref{eq:3.2}), and therefore also form a representation of the group
${\rm Sp}(8Nn,{\cal C})$. 

Now we consider a specific transformation, namely a rotation in frequency
space given by,
\begin{eqnarray}
^i_r{\hat T}_{nm}^{\alpha\beta}&=&\delta_{i0}\,\delta_{r0}\,
                                             \delta_{\alpha\beta}\,
   \biggl\{\delta_{nm}\Bigl[1 + \left(\delta_{n n_1} + \delta_{n n_2}\right)
                                                 (\cos\theta - 1)\Bigr]
\nonumber\\
  &&\qquad\qquad + \left(\delta_{n n_1}\delta_{m n_2} 
                      - \delta_{n n_2}\delta_{m n_1}\right) \sin\theta\biggr\}
\nonumber\\
  &\equiv&\delta_{i0}\,\delta_{r0}\,\delta_{\alpha\beta}\,{\hat t}_{nm}\quad.
\label{eq:3.3}
\end{eqnarray}
This transforms a pair of spinors with frequency labels $n_1$ and $n_2$,
respectively, into linear combinations of the same pair, with a mixing
angle $\theta$. These transformations obey Eq.\ (\ref{eq:3.2}), and thus are 
elements of ${\rm Sp}(8Nn,{\cal C})$. For fixed $n_1$ and $n_2$ they represent
an ${\rm SO}(2)$ subgroup of ${\rm Sp}(8Nn,{\cal C})$. The corresponding 
transformations $T$ of the $Q$ and $\tilde\Lambda$-matrices are identical with
the ${\hat T}$, $T = {\hat T}$. Under an infinitesimal element of this
subgroup, the $Q$-matrices transform like
\begin{mathletters}
\label{eqs:3.4}
\begin{equation}
Q_{nm} \rightarrow Q_{nm} + \delta Q_{nm}\quad,
\label{eq:3.4a}
\end{equation}
with
\begin{eqnarray}
\delta Q_{nm} = \bigl(\delta_{n n_1} Q_{n_2 m} - \delta_{n n_2} Q_{n_1 m}
                    + \delta_{m n_1} Q_{n n_2}
\nonumber\\
    - \delta_{m n_2} Q_{n n_1}\bigr)\ \theta + O(\theta^2)\quad.
\label{eq:3.4b}
\end{eqnarray}
\end{mathletters}%
Here we have shown only the frequency indices, since all other degrees of
freedom are unaffected by the transformation. The $\tilde\Lambda$-matrices 
transform accordingly.

Of the various pieces of the action in $Q$-$\tilde\Lambda$ formulation,
Eq.\ (\ref{eq:2.20}), ${\cal A}_{dis}$ and 
$\int d{\bf x}\ \tr (\tilde\Lambda Q)$ are invariant under the above 
transformation, but $\Tr \ln (G_0^{-1} - i\tilde\Lambda)$ and 
${\cal A}_{int}$ are not. For now, let us consider the transformation 
of the former:
\begin{mathletters}
\label{eqs:3.5}
\begin{equation}
\Tr \ln \left(G_0^{-1} - i\tilde\Lambda\right) \rightarrow
   \Tr \ln \left(G_0^{-1} - i\tilde\Lambda\right) 
        + \theta\,\Tr (G\,\delta i\omega)\quad,
\label{eq:3.5a}
\end{equation}
with $G \equiv (G_0^{-1} - i\tilde\Lambda)^{-1}$, and
\begin{equation}
^i_r(\delta i\omega)_{nm}^{\alpha\beta} = \delta_{i0}\,\delta_{r0}\,
   \delta_{\alpha\beta}\,\left(\delta_{n n_1} \delta_{m n_2} + \delta_{n n_2} 
      \delta_{m n_1}\right)\,i\Omega_{n_1 - n_2}\,.
\label{eq:3.5b}
\end{equation}
\end{mathletters}%
The action of the noninteracting Fermi system therefore transforms like
\begin{equation}
{\cal A}_0 = {\cal A} - {\cal A}_{int} \rightarrow {\cal A}_0 + 
                                                    \delta {\cal A}_0 
           = {\cal A}_0 + \frac{\theta}{2}\,\Tr
                          (G\,\delta i\omega)\quad.
\label{eq:3.6}
\end{equation}
We will consider the transformation properties of ${\cal A}_{int}$ in 
Sec.\ \ref{subsubsec:III.A.3} below.

\subsubsection{A Ward identity for noninteracting electrons}
\label{subsubsec:III.A.2}

Because of the complexity of the interacting case, and for the reasons
noted at the beginning of the previous subsection, we first discuss
the noninteracting system. Let us introduce a source $J$ for the
$Q$-fields, and consider the partition function
\begin{equation}
Z[J] = \int D[Q]\,D[\tilde\Lambda]\ e^{{\cal A}_0 + \int d{\bf x}\ 
                                 \tr\bigl( J({\bf x})\,Q({\bf x})\bigr)}\quad.
\label{eq:3.7}
\end{equation}
By performing an infinitesimal transformation, Eq.\ (\ref{eq:3.4b}), on
the $Q$ and $\tilde\Lambda$-fields, one obtains from Eq.\ (\ref{eq:3.7}) the
identity,
\begin{mathletters}
\label{eqs:3.8}
\begin{eqnarray}
0 = \int D[Q]\,D[\tilde\Lambda]\ \left[\delta {\cal A}_0 + \int d{\bf x}\ 
       \tr\bigl(J({\bf x})\,\delta Q({\bf x})\bigr)\right]
\nonumber\\
   \times e^{{\cal A}_0 + \int d{\bf x}\ 
            \tr\bigl(J({\bf x})\,Q({\bf x})\bigr)}\quad.
\label{eq:3.8a}
\end{eqnarray}
Differentiating this identity with respect to $^0_0J_{n_2 n_1}^{\beta\alpha}$
and putting $J=0$ yields
\begin{equation}
0 = \Bigl\langle\delta {\cal A}_0\,{^0_0Q}_{n_1 n_2}^{\alpha\beta}({\bf x})
                                                              \Bigr\rangle
     + \Bigl\langle{^0_0(\delta Q)}_{n_1 n_2}^{\alpha\beta}({\bf x})
                                                              \Bigr\rangle\quad,
\label{eq:3.8b}
\end{equation}
\end{mathletters}%
where $\langle\ldots\rangle$ denotes an average by means of the action
${\cal A}_0$. From the explicit expression for $\delta {\cal A}_0$, 
Eq.\ (\ref{eq:3.6}),
we see that the first average is essentially a frequency times
$\langle G\,Q\rangle$. By using the identity\cite{GQIdentityFootnote}
$\langle G\,Q\rangle = -2i\langle Q^2\rangle$ we finally obtain
\begin{eqnarray}
8\Omega_{n_1 - n_2}\sum_{\gamma}\int d{\bf y}\ 
      \left\langle{^{0}_{0}Q}_{n_1 n_2}^{\gamma\gamma}({\bf y})\ 
              {^{0}_{0}Q}_{n_1 n_2}^{\alpha\beta}({\bf x})\right\rangle
\nonumber\\
 = \left\langle{^{0}_{0}Q}_{n_1 n_1}^{\alpha\beta}({\bf x})\right\rangle
  - \left\langle {^{0}_{0}Q}_{n_2 n_2}^{\alpha\beta}({\bf x})\right\rangle\quad.
\label{eq:3.9}
\end{eqnarray}

This is the desired Ward identity for noninteracting 
electrons,\cite{MaleevToperverg,SchaferWegner,PruiskenSchafer} which 
relates a two--point
$Q$-correlation function to the one--point function. For $\sgn n_1 = \sgn n_2$,
the right--hand side vanishes as $n_1 - n_2 \rightarrow 0$, and the identity
tells us nothing interesting. However, for $\sgn n_1 \neq \sgn n_2$ we see
from Eq.\ (\ref{eq:2.42}) that the right--hand side
approaches $\pi N(\epsilon_F)/2$, 
which is nonzero everywhere inside the
band. The correlation function on the left--hand side must therefore diverge.
Due to rotational invariance in replica space we have
$\langle Q^{\alpha\beta}\rangle \sim \delta_{\alpha\beta}$, and
$\langle Q^{\alpha\beta}\,Q^{\gamma\delta}\rangle \sim \delta_{\alpha\gamma}
\delta_{\beta\delta}$. We finally obtain, for $n_1 n_2 <0$ and
$\vert\Omega_{n_1 - n_2}\vert \rightarrow 0$,
\begin{equation}
\left\langle{^{0}_{0}Q}_{n_1 n_2}^{\alpha\alpha}({\bf k})\,
   {^{0}_{0}Q}_{n_1 n_2}^{\alpha\alpha}({\bf -k})\right\rangle
                                                    \bigg\vert_{{\bf k}=0}
 = \frac{\pi N(\epsilon_F)}{16\vert\Omega_{n_1 - n_2}\vert}\quad.
\label{eq:3.10}
\end{equation}
Here $N(\epsilon_F)$ is the exact density of states at the Fermi level of the
noninteracting electron system.
The salient point is that, as long as $N(\epsilon_F)>0$, 
the $Q$-$Q$ correlation function at zero momentum
diverges like $1/\vert\Omega_{n_1 - n_2}\vert$.
Equations\ (\ref{eq:2.34a}), (\ref{eq:2.36a}), and (\ref{eq:2.41}) show 
that the Gaussian propagator has this
property. The Ward identity ensures that it holds to all orders in
perturbation theory. We have therefore identified 
$^{0}_{0}Q_{n_1 n_2}^{\alpha\alpha}$ for $n_1 n_2 <0$ as a soft mode.
Rotational symmetry in
replica space implies that the ${^{0}_{0}Q}_{nm}^{\alpha\beta}$,
$nm<0$, $\alpha\neq\beta$, are also soft.
This can also be seen by applying the above
rotation procedure in frequency space directly to spinor pairs with
different replica indices.

From the Gaussian propagator, Eqs.\ (\ref{eq:2.34a}), (\ref{eqs:2.36}),
as well as from
physical considerations, one expects all channels to be soft, not just
the $r=0, i=0$ channel. This is indeed the case. These additional soft
modes are not controlled by separate Ward identities, but rather are
related to Eq.\ (\ref{eq:3.9}) by additional symmetries of the action.
In order to see this, we consider transformations
\begin{mathletters}
\label{eqs:3.11}
\begin{equation}
T_{nm}^{\alpha\beta} = \delta_{\alpha\beta}\,\delta_{nm}\,\bigl[\delta_{n n_2}
   x_r\left(\tau_r\otimes s_i\right) + \left(1-\delta_{n n_2}\right)
        \left(\tau_0\otimes s_0\right)\bigr]\ ,
\label{eq:3.11a}
\end{equation}
with $x_0 = x_3 = 1$, and $x_1 = x_2 = i$, and $n_2$ some fixed frequency
index. These $T$
obey Eq.\ (\ref{eq:3.2}), and in addition they are unitary.
It is easy to check that under these transformations $Q$ transforms like
\begin{equation}
^{0}_{0}Q_{n n_2} \rightarrow{^{i}_{r}Q}_{n n_2}\quad (n\neq n_2)\quad.
\label{eq:3.11b}
\end{equation}
\end{mathletters}%
If the action is invariant under such a transformation for given $i,r$,
then $\langle ^{i}_{r}Q\,^{i}_{r}Q\rangle=\langle ^{0}_{0}Q\,^{0}_{0}Q\rangle$,
and hence the two--point $Q$-correlations in the $i,r$ channel are also
soft. Obviously, the noninteracting actions is indeed invariant under
these transformations, so the two--point correlations are
soft in all channels.

Note that Eqs.\ (\ref{eqs:3.11}) imply that the spin--singlet ($i=0$) and
spin--triplet ($i=1,2,3$) channels are equal, and so are the particle--hole
($r=0,3$) and particle--particle ($r=1,2$) channels. In particular, they
provide a technical explanation for why the so--called Cooperons, i.e. the
two--point $Q$-correlation functions in the particle--particle channel,
are soft modes. We also mention that in the clean case, there still is
only one basic Ward identity, but many more additional symmetries than
in a disordered system.\cite{ustbp} These additional symmetries result in
many additional soft modes. This feature leads to the somewhat paradoxical
fact that it is easier to derive an effective field theory for disordered
electron systems than for clean ones.

\subsubsection{A Ward identity for interacting electrons}
\label{subsubsec:III.A.3}

Now we add the electron--electron interaction to our considerations. For
simplicity, we restrict ourselves to a discussion of the particle--hole
spin--singlet interaction; the discussion of the remaining interaction
channels proceeds analogously.

Let us consider again an infinitesimal element of the rotations in frequency 
space, Eq.\ (\ref{eq:3.3}).
From Eqs.\ (\ref{eq:2.23b}) and (\ref{eq:3.4b}) we find that the particle--hole
spin--singlet part of the action transforms like 
${\cal A}_{int}^{\,(s)}\rightarrow {\cal A}_{int}^{\,(s)} 
+ \delta {\cal A}_{int}^{\,(s)}$, with
\begin{eqnarray}
\delta {\cal A}_{int}^{\,(s)} = 16 \Gamma^{(s)} \int d{\bf x} \sum_{\gamma}
  \sum_{r=0,3}\ T\sum_{n_1' n_2'}{^{0}_{r}Q}_{n_1' n_2'}^{\gamma\gamma}({\bf x})
\nonumber\\
     \times\Bigl[{^{0}_{r}Q}_{n_1+(n_1'-n_2'),n_2}^{\gamma\gamma}({\bf x})
                 -{^{0}_{r}Q}_{n_1,n_2-(n_1'-n_2')}^{\gamma\gamma}({\bf x})
\nonumber\\
                 -{^{0}_{r}Q}_{n_2+(n_1'-n_2'),n_1}^{\gamma\gamma}({\bf x})
                 +{^{0}_{r}Q}_{n_2,n_1-(n_1'-n_2')}^{\gamma\gamma}({\bf x})
                                                        \Bigr]\ \theta\quad.
\label{eq:3.12}
\end{eqnarray}
Now we follow the same steps that led to Eq.\ (\ref{eq:3.8b}), except that
we differentiate with respect to ${^0_0J}_{n_4 n_3}^{\beta\alpha}$ with
$n_3>0, n_4<0$. This yields
\begin{eqnarray}
-\left\langle{^0_0\left(\delta Q({\bf x})\right)}_{n_3 n_4}^{\alpha\beta}
                                                               \right\rangle
  &=& \left\langle\delta {\cal A}_0\,{^0_0Q}_{n_3 n_4}^{\alpha\beta}({\bf x})
                                                               \right\rangle
\nonumber\\
   &&+ \left\langle\delta {\cal A}_{int}\,{^0_0Q}_{n_3 n_4}^{\alpha\beta}
                                                                   ({\bf x})
                                                         \right\rangle \quad.
\label{eq:3.13}
\end{eqnarray}
Choosing $n_1>0, n_2<0$, and using $\langle Q_{nm}\rangle \sim \delta_{nm}$
as well as Eq.\ (\ref{eq:3.12}), we obtain the Ward identity in
the form,
\begin{mathletters}
\label{eqs:3.14}
\begin{eqnarray}
W_{int}&+&8\Omega_{n_1 - n_2}\sum_{\gamma}\int d{\bf y}\
      \left\langle{^{0}_{0}Q}_{n_1 n_2}^{\gamma\gamma}({\bf y})\
              ^{0}_{0}Q_{n_3 n_4}^{\alpha\beta}({\bf x})\right\rangle
\nonumber\\
&=& \delta_{n_1 n_3}\delta_{n_2 n_4}\left(\left\langle{^{0}_{0}Q}_{n_1 n_1}
                                                               ^{\alpha\beta}
  ({\bf x})\right\rangle - \left\langle{^{0}_{0}Q}_{n_2 n_2}^{\alpha\beta}
                 ({\bf x})\right\rangle\right)\quad,
\nonumber\\
\label{eq:3.14a}
\end{eqnarray}
where,
\begin{eqnarray}
W_{int}&=&
-16\Gamma^{(s)}\sum_{\gamma}\int d{\bf y}\sum_{r=0,3}\ 
                                                     T\sum_{n_1' n_2'}
   \biggl\langle{^{0}_{0}Q}_{n_3 n_4}^{\alpha\beta}({\bf x})
\nonumber\\
&\times&{^{0}_{r}Q}_{n_1' n_2'}^{\gamma\gamma}({\bf y})\,
   \Bigl[{^{0}_{r}Q}_{n_1,n_2-(n_1'-n_2')}^{\gamma\gamma}({\bf y})
\nonumber\\
&&-{^{0}_{r}Q}_{n_1+(n_1'-n_2'),n_2}^{\gamma\gamma}({\bf y})
                 +{^{0}_{r}Q}_{n_2+(n_1'-n_2'),n_1}^{\gamma\gamma}({\bf y})
\nonumber\\
&& -{^{0}_{r}Q}_{n_2,n_1-(n_1'-n_2')}^{\gamma\gamma}({\bf y})
                                                   \Bigr]\biggr\rangle\quad,
\nonumber\\
\label{eq:3.14b}
\end{eqnarray}
with $n_1,n_3>0$, and $n_2,n_4<0$.

We now need to analyze the 3-point correlation functions in 
Eq.\ (\ref{eq:3.14b}), and ascertain that the $Q_{n_1 n_2}$ with
$n_1 n_2<0$ remain soft modes in their presence. To this end we first
split up the $Q$ into their averages, and fluctuations: 
$Q_{nm} = \langle Q_{nm}\rangle + (\Delta Q)_{nm} 
= \delta_{nm}\langle Q_{nn}\rangle + (\Delta Q)_{nm}$. Since $n_1\neq n_2$, and
$n_3\neq n_4$, we have 
$\langle Q_{n_1 n_2}\rangle = \langle Q_{n_3 n_4}\rangle = 0$. Furthermore,
if we put $n_1' = n_2'$ in the 3-point functions, then the expression in
square brackets in Eq.\ (\ref{eq:3.14b}) vanishes, so effectively we also
have $\langle Q_{n_1' n_2'}\rangle = 0$. Equation\ (\ref{eq:3.14a}) then
takes the form,
\begin{eqnarray}
W_{int}&+&8\Omega_{n_1 - n_2}\sum_{\gamma}\int d{\bf y}\
      \biggl\langle{^{0}_{0}(\Delta Q)}_{n_1 n_2}^{\gamma\gamma}({\bf y})
\nonumber\\
&&\qquad\qquad\qquad\qquad\qquad\times{^{0}_{0}(\Delta Q)}_{n_3 n_4}
                                      ^{\alpha\beta}({\bf x})\biggr\rangle
\nonumber\\
&=& \delta_{n_1 n_3}\delta_{n_2 n_4}\left(\Bigl\langle{^{0}_{0}Q}_{n_1 n_1}
           ^{\alpha\beta}({\bf x})\Bigr\rangle
    - \Bigl\langle{^{0}_{0}Q}_{n_2 n_2}^{\alpha\beta}({\bf x})\Bigr\rangle
                                               \right)\quad,
\nonumber\\
\label{eq:3.14c}
\end{eqnarray}
and $W_{int}$ can be written,
\begin{eqnarray}
W_{int}&=&-16\Gamma^{(s)}\sum_{\gamma}\left(
         \Bigl\langle{^{0}_{0}Q}_{n_1 n_1}^{\gamma\gamma}({\bf x})\Bigr\rangle
           - \Bigl\langle{^{0}_{0}Q}_{n_2 n_2}^{\gamma\gamma}({\bf x})
                                                         \Bigr\rangle\right)
\nonumber\\
&&\qquad\times T\sum_{n_1' n_2'}\left(\delta_{n_1'-n_2',n_2-n_1}
                                    +\delta_{n_1'-n_2',n_1 - n_2}\right)\ 
\nonumber\\
&&\qquad\quad\times\int d{\bf y}\ 
        \Bigl\langle{^{0}_{0}(\Delta Q)}_{n_1' n_2'}^{\gamma\gamma}
  ({\bf y})\ {^{0}_{0}(\Delta Q)}_{n_3 n_4}^{\alpha\beta}({\bf x})\Bigr\rangle
\nonumber\\
&&-16\Gamma^{(s)}\sum_{\gamma}\int d{\bf y}\sum_{r=0,3}\ T
                     \sum_{n_1' n_2'}
\nonumber\\
 &&\qquad  \times\biggl<{^{0}_{0}(\Delta Q)}_{n_3 n_4}^{\alpha\beta}({\bf x})\
                 {^{0}_{r}(\Delta Q)}_{n_1' n_2'}^{\gamma\gamma}({\bf y})
\nonumber\\
&&\qquad\qquad\times\biggl[{^{0}_{r}(\Delta Q)}_{n_1,n_2-(n_1'-n_2')}
                                                   ^{\gamma\gamma}({\bf y})
\nonumber\\
 &&\qquad\qquad\qquad -{^{0}_{r}(\Delta Q)}_{n_1+(n_1'-n_2'),n_2}
                                                  ^{\gamma\gamma}({\bf y})
\nonumber\\
 &&\qquad\qquad\qquad
            +{^{0}_{r}(\Delta Q)}_{n_2+(n_1'-n_2'),n_1}^{\gamma\gamma}({\bf y})
\nonumber\\
 &&\qquad\qquad\qquad
            -{^{0}_{r}(\Delta Q)}_{n_2,n_1-(n_1'-n_2')}^{\gamma\gamma}({\bf y})
                                                 \biggr]\biggr>\quad,
\nonumber\\
\label{eq:3.14d}
\end{eqnarray}
\end{mathletters}%
Next we remember that the composite variables $Q$ correspond to products
of fermions variables, e.g. $Q\sim{\bar\psi}\psi$ in the particle--hole channel.
Specifically,
\begin{mathletters}
\label{eqs:3.15}
\begin{eqnarray}
^{0}_{0}Q_{n_1 n_2}&\sim&\frac{i}{2}\sum_{\sigma} \left({\bar\psi}_{n_1,\sigma}
   \psi_{n_2,\sigma} + {\bar\psi}_{n_2,\sigma}\psi_{n_1,\sigma}\right)\quad,
\label{eq:3.15a}\\
^{0}_{3}Q_{n_1 n_2}&\sim&\frac{-1}{2}\sum_{\sigma} \left({\bar\psi}_{n_1,\sigma}
   \psi_{n_2,\sigma} - {\bar\psi}_{n_2,\sigma}\psi_{n_1,\sigma}\right)\quad,
\label{eq:3.15b}
\end{eqnarray}
\end{mathletters}
Therefore, writing the correlation functions in terms of
$\Delta Q \sim {\bar\psi}\psi - \langle{\bar\psi}\psi\rangle$ makes them
$Q$-irreducible, but not $\psi$-irreducible. For instance, 
the $2$-point function in Eqs.\ (\ref{eq:3.14c}) and (\ref{eq:3.14d})
has the structure,
\begin{eqnarray}
&\sum_{\gamma}&\int d{\bf y}\
      \Bigl\langle{^{0}_{0}(\Delta Q)}_{n_1 n_2}^{\gamma\gamma}({\bf y})\
             {^{0}_{0}(\Delta Q)}_{n_3 n_4}^{\alpha\beta}({\bf x})\Bigr\rangle
\nonumber\\
&=& \delta_{\alpha\beta}\left[\delta_{n_1 n_3}\delta_{n_2 n_4} X_{n_1 n_2}^{(1)}
   + \delta_{n_1-n_2,n_3-n_4} X_{n_1 n_2,n_3 n_4}^{(2)}\right]\ .
\nonumber\\
\label{eq:3.16}
\end{eqnarray}
The replica structure of this identity is such that $X^{(1)}$ and $X^{(2)}$
are replica independent.\cite{ReplicaIndepenceFootnote}
We can therefore suppress the replica index in what follows,
as we have done already in Eqs.\ (\ref{eqs:3.15}). The frequency structure
of Eq.\ (\ref{eq:3.16}) is very general: By virtue of Eq.\ (\ref{eq:3.15a}),
the full 2-point function is automatically proportional to 
$\delta_{n_1-n_2,n_3-n_4}$ due to time translational invariance, and
the disconnected part is in addition proportional to
$\delta_{n_1 n_3}\delta_{n_2 n_4}$.
Using Eqs.\ (\ref{eqs:3.15}) in the 3-point $Q$-correlations shows that they 
also contain
parts that are proportional to $\delta_{n_1 n_3}\delta_{n_2 n_4}$, and
others that are proportional to $\delta_{n_1-n_3,n_2-n_4}$. This suggests
to break up the Ward identity into two separate pieces with these two
frequency structures, respectively.

By collecting the above arguments, we find that the coefficients of
$\delta_{n_1 n_3}\delta_{n_2 n_4}$ in the Ward identity obey the
relation,
\begin{mathletters}
\label{eqs:3.17}
\begin{eqnarray}
\Bigl<{^{0}_{0}Q}_{n_1 n_1}^{\alpha\beta}({\bf x})\Bigr>
            &-&\Bigl<{^{0}_{0}Q}_{n_2 n_2}^{\alpha\beta}({\bf x})\Bigr>
             =\delta_{\alpha\beta}\Bigl[8\Omega_{n_1-n_2} X^{(1)}_{n_1 n_2} 
\nonumber\\
&&\qquad- N_F\,(N_F\Gamma^{(s)})^2 Y^{(1)}_{n_1 n_2}\Bigr]\ .
\label{eq:3.17a}
\end{eqnarray}
Here $Y^{(1)}$ is related to the piece of the 3-point correlation functions
that is proportional to $\delta_{n_1 n_3}\delta_{n_2 n_4}$. An explicit
calculation yields,
\begin{eqnarray}
Y^{(1)}_{n_1 n_2}&=&\frac{-8}{N_F^3\Gamma^{(s)}} \int d{\bf y} 
                                                 T\sum_{n_1' n_2'}\qquad
\nonumber\\
&&\times\biggl\{
   G({\bf x},{\bf y};\omega_{n_1})\biggl< \left({\bar\psi}_{n_1'}({\bf y})
                            s_0 \psi_{n_2'}({\bf y})\right)
\nonumber\\
&&\qquad\times\left(({\bar\psi}_{n_2-n_1'+n_2'}({\bf y}) s_0 \psi_{n_2}
    ({\bf x}) \right)\biggr>^c
\nonumber\\
&& - G({\bf x},{\bf y};\omega_{n_2})\biggl< \left({\bar\psi}_{n_1'}({\bf y})
                           s_0 \psi_{n_2'}({\bf y})\right)
\nonumber\\
&&\qquad\times
   \left(({\bar\psi}_{n_1-n_1'+n_2'}({\bf y}) s_0 \psi_{n_1}({\bf x})
       \right)\biggr>^c\biggr\}_{dis}\quad.
\nonumber\\
\label{eq:3.17b}
\end{eqnarray}
\end{mathletters}%
Here the $\langle {\bar\psi}\psi{\bar\psi}\psi\rangle^c$ are $\psi$-connected
correlation functions, with the cumulant taken with respect to the quantum
mechanical expectation value only, and 
$G({\bf x},{\bf y};\omega_n) = \langle{\bar\psi}_{n,\sigma}({\bf x})\,
\psi_{n,\sigma}({\bf y})\rangle$  is the Green
function for a given disorder configuration.
For $Tn_1\rightarrow +0$ and $Tn_2\rightarrow -0$,
so that $\Omega_{n_1-n_2}\rightarrow 0$, the left--hand side
of Eq.\ (\ref{eq:3.17a})
approaches again $\pi N(\epsilon_F)/2$, with
$N(\epsilon_F)$ the single--particle density of states at the Fermi level.
However, since we are now dealing with interacting electrons,
$N(\epsilon_F)$ includes interaction as well as disorder effects. 
$Y^{(1)}_{n_1 n_2}$ in that limit approaches some
number $\pi Y^{(1)}/2$. We then obtain, for small $\Omega_{n_1-n_2}$,
\begin{equation}
X^{(1)}_{n_1 n_2} = \frac{\pi}{16\vert\Omega_{n_1-n_2}\vert}
   \left[N(\epsilon_F) + N_F\,(N_F\Gamma^{(s)})^2 Y^{(1)}\right]
                                                                     \quad.
\label{eq:3.18}
\end{equation}
Notice that the $\psi$-connected correlation functions vanish for
noninteracting electrons, so that in an expansion in powers of the
interaction constant, $Y^{(1)}$ is of $O(1)$. The term in brackets on the
right--hand side
of Eq.\ (\ref{eq:3.18}) is equal to $\pi\bigl(N(\epsilon_F)\bigr)^2/8H$,
with $H$ the quasi-particle or specific heat density of 
states.\cite{CastellaniDiCastro,CastellaniKotliarLee,R}

Similarly, one finds for the coefficients of $\delta_{n_1-n_3,n_2-n_4}$
in the Ward identity,
\begin{mathletters}
\label{eqs:3.19}
\begin{eqnarray}
\Omega_{n_1-n_2}&\,& X^{(2)}_{n_1n_2,n_3n_4} =
          N_F(N_F\Gamma^{(s)})^2\ Y^{(2)}_{n_1n_2,n_3n_4}
\nonumber\\
&& - 16 \Gamma^{(s)}T\,\Bigl\{
\bigl[G({\bf x},{\bf x};\omega_{n_1}) - G({\bf x},{\bf x};\omega_{n_2})\bigr]\,
\nonumber\\
&&\qquad\quad\times\phi_{n_3 n_4}({\bf x},{\bf x})\Bigr\}_{dis}\quad,
\label{eq:3.19a}
\end{eqnarray}
where
\begin{equation}
\phi_{n_3 n_4}({\bf x},{\bf x}) = 
\int d{\bf y}\,G({\bf y},{\bf x};\omega_{n_3})\,G({\bf x},{\bf y};\omega_{n_4})
                                                                         \quad,
\label{eq:3.19b}
\end{equation}
\end{mathletters}%
and $Y^{(2)}$ is again given in terms of connected 4-$\psi$ correlation
functions. If desired, it can be calculated in perturbation theory in
$\Gamma^{(s)}$.
For $\Omega_{n_1 n_2}\rightarrow 0$, Eqs.\ (\ref{eqs:3.19}) yield
$X^{(2)}_{n_1n_2,n_3n_4} \sim T/\vert\Omega_{n_1-n_2}\vert 
\vert\Omega_{n_3-n_4}\vert$. Again,
the Gaussian propagator has that property, see Eqs.\ (\ref{eq:2.34a}),
(\ref{eqs:2.36}), (\ref{eqs:2.37}).

From Eqs.\ (\ref{eq:3.18}) and (\ref{eqs:3.19}) we see that 
${^{0}_{0}Q}_{nm}^{\alpha\alpha}$ remains a soft mode in the presence
of interactions. This is due to the frequency structure of the interaction 
term, which ensures that the 2-point $Q$-correlation function remains soft
rather than acquiring a mass, as one might naively expect from an
inspection of the action, Eqs.\ (\ref{eq:2.20}) - (\ref{eqs:2.23}). 
Note that this softness
is {\em not} restricted to the particular linear combination of $Q$ that
constitutes the particle number density. It therefore is not related to
the particle number conservation law. Rather, it is a consequence of the
spontaneous breaking of a continuous symmetry in the system, viz. the
invariance, at zero external frequency, under the rotations between positive
and negative frequencies discussed above, or between retarded and advanced
Green functions. The soft modes in question are the corresponding 
Goldstone modes.

As in the noninteracting case, there is a variety of other soft modes
that are related to the above Ward identity by means of additional
symmetries that are not broken. First of all, rotational symmetry in
replica space again implies that the ${^{0}_{0}Q}_{nm}^{\alpha\beta}$,
$nm<0$, are also soft.\cite{ReplicaIndepenceFootnote}
Next we show that the spin--triplet channel
also remains soft. To this end, let us denote the (identical) coupling
constants in the three branches of the spin--triplet channel by 
$\Gamma_i^{(t)}$, $i=1,2,3$. Now we pick from the transformations $T$, 
Eq.\ (\ref{eq:3.11a}), the one that interchanges, say, $i=0$ and $i=1$. 
The action is still invariant under this transformation, provided that we also 
interchange $\Gamma^{(s)}$ and $\Gamma_1^{(t)}$. This shows that ${^{1}_{0}Q}$ 
also obeys Eq.\ (\ref{eq:3.16}), but with the $\Gamma^{(s)}$ in 
Eqs.\ (\ref{eqs:3.17}) - (\ref{eq:3.19a}) replaced by $\Gamma^{(t)}$. Since this
can be done for all branches of the spin--triplet channel, it follows that
they all are soft modes. Alternatively, we can invoke
rotational invariance in spin 
space (an $SU(2)$ subgroup of the large symplectic group), which implies that
all three components of the spin--triplet have identical correlation
functions. This argument also shows that there are no singlet--triplet cross 
correlations. By means of similar arguments one easily convinces oneself
that the particle--particle channel is still soft, with the appropriate
Cooper channel interaction amplitude replacing $\Gamma^{(s)}$ or 
$\Gamma^{(t)}$. For the
action shown in Eq.\ (\ref{eq:2.20}) - (\ref{eqs:2.23}), 
we thus conclude that all channels
are still soft, and all the soft modes can be traced to the Ward identity,
Eq.\ (\ref{eq:3.14a}), and additional symmetry relations. If some of
the additional symmetries are broken externally, e.g. by means of an
external magnetic field, or by magnetic impurities, then the respective 
modes acquire a mass, see Refs.\ \onlinecite{EfetovLarkinKhmelnitskii,R}.

\subsubsection{Separation of soft and massive modes}
\label{subsubsec:III.A.4}

From the previous subsection we know that the correlation functions of
the $Q_{nm}$ with $nm<0$ are
soft, while those with $nm>0$ are massive. Our next goal is to
separate these degrees if freedom in such a way that the soft modes
remain manifestly soft to all orders in perturbation theory. To give a
simple example, in an $O(N)$ $\phi^4$--theory this is achieved by
writing the $N$-component vector field as ${\vec\phi}({\bf x}) =
\rho({\bf x})\,{\hat\phi}({\bf x})$, with $\rho$ the norm of the
vector, and $\vert{\hat\phi}({\bf x})\vert\equiv 1$. With this separation,
only gradients of ${\hat\phi}$ appear in the theory, and so ${\hat\phi}$
is manifestly soft, see Ref.\ \onlinecite{ZJ} and Appendix \ref{app:B}.
We are looking for an analogous separation
of our matrix degrees of freedom. For noninteracting electrons this has
been done in Refs.\ \onlinecite{SchaferWegner} and
\onlinecite{PruiskenSchafer}, and we will
construct a suitable generalization of the procedure used by these authors.

The matrices $Q$ that we are considering are complex $8Nn\times 8Nn$
matrices that obey Eqs.\ (\ref{eqs:2.24'}). Alternatively, 
we can consider the $Q$ as
$2Nn\times 2Nn$ matrices whose elements are spin--quaternions, or
$4Nn\times 4Nn$ matrices whose elements are quaternions. For our purposes
the last choice will be most convenient. We thus study a set of matrices
$Q$ with quaternion valued matrix elements
$Q_{nm,\sigma\sigma'}^{\alpha\beta}$ 
($n,m=1,\ldots,N;\ \alpha,\beta=1,\ldots,n;\ 
\sigma,\sigma'=\uparrow,\downarrow$),\cite{QuaternionsFootnote}
\begin{mathletters}
\label{eqs:3.20}
\begin{equation}
Q_{nm,\sigma\sigma'}^{\alpha\beta} 
   = \sum_{r=0}^{3} {_{r}Q}_{nm,\sigma\sigma'}^{\alpha\beta}\ \tau_r\quad,
\label{eq:3.20a}
\end{equation}
that obey
\begin{eqnarray}
Q&=&C^{T} Q^T C\quad,
\label{eq:3.20b}\\
Q^{\dagger}&=&-\Gamma Q \Gamma^{-1}\quad,
\label{eq:3.20c}
\end{eqnarray}
\end{mathletters}%
with $\Gamma$ from Eq.\ (\ref{eq:2.24'c}), and $C = \openone\otimes\tau_2$, 
where $\openone$ denotes the $4Nn\times 4Nn$ unit matrix.
This set of matrices is invariant under transformations $T$ that obey
$T C T^T = C$, i.e. under symplectic transformations.\cite{SymplecticFootnote}

For what follows, the constraint expressed by Eq.\ (\ref{eq:3.20c}) is somewhat 
awkward to handle. We therefore analytically continue the matrix elements
of $Q$, considered as functions of the imaginary frequencies, $i\omega_n$ and
$i\omega_m$, to the complex plane: $i\omega_n \rightarrow z_1$, and
$i\omega_m \rightarrow z_2$. As functions of $z_1$ or $z_2$, the $Q$ have
a branch cut on the real axis, and from Sec.\ \ref{subsubsec:III.B.3} we
know that those matrix elements with $z_1$ and $z_2$ approaching the real
axis from opposite sides are soft modes. The analytic continuation of
Eq.\ (\ref{eq:3.20c}) reads
\begin{mathletters}
\label{eqs:3.20'}
\begin{equation}
(Q^{\dagger})_{z_1 z_2} = - Q_{z_1^* z_2^*}\quad.
\label{eq:3.20'a}
\end{equation}
In particular, for real frequencies, $\omega$ and $\omega'$, we have
\begin{equation}
(Q^{\dagger})_{\omega\pm i0,\omega'\pm i0} = - Q_{\omega\mp i0,\omega'\mp i0}
                                                                   \quad.
\label{eq:3.20'b}
\end{equation}
\end{mathletters}%
If we analytically continue $Q$ onto the unphysical Riemann sheet, 
$Q_{\omega,\omega'}$ is thus formally anti--hermitian. It therefore has
$4Nn$ imaginary eigenvalues $\lambda_j\tau_0$
($\lambda_j\in{i\cal R};\ j=1,\ldots ,4Nn$), and can be diagonalized by means
of unitary transformations ${\tilde{\cal S}}\in {\rm U}(4Nn,{\cal Q})$ 
(i.e. by unitary
$4Nn\times 4Nn$ matrices whose elements are quaternions). Now
${\rm U}(4Nn,{\cal Q})$ is isomorphic to the unitary symplectic group
${\rm USp}(8Nn,{\cal C}) \equiv 
{\rm U}(8Nn,{\cal C})\cap {\rm Sp}(8Nn,{\cal C})$.\cite{Wybourne}
This means that the $Q$
can be diagonalized by means of unitary matrices that are also symplectic,
and hence leave the set of $Q$ invariant. That is, the most general $Q$
can be written
\begin{equation}
Q = {\tilde {\cal S}}\ D\ {\tilde {\cal S}}^{-1}\quad,
\label{eq:3.21}
\end{equation}
where $D$ is diagonal, and ${\tilde {\cal S}}\in {\rm USp}(8Nn,{\cal C})$.

However, diagonalization is more than we want. Since we know that the
$Q_{nm}$ with $nm<0$ are soft, while those with $nm>0$ are massive,
we are interested in generating the most general $Q$ from a matrix $P$ that
is block--diagonal in Matsubara frequency space,
\begin{equation}
P = \left( \begin{array}{cc}
       P^> & 0   \\
       0   & P^< \\
    \end{array} \right)\quad,
\label{eq:3.22}
\end{equation}
where $P^>$ and $P^<$ are matrices with elements $P_{nm}$ where $n,m>0$ and
$n,m<0$, respectively. This can easily be achieved. Since the analytic
continuations of $P^>$ and $P^<$
are anti--hermitian, the most general $P$ can be obtained from $D$ by an
element $U$ of ${\rm USp}(4Nn,{\cal C})\times {\rm USp}(4Nn,{\cal C})$.
The most general $Q$ can therefore be written
\begin{equation}
Q = {\cal S}\ P\ {\cal S}^{-1}\quad,
\label{eq:3.23}
\end{equation}
with ${\cal S}={\tilde {\cal S}}U^{-1}$. The set of transformations ${\cal S}$ 
is the set of
all cosets of ${\rm USp}(8Nn,{\cal C})$ with respect to
${\rm USp}(4Nn,{\cal C})\times {\rm USp}(4Nn,{\cal C})$, i.e. the
${\cal S}$ form the homogeneous space
${\rm USp}(8Nn,{\cal C})/{\rm USp}(4Nn,{\cal C})\times {\rm USp}(4Nn,{\cal C})$.
The corresponding most general $Q$ on the imaginary frequency axis is
generated from the most general $P$ by a set of transformations
that is isomorphic to
this coset space, and that can be explicitly constructed by reversing the
Wick rotation that led us from  imaginary frequencies to real ones. For our
purposes, we will not need this explicit construction, and we will not 
distinguish between the two isomorphic spaces.

This achieves the desired separation of our degrees of freedom into soft
and massive ones. The massive degrees of freedom are represented by the
matrix $P$, while the soft ones are represented by the transformations
${\cal S}\in {\rm USp}(8Nn,{\cal C})/{\rm USp}(4Nn,{\cal C})\times 
{\rm USp}(4Nn,{\cal C})$. To come back to the $O(N)$ example at the
beginning of this subsection, the analogy is as follows. For the $N$-component
vector, the direction ${\hat\phi}$ is a point on the $(N-1)$-sphere, which
is isomorphic to the homogeneous space $O(N)/O(N-1)\times O(1)$, see Appendix
\ref{app:B}. The unitary--symplectic coset space identified above is a
matrix generalization of this.

In order to formulate the field theory in terms of the soft and massive
modes, one also needs the invariant measure $I[P]$, or the Jacobian
of the transformation from the $Q$ to the $P$ and the $\cal S$, defined by
\begin{equation}
\int D[Q]\,\ldots = \int D[P]\,I[P] \int D[\cal S]\,\ldots\quad.
\label{eq:3.24}
\end{equation}
Since we will not need to know the explicit form of $I[P]$ in what follows, 
we relegate its derivation to Appendix \ref{app:C}.

\subsubsection{A Ward identity based on a local symmetry}
\label{subsubsec:III.A.5}

In addition to the Ward identity derived on the basis of a global symmetry
in Secs.\ \ref{subsubsec:III.A.2} and \ref{subsubsec:III.A.3} above, for
the derivation of an effective field theory in the next subsection we will
also need an identity that is based on a local symmetry.
Such relations with the structure of a Ward identity or a Noether 
theorem can be derived on the basis of either the replica or the gauge 
structure of the theory.
For our present purposes, we consider a
local $U(1)$ gauge transformation of the Grassmannian field theory,
Eqs.\ (\ref{eqs:2.2}), that is independent of imaginary time $\tau$,
\begin{equation}
\psi_{\sigma}(x)\rightarrow e^{-i\alpha ({\bf x})}\,\psi_{\sigma}(x)\quad.
\label{eq:3.28}
\end{equation}
We define a $(d+1)$-component current vector $q_{\mu} = (q_0,{\bf q})$ with
`spatial' components $q_i$ ($i=1,\ldots d$) as
\begin{mathletters}
\label{eqs:3.29}
\begin{equation}
q_{0}(x) = \sum_{\sigma} \bar\psi_{\sigma}(x)\,\psi_{\sigma}(x)\quad,
\label{eq:3.29a}
\end{equation}
\begin{equation}
{\bf q}(x) = \frac{-1}{2m}\sum_{\sigma}\left(\psi_{\sigma}(x)\,
               \nabla{\bar\psi}_{\sigma}(x)
   + {\bar\psi}_{\sigma}(x)\,\nabla\psi_{\sigma}(x)\right)\quad,
\label{eq:3.29b}
\end{equation}
\end{mathletters}%
and add to the action a term $S_A$ that describes a coupling of the current
to a $\tau$-independent vector potential $A_{\mu} = (A_0,{\bf A})$,
\begin{equation}
S_A = -i\int dx\,A_{\mu}({\bf x})\,q^{\mu}(x) - \frac{1}{2m}\int dx\,
     {\bf A}^2({\bf x})\,q_0(x)\quad.
\label{eq:3.30}
\end{equation}
It is then straightforward to establish that the partition function is
gauge invariant, i.e. it is
invariant under the local gauge tranformation given by Eq.\ (\ref{eq:3.28}), 
supplemented by a transformation of the vector potential,
\begin{equation}
A_{\mu}({\bf x}) \rightarrow A_{\mu}({\bf x}) + \partial_{\mu}\alpha ({\bf x})
                                                                \quad,
\label{eq:3.31}
\end{equation}
with $\partial_{\mu} = (\partial_{\tau},\nabla)$.
The invariance statement is
\begin{equation}
Z[A_{\mu}] = Z[A_{\mu} + \partial_{\mu}\alpha]\quad,
\label{eq:3.32}
\end{equation}
A Taylor expansion of Eq. (\ref{eq:3.32}) immediately leads to the 
conservation law
\begin{equation}
\partial_{\mu} \left( \langle q_{\mu}(x)\rangle - \frac{i}{m}\,A_{\mu}(x)
   \langle q_{0}(x)\rangle \right) = 0 \quad,
\label{eq:3.33}
\end{equation}
where $\langle\ldots\rangle$ denotes an average with respect to the action 
$S + S_A$. By putting $A_{\mu}=0$, one simply obtains the continuity
equation for the fermion number density. Another useful identity
can be obtained by integrating Eq.\ (\ref{eq:3.33}) over space and time, 
differentiating with respect to $A_{i}({\bf y})$, and then setting $A_{\mu}=0$.
The result is,
\begin{eqnarray}
\frac{-1}{m}\,\sum_{{n,m}\atop{\sigma,\sigma^{^{\prime}}}}\int&d{\bf y}&
   \biggl\langle\frac{\partial \bar\psi_{n,\sigma}({\bf x})}{\partial x^{i}}\,
   \psi_{n,\sigma}({\bf x})\,\frac{\partial{\bar\psi}_{m,\sigma^{^{\prime }}}
   ({\bf y})}{\partial y^{j}}\,\psi_{m,\sigma^{^{\prime }}}({\bf y})
                                                              \biggr\rangle 
\nonumber\\
   &&= \delta_{ij}\sum_{n,\sigma}\bigl\langle{\bar\psi}_{n,\sigma}({\bf x})
     \,\psi_{n,\sigma}({\bf x})\bigr\rangle\quad.
\label{eq:3.34}
\end{eqnarray}
This identity holds for a particular realization of the disorder. If we
perform the ensemble average it still holds, with the brackets now
including the disorder average in addition to the quantum mechanical
expectation value. One then has a relation between a homogeneous 
current--current correlation function and
the particle number density that is known as the f-sum rule. The above
derivation makes it clear that it is closely related to the particle
number conservation law. Within models or approximations that describe an
effective single-particle problem, the quantum mechanical average on the
left--hand side
of Eq.\ (\ref{eq:3.34}) factorizes. Furthermore, Eq.\ (\ref{eq:3.34})
then holds for each term in the $n$-summation separately, since in a
single-particle problem there is no frequency mixing.
Within approximations where the disorder average factorizes as well,
the identity can be written in terms of disorder averaged
one--particle Green functions as,
\begin{mathletters}
\label{eqs:3.35}
\begin{eqnarray}
\delta_{ij} G({\bf x}=0,\omega_n)&=&\frac{1}{m} \int d{\bf y}\,\left[ 
   \partial_{i} G({\bf x}-{\bf y},\omega_n)\right]\,
\nonumber\\
&&\qquad\quad\times\left[\partial_{j} G({\bf y}-{\bf x},\omega_n)\right]\ ,
\label{eq:3.35a}
\end{eqnarray}
or, in Fourier space,
\begin{equation}
\delta_{ij} \sum_{\bf q} G({\bf q},\omega_n) = \frac{-1}{m} \sum_{\bf q}
      q_i\,q_j\,\bigl(G({\bf q},\omega_n)\bigr)^2\quad.
\label{eq:3.35b}
\end{equation}
\end{mathletters}%
This identity will be useful later.

\subsection{Effective field theory for disordered fermions}
\label{subsec:III.B}

We are now in a position to formulate the theory in such a way that the
soft and massive modes, respectively, are separated, and that the former
remain manifestly soft to all orders in perturbation theory. This formulation
also provides a technically satisfactory derivation of the generalized
nonlinear sigma--model for interacting, disordered fermions. We will first
discuss this derivation, as well as corrections to the sigma--model. Then
we will discuss the Fermi--liquid FP and the critical FP
that describes the Anderson--Mott metal--insulator transition, and show that
the sigma--model provides an adequate description of either one.

\subsubsection{The nonlinear sigma--model}
\label{subsubsec:III.B.1}

We return to Eqs.\ (\ref{eq:2.19}), (\ref{eq:2.20}), and use the 
representation, Eq.\ (\ref{eq:3.23}), of $Q$ in terms of $\cal S$ and $P$.
It is convenient to define a transformed ghost field $\Lambda$ by
\begin{equation}
\Lambda ({\bf x}) = {\cal S}^{-1}({\bf x})\,{\tilde\Lambda}({\bf x})\,{\cal S}
                        ({\bf x})\quad.
\label{eq:3.36}
\end{equation}
In terms of these variables, the partition function for the replicated
theory, Eq.\ (\ref{eq:2.19}), reads
\begin{equation}
{\tilde Z} = \int D[P]\,D[{\cal S}]\,D[\Lambda]\,\exp\bigl(
              {\cal A}\left[{\cal S},P,\Lambda\right] + \Tr\ln I[P]\bigr)\quad.
\label{eq:3.37}
\end{equation}
Here $I[P]$ is the invariant measure, Eq.\ (\ref{eq:3.24}), and the
action ${\cal A}$ reads,
\begin{eqnarray}
{\cal A}\left[{\cal S},P,\Lambda\right]&=&{\cal A}_{dis}[P] 
            + {\cal A}_{int}[{\cal S}P{\cal S}^{-1}]
\nonumber\\
&&+ \frac{1}{2}\,\Tr\ln \left( G_0^{-1} - i{\cal S}\Lambda {\cal S}^{-1}\right)
\nonumber\\
&&+ \int d{\bf x}\,\tr\bigl(\Lambda({\bf x}) P({\bf x})\bigr)\ ,
\label{eq:3.38}
\end{eqnarray}
with ${\cal A}_{dis}$ and ${\cal A}_{int}$ defined by Eqs.\ (\ref{eqs:2.22}) 
and (\ref{eqs:2.23}).

Now we expand $\cal S$, $P$, and $\Lambda$ about their saddle point values.
Alternatively, one could expand about the exact expectation values. However,
the `equation of state' for the latter is just the saddle point equation of
state plus loop corrections, and for our purposes the difference is irrelevant.
We therefore do not distinguish between saddle point values and expectation
values, and denote both by $\langle {\cal S}\rangle$, $\langle P\rangle$, and
$\langle\Lambda\rangle$, respectively. As a result, we do not distinguish either
between the saddle point Green function as defined in Eq.\ (\ref{eq:2.33b}),
and the Green function that contains the full expectation value of $\Lambda$ 
as a self energy. From Eqs.\ (\ref{eqs:2.30}) it follows that
\begin{mathletters}
\label{eqs:3.39}
\begin{equation}
\langle {\cal S}\rangle = \openone\otimes\tau_0\quad,
\label{eq:3.39a}
\end{equation}
\begin{eqnarray}
\langle P\rangle_{12}&=&\delta_{12}\,\left(\tau_0\otimes s_0\right)\,
   \frac{i}{2V}\sum_{\bf p}\Bigl[i\omega_{n_1} - {\bf p}^2/2m + \mu
\nonumber\\
&&\qquad\qquad\qquad\qquad\qquad\qquad - i\langle\Lambda_{12}\rangle\Bigr]^{-1}
\nonumber\\
&\equiv&\frac{\pi}{4}\,N_F\,(\pi_{12} + i\gamma_{12})\quad,
\nonumber\\
&\equiv&\delta_{12}\,
   \left(\tau_0\otimes s_0\right)\,\frac{\pi}{4}\,N_F\,(\pi_1 + i\gamma_1)
                                                                      \quad,
\label{eq:3.39b}
\end{eqnarray}
\begin{equation}
\langle\Lambda_{12}\rangle = \frac{-2}{\pi N_F\tau_{rel}}\,\langle P_{12}
     \rangle - 4\Gamma^{(s)} T\sum_n e^{i\omega_n 0} \langle 
                                          P_{nn}^{\alpha\alpha}\rangle\quad.
\label{eq:3.39c}
\end{equation}
and we write 
\begin{equation}
P = \langle P\rangle + \Delta P\quad,\quad \Lambda = \langle\Lambda\rangle
                                            + \Delta\Lambda\quad.
\label{eq:3.39d}
\end{equation}
Here we have defined two new matrix fields, $\pi$ and $\gamma$, 
for later reference. Their diagonal elements, $\pi_1$ and $i\gamma_1$,
are odd and even functions, respectively, of their Matasubara frequency
arguments. For small frequencies,
\begin{equation}
\pi_1 = \sgn\omega_{n_1}\quad,
\label{eq:3.39e}
\end{equation}
\end{mathletters}%
while $\gamma_1$ is a real constant that depends on the momentum cutoff in
Eq.\ (\ref{eq:3.39b}).

Now we use the cyclic property of the trace to write the $\Tr\ln$-term in
the action, Eq.\ (\ref{eq:3.38}), in the form
\begin{eqnarray}
\Tr\ln\bigl(G_{0}^{-1}&-&{\cal S}i\Lambda {\cal S}^{-1}\bigr) 
     = \Tr\ln\left({\cal S}^{-1}G_0^{-1} {\cal S} - i\Lambda\right)\qquad\qquad
\nonumber\\
&=&\Tr\ln\left(G_{sp}^{-1}\right) 
  + \Tr\ln\biggl(1 + G_{sp}{\cal S}^{-1}\left(\partial_{\tau} {\cal S}\right)
\nonumber\\
&&+ \frac{1}{m}\,G_{sp}{\cal S}^{-1}\left(\nabla {\cal S}\right)\nabla
               + \frac{1}{2m}\,G_{sp}{\cal S}^{-1}\left(\nabla^2 {\cal S}\right)
\nonumber\\
&&\qquad\qquad\qquad\qquad\qquad\quad - G_{sp}i(\Delta\Lambda)\biggr)\quad,
\nonumber\\
\label{eq:3.40}
\end{eqnarray}
Notice that now the transformation matrix $\cal S$ always appears in conjuction
with some derivative, while the fluctuations $\Delta\Lambda$ are 
massive.\cite{LambdaFootnote} We proceed to expand in powers of the derivatives
of $\cal S$, and in powers of $\Delta\Lambda$, in analogy to
the procedure used in Ref.\ \onlinecite{PruiskenSchafer}. The simplest term
in this expansion is
\begin{eqnarray}
&\Tr&\,\bigl(G_{sp}{\cal S}^{-1}(\partial_{\tau} {\cal S})\bigr)=\int d{\bf x}\,
                                                                  \tr\bigl(
       i\Omega\, {\cal S}({\bf x})\, G_{sp}({\bf x}=0)\,
\nonumber\\
&&\qquad\qquad\qquad\qquad\qquad\qquad\qquad \times{\cal S}^{-1}({\bf x})\bigr)
\nonumber\\
&=&\qquad\frac{\pi N_F}{2}\int d{\bf x}\,\tr\left(\Omega 
                                                   {\hat Q}({\bf x})\right)
    + O\left(\Omega^2 {\hat Q}\right)\quad.
\label{eq:3.41}
\end{eqnarray}
Here we have defined a frequency matrix,
\begin{mathletters}
\label{eqs:3.42}
\begin{equation}
\Omega_{12} = \left(\tau_0\otimes s_0\right)\,\delta_{12}\,\omega_{n_1}\quad,
\label{eq:3.42a}
\end{equation}
and a new field, ${\hat Q}({\bf x}) = {\cal S}({\bf x})\,(\pi + i\gamma)
\,{\cal S}^{-1}({\bf x})$. Since the contribution of $\gamma_{12}$ to
$\langle P\rangle_{12}$ is, in the low--frequency limit, just an imaginary
constant, it leads only to a constant contribution to $\hat Q$, which
does not contribute to the lowest order terms in the present expansion.
We can therefore neglect it, and write
\begin{equation}
{\hat Q}({\bf x}) = {\cal S}({\bf x})\,\pi\,{\cal S}^{-1}({\bf x})\quad,
\label{eq:3.42b}
\end{equation}
\end{mathletters}%
with $\pi$ from Eq.\ (\ref{eq:3.39b}). Notice that $\hat Q$ is hermitian,
and has the
properties ${\hat Q}^2 = 1$, and $\tr {\hat Q} = 0$. $\hat Q$ will turn
out to be a convenient parametrization of the soft modes, in analogy to
the unit vector $\hat\phi$ in $\phi^4$--theory, see Appendix \ref{app:B}.

Now consider the gradient terms in the expansion of Eq.\ (\ref{eq:3.40}).
To this end, it is convenient to define the vector field,
\begin{equation}
{\bf s} ({\bf x}) = {\cal S}^{-1}({\bf x})\,(\nabla {\cal S})({\bf x})\quad.
\label{eq:3.43}
\end{equation}
with matrix valued components $s^i({\bf x})$, ($i=1,\ldots d$).
The expansion now proceeds in powers of ${\bf s}$, or $\nabla {\cal S}$.
The term linear in ${\bf s}$ vanishes by symmetry. To quadratic order in
${\bf s}$, both the next--to--last term under the $\Tr\ln$ (after a partial
integration) and the square of the preceding term contribute. The result
is a contribution to the action,
\begin{mathletters}
\label{eqs:3.44}
\begin{equation}
\frac{1}{4m}\sum_{12}\frac{1}{V}\sum_{\bf q} \eta_{12,ij}({\bf q}) 
    \int d{\bf x}\ s^i_{12}({\bf x})\,s^j_{21}({\bf x})\quad,
\label{eq:3.44a}
\end{equation}
with
\begin{eqnarray}
\eta_{12,ij}({\bf q})&=&\delta_{ij}\,\frac{1}{2}\,\left[
   G_{sp}({\bf q},\omega_{n_1}) + G_{sp}({\bf q},\omega_{n_2})\right] 
\nonumber\\
&&+ \frac{1}{m}\,q_i\,q_j\,G_{sp}({\bf q},\omega_{n_1})
       G_{sp}({\bf q},\omega_{n_2})\quad.
\label{eq:3.44b}
\end{eqnarray}
\end{mathletters}%
The approximation that neglects all fluctuations of $P$ and $\Lambda$ implies
a factorization of four--point functions into products of Green functions,
and hence Eq.\ (\ref{eq:3.35b}) applies. We thus have\cite{GIdentityFootnote}
\begin{mathletters}
\label{eqs:3.45}
\begin{equation}
\lim_{n_1,n_2 \rightarrow 0}\frac{1}{V} \sum_{\bf q} \eta_{12,ij}({\bf q}) =
   \frac{1}{2}\,\left(1 - \pi_1 \pi_2\right)\,\delta_{ij}\,\eta\quad,
\label{eq:3.45a}
\end{equation}
where
\begin{eqnarray}
\eta&=&\frac{1}{2V} \sum_{\bf q} \bigl( G_+({\bf q}) + G_-({\bf q})\bigr)
\nonumber\\
   &&+ \frac{1}{dmV} \sum_{\bf q} {\bf q}^2\,G_+({\bf q})\,G_-({\bf q})\quad,
\label{eq:3.45b}
\end{eqnarray}
\end{mathletters}%
with $G_{\pm} = G_{sp}(\omega_n\rightarrow \pm 0)$. 
The structure of Eq.\ (\ref{eq:3.45a})
allows one to write the low--frequency limit of the term quadratic in $\bf s$ 
in terms of the matrix $\hat Q$ defined in Eq.\ (\ref{eq:3.42b}),
\begin{eqnarray}
\sum_{12}\int&d{\bf x}&\ s^i_{12}({\bf x})\,s^j_{21}({\bf x})\,
   \frac{1}{V}\sum_{\bf q}\eta_{12,ij}({\bf q})
\nonumber\\
&=&\frac{-1}{4}\,\eta \int d{\bf x}\ \tr\left(\nabla{\hat Q}({\bf x})\right)^2
   + O(\nabla^3,\Omega\nabla)\ .
\label{eq:3.46}
\end{eqnarray}
Finally, we can use Eq.\ (\ref{eq:3.35b}) again to rewrite the coupling
constant $\eta$ in a more familiar form,
\begin{equation}
\eta = \frac{-1}{2dmV} \sum_{\bf q} {\bf q}^2\,\bigl[G_+({\bf q}) - G_-({\bf q})
           \bigr]^2
     = \pi\,m\,\sigma_0\quad.
\label{eq:3.47}
\end{equation}
One recognizes $\sigma_0$ as the conductivity in the self--consistent Born 
approximation with the interaction taken into account in Hartree--Fock
approximation.

The remaining task is to deal with the fields $P$ and $\Lambda$. Since
the fluctuations of these fields are massive modes, in order to capture
the leading effects of the soft modes in the system it suffices to
integrate them out in any approximation that preserves the Ward identities
derived in Sec.\ \ref{subsec:III.A}. The simplest way to do so is to
integrate them out in saddle--point approximation, i.e. to neglect 
$\Delta P$ and $\Delta\Lambda$ everywhere. From Eqs.\ (\ref{eq:3.38}),
(\ref{eqs:3.39}) we see that then the entire action can be expressed in
terms of $\hat Q$. Finally, we subtract from $\hat Q$ its saddle--point 
value $\langle{\hat Q}\rangle = \pi$, and define a matrix field 
${\tilde Q} = {\hat Q} - \pi$. Replacing $\hat Q$ by $\tilde Q$
everywhere in the action leads just to an uninteresting constant
contribution. This is obvious for all terms except for the $r=0$ channel
in the spin--singlet interaction term. There a term linear in $\hat Q$
seems to remain, but inspection shows that it is proportional to
$\tr {\hat Q} = 0$. We then obtain the following effectice action,
\begin{mathletters}
\label{eqs:3.48}
\begin{eqnarray}
{\cal A}_{NL\sigma M}&=&\frac{-\pi}{16}\,\sigma_0\,\int d{\bf x}\ 
     \tr\left(\nabla\tilde Q ({\bf x})\right)^2
\nonumber\\
&&+ \frac{\pi N_F}{4} \int d{\bf x}\ \tr\left(\Omega\,{\tilde Q}({\bf x})
     \right) + {\cal A}_{int}[\tilde Q]\quad,
\nonumber\\
\label{eq:3.48a}
\end{eqnarray}
in terms of the matrix field ${\tilde Q} = {\hat Q} - \pi$, where
$\pi$ is the diagonal matrix defined in Eq.\ (\ref{eq:3.39b}), and
$\hat Q$ is subject to the constraints,
\begin{equation}
{\hat Q}^2({\bf x}) \equiv \openone\otimes\tau_0\quad,\quad 
       {\hat Q}^{\dagger} = {\hat Q}\quad,\quad\tr{\hat Q}({\bf x}) = 0\quad.
\label{eq:3.48b}
\end{equation}
\end{mathletters}%
We note that the hermiticity of $\hat Q$ is a consequence of our having
dropped its anti--hermitian part, since it does not contribute to the
leading terms in the long--wavelength, low--frequency effective action.
The effective action given by Eqs.\ (\ref{eqs:3.48}) 
is the generalized nonlinear sigma--model that was
first proposed by Finkel'stein\cite{Finkelstein} as a model for interacting
disordered electrons, and whose properties have been studied in
considerable detail, in particular with respect to the metal--insulator
transition that is described by the model.\cite{R,LandauMIT} In the
next subsection, we will discuss ${\cal A}_{NL\sigma M}$, 
as well as its corrections,
from a RG point of view, and show that it suffices for a correct description
of both the critical FP and the stable Fermi--liquid FP, as well as the
leading corrections to scaling at the latter.

\subsubsection{The disordered Fermi--liquid fixed point}
\label{subsubsec:III.B.2}

In this subsection we discuss the RG properties of the model derived in
the previous section and, in particular, introduce the concept of a
disordered Fermi--liquid FP. Physically, this FP
characterizes a system with a finite density of states at the Fermi surface
and diffusive two--particle excitations.

\paragraph{The fixed point action}
\label{subsubsubsec:III.B.2.a}

To proceed, it is convenient to parametrize the matrix field $\hat Q$
in a way analogous to the $\pi$-field parametrization of the $O(N)$ vector
model, see Appendix \ref{app:B}. We thus write $\hat Q$ in a block matrix
form analogous to that used in Eq.\ (\ref{eq:3.22}) as
\begin{equation}
{\hat Q} = \left( \begin{array}{cc}
                 \sqrt{1-qq^{\dagger}} & q   \\
                    q^{\dagger}        & -\sqrt{1-q^{\dagger} q} \\
           \end{array} \right)\quad,
\label{eq:3.49}
\end{equation}
where the matrix $q$ has elements $q_{nm}$ whose frequency labels are
restricted to $n\geq 0$, $m<0$. This representation builds in the
constraints given by Eq.\ (\ref{eq:3.48b}). While the sigma--model
can be entirely expressed in terms of $\hat Q$, this is not the case for
the corrections to it. We therefore also express $\cal S$ in terms of $q$.
For what follows, we will need only the first terms of an expansion in
powers of $q$. From Eqs.\ (\ref{eq:3.42b}) and (\ref{eq:3.49}) we obtain,
\begin{eqnarray}
{\cal S}&=&\openone\otimes\tau_0 + \Delta{\cal S}
\nonumber\\
        &=& \openone\otimes\tau_0 + \frac{1}{2}\,\left( \begin{array}{cc}
                              0 & -q \\
                              q^{\dagger} & 0 \\
                      \end{array} \right)
                                                     + O(q^2)\quad.
\label{eq:3.50}
\end{eqnarray} 
Now we perform a momentum-shell RG 
procedure.\cite{WilsonKogut,MomentumShellFootnote}
For the rescaling part of this transformation, we need to assign scale 
dimensions to the soft field $q$, and to the massive fields
$\Delta P$, $\Delta\Lambda$ from Eq.\ (\ref{eq:3.39d}). Choosing the scale
dimension of a length $L$ to be $[L] = -1$, we write in analogy to
Eqs.\ (\ref{eqs:B.6}),
\begin{mathletters}
\label{eqs:3.51}
\begin{equation}
[q({\bf x})] = \frac{1}{2}\,(d-2+\eta')\quad,
\label{eq:3.51a}
\end{equation}
\begin{equation}
[\Delta P({\bf x})] = [\Delta\Lambda ({\bf x})] = \frac{1}{2}\,(d-2+\eta)
                                                                     \quad,
\label{eq:3.51b}
\end{equation}
\end{mathletters}%
which defines the exponents $\eta$ and $\eta'$. The stable Fermi--liquid
FP of the theory is characterized by the choice,\cite{MaFootnote}
\begin{mathletters}
\label{eqs:3.52}
\begin{equation}
\eta = 2\quad,\quad \eta' = 0\quad.
\label{eq:3.52a}
\end{equation}
Physically, $\eta'=0$ corresponds to diffusive correlations of the $q$, and
$\eta = 2$ means that the correlations of the $\Delta P$ are of short range.
This is indeed what one expects in a disordered Fermi liquid.
In addition, we must specify the scale dimension of frequency or temperature,
i.e. the dynamical scaling exponent $z = [\omega] = [T]$. In order for the
FP to be consistent with diffusion, that is with frequencies that scale like
the squares of wave numbers, we must choose
\begin{equation}
z = 2\quad.
\label{eq:3.52b}
\end{equation}
\end{mathletters}%

Now we expand the sigma--model action, Eq.\ (\ref{eq:3.48a}), in powers of
$q$. In a symbolic notation that leaves out everything not needed for
power counting purposes, we write
\begin{eqnarray}
{\cal A}_{NL\sigma M}&=&\frac{-1}{G} \int d{\bf x}\ 
     (\nabla q)^2 + H \int d{\bf x}\ \omega \,q^2 
\nonumber\\
&&+ \Gamma\, T \int d{\bf x}\ q^2 + O(\nabla^2\,q^4,\omega\,q^4, T\,q^3) \quad,
\label{eq:3.53}
\end{eqnarray}
with the bare coupling constants $G\sim 1/\sigma_0$, and $H \sim N_F$.
$\Gamma$ can stand for any of the three interaction coupling constants.
Power counting shows that, with the above choices for the exponents, 
all of these coupling constants have
vanishing scale dimensions with respect to our FP,
\begin{equation}
[G] = [H] = [\Gamma] = 0\quad.
\label{eq:3.54}
\end{equation}
These terms therefore make up part of the FP action. 

Now consider first the corrections that arise within the sigma--model.
The leading ones have been indicated in Eq.\ (\ref{eq:3.53}). We denote
the corresponding coupling constants by $c_{\nabla^2 q^4}$, etc., with
a subscript that identifies the structure of the respective contribution
to the action. One finds
\begin{mathletters}
\label{eqs:3.55}
\begin{equation}
\left[c_{\nabla^2 q^4}\right] = \left[c_{\omega q^4}\right] = -(d-2)\quad.
\label{eq:3.55a}
\end{equation}
With respect to the remaining term we note that it is of order $q^3$. Any
contribution to physical $q$-correlation functions therefore contain this
term squared, and therefore the relevant scale dimension is
\begin{equation}
\left[\left(c_{T q^3}\right)^2\right] = 2\,\left[c_{T q^3}\right] = 
                                                                 -(d-2)\quad.
\label{eq:3.55b}
\end{equation}
\end{mathletters}%
We see that all of these operators are irrelevant with respect to the 
Fermi--liquid FP for all dimensions $d>2$, and 
that they become marginal in $d=2$
and relevant in $d<2$. All other terms that are contained in the sigma--model
are more irrelevant than the ones considered.

We next consider corrections to the sigma--model action.
From Eq.\ (\ref{eq:3.38}) we see that there are five classes of such terms:
(1) ${\cal A}_{dis}$, (2) the term $\tr\,(\Lambda P)$, (3) those parts of the
$\Tr\ln$-term that contain the massive fluctuation $\Delta\Lambda$,
(4) the parts of ${\cal A}_{int}$ 
that contain the massive fluctuation $\Delta P$, and (5) contributions to
$\langle P\rangle$ that were neglected in writing Eqs.\ (\ref{eq:3.41}) and
(\ref{eq:3.42b}). Since all
terms linear in the fluctuations vanish, class (1) just contributes
\begin{eqnarray}
{\cal A}_{dis}[\Delta P]&=&-\frac{1}{\tau_1}\int d{\bf x}\ 
     \left(\tr\,\Delta P({\bf x})\right)^2 
\nonumber\\
&&+ \frac{1}{\tau_{rel}} \int d{\bf x}\ \tr\left(
                           \Delta P({\bf x})\right)^2\quad,
\label{eq:3.56}
\end{eqnarray}
where we have scaled the relaxation times $\tau_1$ and $\tau_{rel}$
from Eqs.\ (\ref{eqs:2.22}) with $1/2\pi N_F$ and $1/\pi N_F$,
respectively. Power counting yields
\begin{equation}
\left[\tau_1\right] = \left[\tau_{rel}\right] = 0\quad,
\label{eq:3.57}
\end{equation}
so ${\cal A}_{dis}$ is part of the FP action. 
Class (2) also contributes a marginal
term to the action, and so does the leading contribution of class (3), which
has the structure $\Tr (G\,\Delta\Lambda)^2$. Combining these
two marginal terms, we have a contribution to the FP action,
\begin{eqnarray}
{\cal A}_{\Lambda P}&=&\int d{\bf x}\ \tr\,\left(\Delta\Lambda ({\bf x})\,
     \Delta P ({\bf x})\right) 
\nonumber\\
&+& \frac{1}{4} \int d{\bf x}\,d{\bf y}\ \tr\big(G({\bf x}-{\bf y})\,
   \Delta\Lambda({\bf y})\,G({\bf y}-{\bf x})\,\Delta\Lambda ({\bf x})\bigr),
\nonumber\\
\label{eq:3.58}
\end{eqnarray}
apart from an uninteresting constant. The remaining terms of class (3) are
irrelevant contributions that couple $q$ and $\Delta\Lambda$. The least
irrelevant terms have the structures $\Tr\,(\Delta\Lambda\,\omega\,q)$, and
$\Tr\,(\Delta\Lambda\,\nabla^2\,q)$. For contributions to physical correlation
functions we must again consider the squares of these structures, and the 
relevant scale dimensions are
\begin{equation}
\left[\left(c_{\Delta\Lambda\,\omega\,q}\right)^2\right] =
  \left[\left(c_{\Delta\Lambda\,\nabla^2\,q}\right)^2\right] = -2\quad.
\label{eq:3.59}
\end{equation}
These terms are more irrelevant than the corrections contained 
in the sigma--model
for all $d<4$. Moreover, due to the even integer value of the scale dimension,
they lead only to analytic, and therefore uninteresting, corrections to scaling
near the FP. The leading term of class (4) is
one that couples $q$ (from one of the $\cal S$) and $\Delta P$. The
corresponding coupling constant has a scale dimension
\begin{equation}
\left[\left(c_{Tq\,\Delta P}\right)^2\right] = -2\quad.
\label{eq:3.60}
\end{equation}
Again, these operators lead only to analytic corrections to scaling. Cross
correlations between these various terms also appear, and the corresponding
operators have the same scale dimension, namely $-2$, as the ones shown.
Finally, it is easily checked that the least irrelevant terms of class (5)
are of the structure $\omega^2 q^2$, and they also have a scale dimension
\begin{equation}
\left[c_{\omega^2 q^2}\right] = -2\quad.
\label{eq:3.60'}
\end{equation}
 
The above RG arguments show that the theory contains a Fermi--liquid FP that
is stable for all $d>2$, and analogous to the low--temperature FP of an
$O(N)$ Heisenberg model. The FP action consists of the nonlinear
sigma--model action ${\cal A}_{NL\sigma M}$, Eq.\ (\ref{eq:3.48a}), expanded to
second order in $q$, and of ${\cal A}_{dis}$ and ${\cal A}_{\Lambda P}$, Eqs.\ 
(\ref{eq:3.56}) and (\ref{eq:3.58}).
We have further shown that the leading
nonanalytic corrections to scaling near the Fermi--liquid FP are contained
in the nonlinear sigma--model. An effective action that describes both the
FP and the leading corrections to scaling is therefore
\begin{equation}
{\cal A}_{eff} = {\cal A}_{NL\sigma M}[q] + {\cal A}_{dis}[\Delta P] 
       + {\cal A}_{\Lambda P}[\Delta\Lambda,\Delta P]\quad.
\label{eq:3.61}
\end{equation}
Notice that the soft and massive modes do not couple in this effective
action. Furthermore, the massive fields $\Delta P$ and $\Delta\Lambda$ appear 
only quadratically. If desired, then the latter can be integrated out to
obtain an action entirely in terms of the physical fields $q$ and $\Delta P$.


\paragraph{Scaling behavior of observables}
\label{subsubsubsec:III.B.2.b}

We now discuss the physical meaning of the corrections to scaling induced
by the irrelevant operators that we have identified above. Let us denote by
the generic name $u$ any of the least irrelevant operators whose scale
dimension is $[u] = -(d-2)$, and let us discuss various observables, viz.
the conductivity $\sigma$, the specific heat coefficient $\gamma_V$, the
single--particle density of states $N$, and the spin susceptibility $\chi_s$.
Which of the various operators with scale dimension $-(d-2)$
is the relevant one depends on the quantity under consideration.

Let us first consider the dynamical conductivity, $\sigma (\omega)$. 
Its bare value is proportional to
$1/G$, and according to Eq.\ (\ref{eq:3.54}) its scale dimension is zero.
We therefore have the scaling law,
\begin{mathletters}
\label{eqs:3.62}
\begin{equation}
\sigma (\omega,u) = \sigma (\omega\,b^z,ub^{-(d-2)})\quad,
\label{eq:3.62a}
\end{equation}
where $b$ is an arbitrary RG scale factor. By putting $b=1/\omega^{1/z}$,
and using $z=2$, Eq.\ (\ref{eq:3.52b}), as well as the fact that $\sigma (1,x)$ 
is an analytic function of $x$,
we find that the conductivity has a singularity at zero frequency, or a
long--time tail, of the form
\begin{equation}
\sigma (\omega) = {\rm const.} + \omega^{(d-2)/2}\quad.
\label{eq:3.62b}
\end{equation}
\end{mathletters}%
This nonanalyticity is well known from perturbation theory for both
noninteracting\cite{GLK} and interacting\cite{AAKL} electrons. 
This shows that in this
case either one of the coupling constants in Eqs.\ (\ref{eq:3.55a}) and
(\ref{eq:3.55b}) can play the role of $u$.
The present analysis proves that the $\omega^{(d-2)/2}$ is the exact 
leading nonanalytic behavior.

The specific heat coefficient, $\gamma_V = C_V/T$, is proportional to
the quasiparticle density of states
$H$,\cite{CastellaniDiCastro,CastellaniKotliarLee,Castellanietal}
whose scale dimension vanishes according to Eq.\ (\ref{eq:3.54}).
We thus have a scaling law
\begin{mathletters}
\label{eqs:3.63}
\begin{equation}
\gamma_V(T,u) = \gamma_V(Tb^z,ub^{-(d-2)})\quad,
\label{eq:3.63a}
\end{equation}
which leads to a low--temperature behavior
\begin{equation}
\gamma_V(T) = {\rm const.} + T^{(d-2)/2}\quad.
\label{eq:3.63b}
\end{equation}
\end{mathletters}%
From perturbation theory it is known\cite{AAKL}
that $\gamma_V$ shows this behavior 
only for interacting electrons, while for noninteracting systems the
prefactor of the nonanalyticity vanishes. In this case, therefore, $u$
must be equated with $(c_{T q^3})^2$. This can not be seen by our simple
counting arguments.

The single--particle density of states, $N$, is proportional to the 
expectation value of
$Q$, and to study the leading correction to the finite FP value of $N$
it suffices to replace $Q$ by $\hat Q$. Then we have, in symbolic notation,
$N \sim 1 + \langle q\,q^{\dagger}\rangle + \ldots = 1 + \Delta N$. The
scale dimension of $\Delta N$ is $[\Delta N] = 2 [q] = d-2$. We find the
scaling law
\begin{mathletters}
\label{eq:3.64}
\begin{equation}
\Delta N(\omega) = b^{-(d-2)}\,\Delta N(\omega b^{z})\quad,
\label{eq:3.64a}
\end{equation}
which leads to the so--called Coulomb anomaly,\cite{AA}
\begin{equation}
N(\omega) = {\rm const.} + \omega^{(d-2)/2}\quad,
\label{eq:3.64b}
\end{equation}
\end{mathletters}%
Again, this behavior is known to occur only in the presence of 
electron--electron interactions.

Finally, we consider the wave vector dependent spin susceptibility,
$\chi_s({\bf q})$. $\chi_s$ is given by a $Q$-$Q$ correlation function,
and the leading correction to the finite Fermi--liquid value is obtained
by replacing both of the $Q$ by $q$. Then we have a term of the
structure $\chi_s \sim T\int d{\bf x}\ \langle q^{\dagger}q\rangle$,
with scale dimension $[\chi_s] = 0$. The relevant scaling law is
\begin{mathletters}
\label{eqs:3.65}
\begin{equation}
\chi_s ({\bf q},u) = \chi_s ({\bf q}b,ub^{-(d-2)})\quad,
\label{eq:3.65a}
\end{equation}
which leads to a nonanalytic dependence on the wave number,
\begin{equation}
\chi_s ({\bf q}) = {\rm const.} + \vert {\bf q}\vert^{(d-2)}\quad.
\label{eq:3.65b}
\end{equation}
\end{mathletters}%
This behavior is also known from perturbation theory, and holds only for
interacting electrons. It has recently been shown to have interesting
consequences for the theory of ferromagnetism.\cite{fm_dirty}

To summarize, we see from the above arguments that all of the so--called
weak--localization effects,\cite{WeakLocalizationFootnote}
i.e. nonanalytic dependencies of various
observables on frequency, temperature, or wave number in disordered electron
systems that are well known from perturbation theory,
emerge naturally in
the present context as the leading corrections to scaling near the 
Fermi--liquid FP of a general field theory for disordered interacting electrons.
Apart from providing an aesthetic, unifying, and very simple explanation
for these effects, our arguments also prove that they do indeed constitute
the leading nonanalytic behavior, a conclusion that cannot be drawn from
perturbation theory alone. We emphasize the ease with which these results
are obtained within the present framework. This reflects the judicious choice
of our starting point for the RG analysis. The transformation from fermionic
variables to composite ones, together with the symmetry analysis that
identifies the soft modes, provides for a formulation of the theory that
allows for rather nontrivial results to be obtained from a simple RG analysis
at the Gaussian level. Similar benefits can be obtained from analogous
treatments of other systems, see Appendix \ref{app:B}.

Finally, it is interesting to note that similar 
nonanalyticities occur in very different systems, for example, in
classical fluids. In that context they are known as long--time tail effects, and
were first discussed theoretically by using many body perturbation theory and
mode coupling theory.\cite{DorfmanTRKSengers} Later, they were examined using 
RG ideas, and they
were shown to be related to corrections to the scaling behavior near a
hydrodynamic FP.\cite{ForsterNelsonStephen}

\subsubsection{The critical fixed point}
\label{subsubsec:III.B.3}

In addition to the stable Fermi--liquid FP discussed above, the
generalized matrix nonlinear sigma--model given by Eq.\ (\ref{eq:3.48a}) 
also possesses another
FP that in general has the properties of a critical FP.
It characterizes a zero--temperature metal--insulator transition in
dimensions greater than two. For a review of this FP,
and of the various universality classes for the metal--insulator transition, we
refer the reader elsewhere.\cite{R} Here we briefly comment on the differences
between the Fermi--liquid FP and the critical one, and on the
justification for using the sigma--model to describe the critical behavior.

First of all, at the critical point the spontaneously broken symmetry is
restored, so the distinction between positive and negative
frequencies vanishes as the magnitude of the frequencies goes to 
zero.\cite{AndersonTransitionFootnote}
Physically this corresponds to a vanishing single--particle of density of
states at the Fermi level, which plays the role of an order parameter. 
This characterizes the metal--insulator transition as being of
Anderson--Mott type. For example, the left--hand side of 
Eq.\ (\ref{eq:3.17a}) vanishes at the transition 
as $n_1,n_2\rightarrow 0$. Therefore, 
matrix elements $\hat{Q}_{n_1n_2}$ with $n_1n_2>0$ 
and $n_1n_2<0$, respectively, should both scale in the same way. This implies
that at the critical FP, $\eta'$ in Eq.\ (\ref{eq:3.51a}) is given
by
\begin{equation}
\eta' = 2-d \equiv - \epsilon\quad,
\label{eq:3.66}
\end{equation}
so that $q$ is dimensionless. The physical critical exponent, $\eta$,
on the other hand, is given by
\begin{equation}
\eta = -\epsilon + 2\beta/\nu\quad.
\label{eq:3.67}
\end{equation}
To see this, we note that $\hat Q$ is proportional to $\langle P\rangle$,
which in turn is proportional to the density of states which vanishes at
the transition with a critical exponent $\beta$.\cite{etaFootnote1}
Here $\nu$ is the critical exponent that describes the divergence of the
localization or correlation length $\xi$. It occurs in the
scaling equality, Eq.\ (\ref{eq:3.67}), since $\eta$
characterizes momentum singularities at criticality.

We add some comments on the scaling of $\Delta P$ fluctuations.
At the critical point, the $\Delta P$ fluctuations are actually of
long range, and the sigma--model description breaks down. To avoid this
problem, one must work at sufficiently small momenta $p$ (or frequencies),
and at sufficiently large distances $t$
from the critical point, so that the $q$ fluctuations still
dominate over the $\Delta P$ fluctuations. Since the $\Delta P$ correlation
function is related to the longitudinal susceptibility, it diverges as 
$t^{-\gamma}$ while the $q$ correlations diverge as 
$p^{-2+\eta}$.\cite{ScaleDimensionFootnote} The relevant inequality is 
therefore
\begin{equation}
p\ll t^{\gamma/(2-\eta)} = t^{\nu} \quad,
\label{eq:3.68}
\end{equation}
where we have used the scaling equality $\nu =\gamma/(2-\eta)$. This can
also be seen more explicitly by considering the $\Delta P$-$\Delta P$
propagator. From Eq.\ (\ref{eq:3.56}), and the corresponding term that
is proportional to $\bigl(\nabla(\Delta P)\bigr)^2$ 
one sees that the $\Delta P$-mass
$m_P$ has a scale dimension of $1$ (see also Appendix\ \ref{app:B.3}). The
$\Delta P$ fluctuations can therefore be neglected for momenta that are
small compared to $m_P \sim \xi^{-1} \sim t^{\nu}$, which is again
Eq.\ (\ref{eq:3.68}). In this
regime the $\Delta P$ fluctuations are effectively of short range, just as
they are at the Fermi--liquid FP, and the sigma--model is valid.
Even though the range of validity of the sigma--model goes to zero as the
critical point is approached, it still can be used to extract the critical
behavior. The salient point is that the RG, and the scaling behavior
derived from it, allows one to extrapolate from the regime given by
Eq.\ (\ref{eq:3.68}) to the critical region. However, no information can
be obtained in this way about the symmetric phase, i.e. the insulator.

Apart from these theories that work near $d=2$, the matrix nonlinear 
sigma--model has also been studied in high dimensions.\cite{LandauMIT}
This work has established $d_c^+ = 6$ as an upper critical dimension,
above which the Anderson--Mott transition is correctly described by a
simple Landau--type theory. This treatment of the problem stresses that
the metal--insulator transition problem is, somewhat counterintuitively,
conceptually simpler in the presence of electron--electron interactions 
than in their absence, since the interacting
problem possesses a simple critical order parameter, viz. the density
of states at the Fermi level. Also, in these papers the presence of
random--field like terms in the renormalized action was discovered. 
These terms are responsible for
the upper critical dimension being $6$ (rather than $4$ as in the $O(N)$
vector model), and they have led to the suggestion that the Anderson--Mott
transition may have features reminiscient of a glass transition.\cite{glass}

\subsubsection{Ferromagnetic transition in the metallic phase}
\label{subsubsec:III.B.4}

It has been known for some time that the matrix nonlinear sigma--model,
Eq.\ (\ref{eq:3.48a}), contains another critical FP that is not 
related to
a metal--insulator transition.\cite{IFS} While originally the nature of the
corresponding phase transition was not clear, we have recently shown that
it is the zero--temperature transition from a paramagnet to a
ferromagnet in a disordered metal in dimensions greater
than two.\cite{fm_dirty} In this reference, a Landau--Ginzburg--Wilson
(LGW) functional for spin
density, or order--parameter, fluctuations has been derived. To obtain this 
functional, we integrated out all excitations or modes other than the order 
parameter, including all of the soft
modes related to weak localization effects, see Sec.\ \ref{subsubsec:III.B.2}
above. In the effective, or
order--parameter, field theory these extra soft modes led to nonanalyticies
in the bare LGW functional. Using renormalization group methods, we were
able to exactly determine the critical behavior at the magnetic phase
transition.\cite{fm_dirty} The earlier approach of Ref.\ \onlinecite{IFS}, 
on the other hand, had
focused entirely on the behavior of the diffusion modes
across the magnetic phase transition, while the order parameter fluctuations
had effectively been integrated out. In some respects, these two approaches
to the problem were therefore complementary to one another.

Even though both of these theories led to the same results, 
their physical underpinnings are very
different, and neither one of them constitutes what one might consider
the physically most obvious approach.
Physically, the most sensible procedure would be to treat all of 
the soft modes at this
quantum phase transition on the same footing. These soft modes
include the diffusive modes discussed above, i.e. the $Q_{n_1n_2}$ with 
$n_1n_2<0$, as well as the fluctuations of the magnetic order parameter,
that is, the spin density or magnetization.
In terms of $Q$-fields, the local spin density at frequency $\omega_n$,
$n_s^i({\bf x},\omega_n)$, is given by (cf. Eq.\ (\ref{eq:2.11b})
\begin{equation}
n_s^i(\omega_n,{\bf x}) = \sqrt{T}\sum_m\sum_{r=0,3}(-1)^r\ 
                 {^i_r Q}_{m,m+n}^{\alpha\alpha}({\bf x})\quad,
\label{eq:3.69}
\end{equation}
and the macroscopic magnetization order parameter, ${\bf M}$, has components,
\begin{eqnarray}
M^i&=&\bigl\langle n_s^i(\omega_n=0,{\bf x})\bigr\rangle 
\nonumber\\
   &=&\sqrt{T}\sum_{m}\sum_{r=0,3}(-1)^r
    \left\langle{^i_r Q}_{mm}^{\alpha\alpha}({\bf x})\right\rangle\quad.
\label{eq:3.70}
\end{eqnarray}
From these equations we see that, as the magnetic transition is
approached, the classification of soft modes that was given
in Section\ \ref{subsec:III.A} above, breaks down. 
New soft modes related to the criticality of $Q_{mm}$ occur. 
Alternatively, the extra soft mode that appears near the
ferromagnetic transition can be related to a new zero eigenvalue in the
Gaussian eigenvalue problem. The matrix $M$ in Eqs.\ (\ref{eq:2.34a}) and 
(\ref{eq:2.36a}) is proportional to,
\begin{eqnarray}
{^i_0 M}_{n_1n_2,n_3n_4}(p) &\sim&\delta_{n_1n_3}\delta_{n_2n_4}
       {\cal D}_{n_1n_2}^{-1}({\bf p}) 
\nonumber\\
&&-2T\Gamma_t\,\delta_{n_1-n_2,n_3-n_4}\quad, 
\label{3.71}
\end{eqnarray}
with $i=1,2,3$ and we have considered the term diagonal in all of
the replica labels since these terms contain the interactions that lead to
magnetism. The corresponding eigenvalue problem is,
\begin{eqnarray}
{\cal D}_{n_1n_2}^{-1}({\bf p}) f^i_{n_1n_2}({\bf p}) 
    - 2\Gamma_tT\sum_n f^i_{n,n-(n_1-n_2)}({\bf p})
\nonumber\\
     = \lambda f^i_{n_1n_2}({\bf p})\quad , 
\label{eq:3.72}
\end{eqnarray}
with $f$ the eigenfunction and $\lambda$ the eigenvalue.
Setting $n_1=n_2$ and integrating over $n_1$ leads to an equation for 
$\lambda$,
\begin{equation}
1=2\Gamma _tT\sum_n\frac{{\cal D}_{nn}({\bf p}=0)}
     {1-\lambda {\cal D}_{nn}({\bf p}=0)}\quad.
\label{eq:3.73}
\end{equation}
Criticality is characterized by a new zero eigenvalue, $\lambda = 0$, and
we have put ${\bf p}=0$, since this is where the first zero eigenvalue
occurs. Eq.\ (\ref{eq:3.73}) reduces to a Stoner criterion for the occurrence
of ferromagnetism, modified by disorder. 
Expanding in powers of $\lambda$ leads to
\begin{equation}
\lambda \sim -t\quad,
\label{eq:3.74}
\end{equation}
with $t$ the dimensionless distance from the critical point. From
Eq.\ (\ref{eq:3.72}), we see that the critical eigenfunction is
\begin{equation}
f^i_{nn}({\bf p}) = {\cal D}_{nn} 2\Gamma_t\sqrt{T}\, n_s^i(\omega_n=0,{\bf p})
   \quad.
\label{eq:3.75}
\end{equation}

The above discussion makes it clear that a physically satisfactory theory
of the ferromagnetic, or any other, phase transition in a disordered metal
should take both the soft modes that were discussed in the preceding
subsections, and the additional, critical, soft modes into account on the
same footing. Technically one needs to extract the critical part from the
$P$-fluctuations. The details of such a theory of the 
paramagnetic--to--ferromagnetic phase transition will be discussed 
elsewhere.\cite{ustbp} Here we just mention why the sigma--model, in
Refs.\ \onlinecite{IFS}, manages to
describe the correct critical behavior, even though it neglects the soft
modes related to the magnetization. The salient point is again that there
is a regime, described by Eq.\ (\ref{eq:3.68}), where the diffusive modes
that are contained in the sigma--model dominate over the magnetization
fluctuations that are not. As in the case of the metal--insulator
transition, the RG allows one to extract the critical behavior from an
analysis of that region.

\section{The clean limit}
\label{sec:IV}

In this section we discuss the clean limit of our field theory. For the
reasons that were pointed out after Eqs.\ (\ref{eqs:2.12}) in 
Sec.\ \ref{subsec:II.A}, this treatment of clean Fermi systems will not
be as complete as our theory for the disordered case. For notational
simplicity we will also suppress the particle--particle degrees of freedom,
but they can be easily restored. Even with
our restriction to two Fermi--liquid parameters (in the absence of the Cooper
channel) instead of infinitely
many, we will be able to study fundamental structural properties of the
clean limit that will have to be included in any more complete theory.
 
\subsection{Effective $Q$-field theory}
\label{subsec:IV.A}
 
Let us return to the action, Eq.\ (\ref{eq:2.20}), and perform the clean
limit, $\tau_1,\tau_{rel}\rightarrow\infty$. 
As pointed out earlier, the theory has
been purposely set up so that this limit can be taken without difficulty.
${\cal A}_{dis}$ then vanishes. The saddle point solution, 
Sec.\ \ref{subsec:II.C}, reduces to the ordinary Hartree--Fock approximation,
and the Gaussian approximation reduces to RPA.
In particular, the saddle point action contains an exact description of
the noninteracting electron gas. This will be important in what follows.

As was pointed out at the end of Section\ \ref{sec:II}, the fluctuations
of $\bar\Lambda$ are not massive in the clean case, since the function
$\varphi$, Eq.\ (\ref{eq:2.35b}), is singular in the long--wavelength,
low--frequency limit, see Eq.\ (\ref{eq:2.40}). Indeed, Eqs.\ (\ref{eqs:2.34})
and (\ref{eqs:2.36}) show that the Gaussian $\bar\Lambda$-propagator is just
minus the noninteracting part of the $Q$-propagator. On the general principle
that one should keep all of the soft modes on the same footing, this seems to
suggest keeping the $\bar\Lambda$ field at the same level as the $Q$-field.
Dropping the constant saddle--point contribution, and
expanding the $\Tr\ln$ in Eq.\ (\ref{eq:2.20}), the action then reads
\begin{mathletters}
\label{eqs:4.1}
\begin{equation}
{\cal A}[Q,\bar\Lambda] = {\cal A}^{clean}_G[Q,\bar\Lambda] 
       + \Delta {\cal A} [Q,\bar\Lambda]\quad,
\label{eq:4.1a}
\end{equation}
where ${\cal A}^{clean}_G = {\cal A}_G - {\cal A}_{dis}$ 
with ${\cal A}_G$ and ${\cal A}_{dis}$ from Eq.\ (\ref{eq:2.33a}), and
\begin{eqnarray}
\Delta {\cal A} [Q,\bar\Lambda] = -\sum_{M=3}^{\infty}\frac{2^{M-1}}{M}
   \frac{i^M}{V^{(M-2)/2}} \sum_{{\bf k}_1,\ldots {\bf k}_M}
                         \sum_{n_1\ldots n_M} 
\nonumber\\
\times\delta_{{\bf k}_1 + \ldots + {\bf k}_M,0}\ 
\chi^{(M)}_{n_1\ldots n_M}({\bf k}_1,\ldots,{\bf k}_{M-1})
\nonumber\\
\times\varphi^{-1}_{n_1n_2}({\bf k}_1)\cdots\varphi^{-1}_{n_Mn_1}({\bf k}_M)\,
\nonumber\\
  \times\tr\left[\left((\delta{\bar\Lambda})_{n_1n_2}({\bf k}_1) 
               - (\delta Q)_{n_1n_2}({\bf k}_1)\right)\cdots\right.
\nonumber\\
   \times\left.\left((\delta{\bar\Lambda})_{n_Mn_1}({\bf k}_M) 
               - (\delta Q)_{n_Mn_1}({\bf k}_M)\right)\right]\quad.
\nonumber\\
\label{eq:4.1b}
\end{eqnarray}
Here $\varphi$ is given by Eq.\ (\ref{eq:2.35b}), and
\begin{eqnarray}
\chi^{(M)}_{n_1\ldots n_M}({\bf k}_1,\ldots,{\bf k}_{M-1}) = \frac{1}{V}
   \sum_{\bf p} G_{sp}({\bf p},\omega_{n_1})
\nonumber\\
\times G_{sp}({\bf p} + {\bf k}_1,
      \omega_{n_2})\cdots G_{sp}({\bf p} + {\bf k}_{M-1},\omega_{n_M})\quad.
\label{eq:4.1c}
\end{eqnarray}
\end{mathletters}%

In this formulation of the theory the vertices are given by the product of the
$M$-point correlation function $\chi^{(M)}$ and the $M$ factors of 
$\varphi^{-1}$
in Eq.\ (\ref{eq:4.1b}). An inspection of Eq.\ (\ref{eq:4.1c}) shows that
the generic behavior of $\chi^{(M)}$, considered as a function of some generic
wave number $k$, is characterized by a divergence $1/k^{M-1}$ for small $k$.
This is the most divergent behavior $\chi^{(M)}$ can display, and it is
realized unless either all of the frequencies $\omega_{n_1}$ through 
$\omega_{n_M}$ are positive, or all of them are negative, in which case
$\chi^{(M)}$ scales like a constant. The behavior of
the product of the $\varphi^{-1}$, on the other hand, depends on the detailed
distribution of the frequency labels. If $n_{i}n_{i+1}>0$, then 
$\varphi_{n_{i}n_{i+1}}^{-1}$ is a number, while for $n_{i}n_{i+1}<0$ it goes
like $k$ for small $k$. If only two of the frequency pairs have elements 
with opposite signs (this is the smallest nonzero number possible), then the 
complete vertex scales like $1/k^{M-3}$. For $M>3$, the vertices are thus not
finite in the limit of long wavelengths and small frequencies. In other
words, the $Q$-$\bar\Lambda$ field theory given by Eqs.\ (\ref{eqs:4.1})
is not local. The reason for this is as follows. As was mentioned in 
Sec.\ \ref{sec:II}, in the clean limit the one--particle excitations are soft,
as is manifested by the massless single--particle Green function. These
soft modes have been integrated out when we integrated out the fermions,
and this is what leads to the nonlocalities in the effective field theory.
The massless single--particle excitations are also indirectly responsible
for the softness of the $\bar\Lambda$ fluctuations, and for the need to
keep infinitely many Landau parameters. All of these difficulties thus have
the same underlying source.

Given that the matrix field theory is nonlocal in any case, we can proceed
and integrate out $\bar\Lambda$, as it will turn out that this does not
lead to further undesirable properties of the theory. The integrating out
of $\bar\Lambda$ can be done exactly in the sense of a prescription for
doing perturbation theory for the resulting $Q$-field theory. To see this,
we recall that the $\bar\Lambda$ propagator is minus the noninteracting
part of the $Q$-propagator. As a result, the integration over $\bar\Lambda$
just cancels the noninteracting parts of any internal $Q$-propagators in a
loop expansion for any $Q$-correlation function. This is easy to see for
the first few terms in the loop expansion, and it can be proved by
induction to be true order by order in perturbation theory. As an illustrative
example, the cancellation scheme is demonstrated diagrammatically
in Figs.\ \ref{fig:2} and \ref{fig:3} for the
two--point vertex function, and the two--point propagator, respectively,
to one--loop order. Notice that, in order to avoid double counting, one effect 
of the $\bar\Lambda$-field {\em must} be the cancellation of the noninteracting
contributions, since the saddle--point contribution to the action already 
contains a complete description of the noninteracing electron system. What is 
remarkable is that it does nothing else.
%
\begin{figure}[t]
\centerline{\psfig{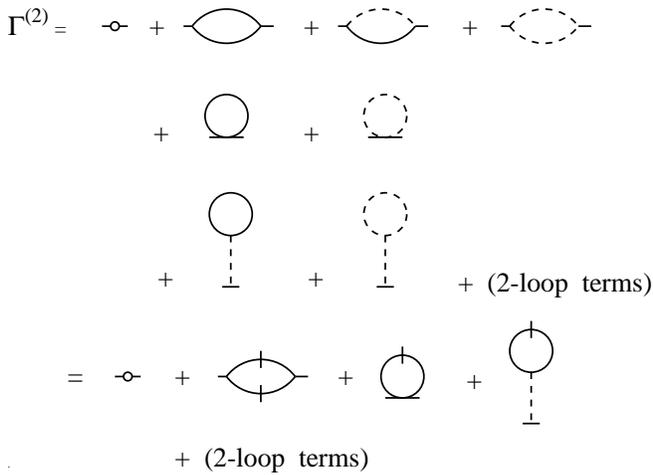}\vspace*{5mm}}
\caption{Perturbation theory to one--loop order for the 2-point $Q$-vertex
 function. The small circle denotes the Gaussian vertex, the solid and
 dashed lines are $Q$-propagators and $\bar\Lambda$-propagators, respectively,
 and the solid lines with a vertical bar denote the interacting part of the
 $Q$-propagator.}
\label{fig:2}
\end{figure}
\begin{figure}[t]
\centerline{\psfig{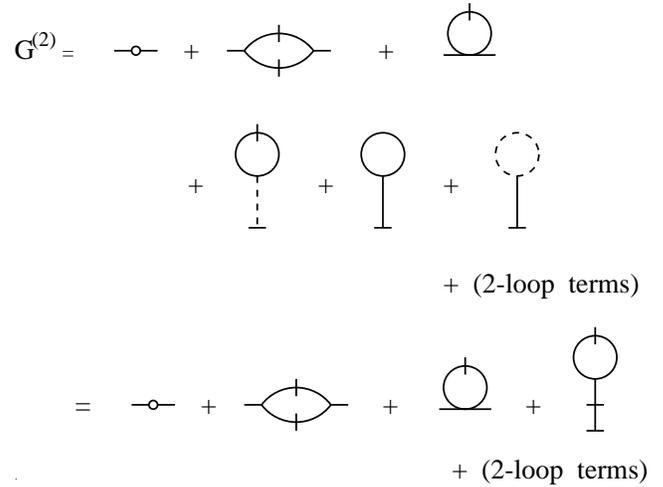}\vspace*{5mm}}
\caption{Perturbation theory to one--loop order for the 2-point $Q$-propagator.
 The notation is the same as in Fig.\ \ref{fig:2}.}
\label{fig:3}
\end{figure}
%
We then obtain
the following effective action entirely in terms of $Q$-fields,
\begin{mathletters}
\label{eqs:4.2}
\begin{equation}
{\cal A}[Q] = {\cal A}_0[Q] + {\cal A}_{int}[Q] + \Delta {\cal A} [Q]\quad,
\label{eq:4.2a}
\end{equation}
where 
\begin{eqnarray}
{\cal A}_0[Q] &=& -4 \sum_{\bf k} \sum_{nm} \sum_{r,i} {^i_r (\delta Q)}_{nm}
   ({\bf k})\,\varphi^{-1}_{nm}({\bf k})\,
\nonumber\\
&&\qquad\qquad\quad \times{^i_r (\delta Q)}_{nm}(-{\bf k})\quad,
\label{eq:4.2b}
\end{eqnarray}
is the noninteracting part of the Gaussian action, and
\begin{eqnarray}
{\cal A}_{int}[Q]&=& -8 \sum_{i=0}^{3} \sum_{r=0,3} \Gamma^{(i)} \sum_{\bf k}
             T\sum_{n_1,n_2,n_3,n_4} {^i_r (\delta Q)}_{n_1n_2}({\bf k})
\nonumber\\
&&\times\delta_{n_1-n_2,n_3-n_4}\,{^i_r (\delta Q)}_{n_3n_4}(-{\bf k})\quad,
\label{eq:4.2c}
\end{eqnarray}
is the interacting one. The $\Gamma^{(i)}$ $(i=0,1,2,3)$ were defined after
Eq.\ (\ref{eq:2.36b}). The non--Gaussian part of the action reads,
\begin{eqnarray}
\Delta {\cal A} [Q] = - \sum_{M=3}^{\infty}\frac{2^{M-1}}{M}
   \frac{(-i)^M}{V^{(M-2)/2}} \sum_{{\bf k}_1,\ldots {\bf k}_M}
                         \sum_{n_1\ldots n_M}
\nonumber\\
\times\delta_{{\bf k}_1 + \ldots + {\bf k}_M,0}\ 
      \chi^{(M)}_{n_1\ldots n_M}({\bf k}_1,\ldots,{\bf k}_{M-1})
\nonumber\\
\times\varphi^{-1}_{n_1n_2}({\bf k}_1)\cdots\varphi^{-1}_{n_Mn_1}({\bf k}_M)\,
\nonumber\\
  \times\tr\bigl[(\delta Q)_{n_1n_2}({\bf k}_1)\cdots (\delta Q)_{n_Mn_1}
                   ({\bf k}_M)\bigr]\quad.
\label{eq:4.2d}
\end{eqnarray}
\end{mathletters}%
This effective action needs to be supplemented by the following rules for
doing perturbation theory: 

{\em rule 1}: For calculating $Q$-propagators, all internal propagators 
must be taken as the interacting part of the Gaussian propagator, i.e.
as the second term on the right--hand side of Eq.\ (\ref{eq:2.36a}).

{\em rule 2}: For calculating $Q$-vertex functions, rule 1 also applies.
In addition, one needs to consider all reducible diagrams (which normally
do not contribute to the vertices), with all reducible propagators replaced
by minus the noninteracting Gaussian $Q$-propagator, i.e. minus the first term
on the right--hand side of Eq.\ (\ref{eq:2.36a}).

Explicit calculations using these rules readily show that the integrals
that correspond to the diagrams in a loop expansion are identical to
integrals that one encounters in standard many--body perturbation theory 
for the same quantity. This allows for a one--to--one correspondence between
many--body diagrams and the loop expansion based on the present field
theory. Nevertheless, even at a calculational level the present formulation
provides advantages compared to standard perturbation theory. For instance,
due to the above rules, the loop expansion is equivalent to an expansion in
powers of the screened Coulomb interaction, with the zeroth order, i.e. 
the Gaussian
theory, reproducing RPA. The loop expansion therefore allows for a systematic
improvement over RPA. Perhaps more importantly, our field--theoretical
formulation allows for an application of the renormalization group to draw
structural conclusions about the theory in analogy to those discussed for 
the disordered case in Sec.\ \ref{subsubsec:III.B.2}. This we will discuss
next.
 
\subsection{The Fermi--liquid fixed point}
\label{subsec:IV.B}
 
We now are looking for a stable RG fixed point that describes the clean
Fermi liquid, in analogy to the disordered Fermi--liquid FP of Sec.\ 
\ref{subsubsec:III.B.2}. For this purpose it is again convenient to
split the matrix $Q$ into blocks in frequency space,
\begin{equation}
Q_{nm} = \cases{P_{nm} & if $nm>0$\cr
                q_{nm} & otherwise}\quad.
\label{eq:4.3}
\end{equation}
As in Sec.\ \ref{subsubsec:III.B.2}, we define the scale dimension of a 
length $L$ to be $[L] = -1$, and we define exponents $\eta$ and $\eta'$ by
writing
\begin{mathletters}
\label{eqs:4.4}
\begin{equation}
[q({\bf x})] = \frac{1}{2}\,(d - 1 + \eta')\quad,
\label{4.4a}
\end{equation}
\begin{equation}
[\Delta P({\bf x})] = \frac{1}{2}\,(d - 1 + \eta)\quad.
\label{eq:4.4b}
\end{equation}
\end{mathletters}%
Here $\Delta P = P - \langle P\rangle$, and as in Sec. \ref{sec:III} we do
not distinguish between $\Delta P$ and $\delta P$.
The FP action has the properties one expects from a Fermi liquid
if we choose
\begin{mathletters}
\label{eqs:4.5}
\begin{equation}
\eta = 1\quad,\quad \eta' = 0\quad,
\label{eq:4.5a}
\end{equation}
and a dynamical exponent
\begin{equation}
z = 1\quad.
\label{eq:4.5b}
\end{equation}
\end{mathletters}%
Power counting shows that with these choices, ${\cal A}_0$ is dimensionless, 
and hence part of the FP action. So is the part of ${\cal A}_{int}$ that is 
quadratic in $q$. The parts of ${\cal A}_{int}$ that couple $q$ with 
$\Delta P$, and $\Delta P$ with itself,
are irrelevant with scale dimensions of $-1/2$ and $-1$, respectively.

Now consider the term of $O(Q^M)$ in the non--Gaussian part of the action,
$\Delta {\cal A}$, Eq.\ (\ref{eq:4.2d}). Let $N$ of the $Q$-fields be $P$'s, 
and $M-N$ be $q$'s. Denote the coupling constant for that term by $u_{N,M-N}$.
Taking into account the properties of $\chi^{(M)}$ and $\varphi$ discussed
after Eqs.\ (\ref{eqs:4.1}), we obtain 
\begin{mathletters}
\label{eqs:4.6}
\begin{equation}
[u_{M,0}] = -(d+1)\,\frac{(M-2)}{2}\quad,
\label{eq:4.6a}
\end{equation}
for $N=M$, and
\begin{equation}
[u_{N,M-N}] = -\frac{1}{2}\,(M-N-2) - (d-1)\,\frac{(M-2)}{2}\quad,
\label{eq:4.6b}
\end{equation}
\end{mathletters}%
for $N\neq M$. Here we have made use of the following observation. 
Because of the two rules in the preceding subsection, all contractions
that are performed in $\Delta {\cal A}$ to obtain renormalizations of the
Gaussian action result in interacting propagators. Since the interaction
carries a factor of temperature, see Eq.\ (\ref{eq:4.2c}), this results
effectively in $(M-2)/2$ factors of $T$ that need to be taken into account
in addition to the terms that show explicitly in Eq.\ (\ref{eq:4.2d}).
This factor of $T^{(M-2)/2}$ corresponds to the $M-2$ internal $Q$-fields that
get contracted in a renormalization of the Gaussian term. We note that this
feature is automatically built into the theory only after the 
$\bar\Lambda$-field has been integrated out. Doing so is therefore 
advantageous. Since $M-N$ is
necessarily even, we see from Eqs.\ (\ref{eqs:4.6}) that all of the 
non--Gaussian terms are RG irrelevant with respect to the Fermi--liquid FP,
provided that $d>1$. For $d=1$ there is an infinite set of marginal operators,
which signals the instability of the Fermi--liquid ground state against the
formation of a Luttinger liquid.\cite{Schulz} For $d<1$ the Fermi liquid is
unstable due to infinitely many relevant operators, as expected.

We mention that there is no consensus on whether or not interacting Fermi
systems in $d>1$ are necessarily Fermi liquids.
In particular, Anderson\cite{Anderson} has
proposed that there exists a stable FP that corresponds to a Luttinger--type
liquid, at least in certain $2$-$d$ models. While we do not find such a FP, we
stress that the above considerations do not constitute a proof that
none exists. All we have shown is that the assumption of a Fermi--liquid FP
in $d>1$ leads self--consistently to the conclusion that this FP is
stable. This is in agreement with a variety of other RG
arguments,\cite{Shankar,BenfattoGallavotti} and with explicit
calculations.\cite{FLCalculations} However, we
do not know what the basin of attraction for the Fermi--liquid FP
is, and we cannot exclude the existence of other fixed points.

Obviously, the discussion of the corrections to scaling and their
consequences for the behavior of thermodynamic quantities that was given for 
the disordered case in Sec.\ \ref{subsubsec:III.B.2} can be carried over.
The only difference is that the scale dimension of the least irrelevant
operators, which we again denote by $u$, is now $[u] = -(d-1)$. As an explicit
example, let us consider the spin susceptibility, $\chi_s$. It obeys a
scaling law in analogy to Eq.\ (\ref{eq:3.65a}),
\begin{mathletters}
\label{eqs:4.7}
\begin{equation}
\chi_s({\bf q},u) = \chi_s ({\bf q}b,ub^{-(d-1)})\quad,
\label{eq:4.7a}
\end{equation}
which leads to a nonanalytic dependence on the wave number,
\begin{equation}
\chi_s ({\bf q}) = {\rm const.} + \vert {\bf q}\vert^{d-1}\quad.
\label{eq:4.7b}
\end{equation}
\end{mathletters}%
This is the leading wave number dependence of $\chi_s$ for $1<d<3$, and the
leading nonanalytic one in all dimensions. In $d=3$ there is a logarithmic
correction to scaling, which our power counting arguments are not sensitive
to, and the behavior is ${\bf q}^2\ln\vert {\bf q}\vert$. This behavior has
recently been obtained by means of explicit perturbative 
calculations,\cite{chi_s} and its implications for the 
paramagnet--to--ferromagnet transition at zero temperature have been
discussed.\cite{fm_clean} The nonanalytic behavior of other thermodynamic
quantities, and of the quasiparticle inelastic life time, can be understood
by means of analogous arguments.

\section{Conclusion}
\label{sec:V}

In this paper we have given a general method to study the long--wavelength,
low--frequency behavior of many--fermion systems, both with and without
quenched disorder. The crucial ideas are to first identify the slow modes
of the system by using a symmetry analysis, then to separate these soft modes
from the massive ones, and, finally, to use renormalization group
ideas to eliminate the degrees of freedom that are irrelevant in the 
long--wavelength limit.

Using these ideas we have accomplished a number of things. We first
established that in a disordered system, a stable disordered Fermi--liquid
FP is possible for $d>2$. We showed that the so--called weak
localization effects in itinerant electronic 
systems\cite{WeakLocalizationFootnote} are, or can be
interpreted as, corrections to scaling near this FP. This
derivation not only reproduces known perturbative results, but also
establishes that their functional form is asymptotically exact in the 
long--wavelength limit. In this respect our achievement is analogous to that
of Ref.\ \onlinecite{ForsterNelsonStephen} for long--time tails in classical
fluids. In Section\ \ref{subsec:III.B} we have given 
a technically satisfactory
derivation of the generalized nonlinear sigma--model that has been used to
describe metal--insulator transitions near two dimensions. Finally, we have
indicated how the theory must be modified near other quantum phase
transitions where additional soft modes, namely the critical modes, appear.

For fermion systems without disorder, analogous results have been obtained.
First, in agreement with others, we find that a clean Fermi--liquid FP
exists for $d>1$. Corrections to scaling near this FP show
that clean electronic systems have nonanalyticities in various correlation
functions that are analogous to weak--localization effects in disordered
systems. This is again in agreement with results obtained on the basis of
either Fermi--liquid theory,\cite{BaymPethick} or many--body
perturbation theory. Our treatment provides a unified description that
reveals a deep connection between the behavior of fermionic systems with
and without disorder, respectively.
In Section\ \ref{subsec:IV.A} we also gave a novel 
perturbation theory method for clean fermion systems.

There are still a number of things to be done. For disordered systems, we
recently suggested an order parameter description of the metal--insulator
transition.\cite{LandauMIT} 
Our theory was based on the nonlinear sigma--model approach,
applied to high dimensions (near $d=6$), where its validity is not clear.
This approach indicated that the metal--insulator transition has features 
in common with the
transition in a classical random--field Ising model. These ideas need to be
reexamined using a more general theory, since the sigma--model approach is
asymptotically exact only for $d<4$. Instanton solutions of the general
field theory should also lead to insights concerning rare events, like local
moment formation, and the effects of local moments on the metal--insulator
transition. Finally, as
already noted, additional quantum phase transitions from the Fermi--liquid
state need to be investigated.

For clean systems, the most important thing to do is to include the effects
of other Fermi--liquid parameters, as was discussed in Section\ \ref{sec:II}. 
Once this
is done, quantum phase transitions in clean itinerant systems can be
properly studied. It will then also be possible to study the crossover from
the clean to the disordered Fermi liquid fixed point in detail. For
non--interaction systems, this latter point has been addressed by
Muzykantskii and Khmelnitskii.\cite{MuzykantskiiKhmelnitskii}

For both clean and disordered systems, it is interesting to ask how a
Fermi--liquid FP can be avoided. This question has been of great interest
recently in connection with high-$T_c$ superconductors, and other systems
that have `strange metal' phases. For clean systems this has recently
been reviewed in Ref.\ \onlinecite{MetznerCastellaniDiCastro}. 
For disordered systems, the situation is less clear. Various types of Kondo 
lattice mechanisms have been proposed.\cite{SantaBarbara}
In these approaches it is unclear how these effects modify the long--wavelength
transport properties. Another possibility is to consider systems with a
vanishing single--particle density of states at the Fermi surface. If this
occurs, then the soft modes discussed in Sec.\ \ref{sec:II} are not as
singular, and a two--dimensional disordered metal phase becomes possible.
Further, it suggests the possibility of an exotic metal--insulator transition
in two dimensions. Clearly, these problems require much more work.

\acknowledgments

This work was supported by the NSF under grant numbers DMR-96-32978 and 
DMR-95-10185. We thank Harmen Bussemaker, Ferdinand Evers, and Brad Marston 
for helpful discussions.

\appendix
\section{Projection onto the density}
\label{app:A}

Here we demonstrate, for the spin--singlet particle--hole interaction term,
the projection onto density modes that was used in
Secs.\ \ref{subsec:II.A} and \ref{subsubsec:II.B.1}.
An analogous procedure can be applied to the other interaction
channels, and to the disorder part of the action.

It is most convenient to go back to a Hamiltonian description of the
system. The part of the interaction Hamiltonian that corresponds to
Eq.\ (\ref{eq:2.9b}) reads
\begin{mathletters}
\label{eqs:A.1}
\begin{equation}
H_{int}^{(s)} = \frac{1}{2} \sum_{{\bf k},{\bf p}} {\sum_{\bf q}}^{\,\prime}\,
                \Gamma_{k,p}^{(s)}(q)\ \Delta f_{\bf k}({\bf q})\,
                \Delta f_{\bf p}({\bf q})\quad,
\label{eq:A.1a}
\end{equation}
where 
\begin{equation}
f_{\bf k}({\bf q}) = \sum_{\sigma}\,c^{\dagger}_{\sigma,{\bf k}}\ 
                       c_{\sigma,{\bf k} + {\bf q}}\quad,
\label{eq:A.1b}
\end{equation}
\end{mathletters}%
is the phase space density operator in terms of electron creation and
annihilation operators $c^{\dagger}$ and $c$, respectively, and
$\Delta (c^{\dagger} c) = c^{\dagger} c - \langle c^{\dagger} c\rangle$.
In the space of products of fermion operators, we define the Kubo
product,\cite{Kubo}
\begin{equation}
\left(A\vert B\right) = \int_{0}^{\beta} d\tau\ \langle\Delta B(\tau)\,
                         \Delta A^*\rangle\quad,
\label{eq:A.2}
\end{equation}
with $A$ and $B$ operators, and $\tau$ the imaginary time
variable. In terms of this scalar product in operator space, the desired
projector reads
\begin{equation}
{\cal P} = \vert n_n({\bf q})\bigr)\,\frac{1}{g({\bf q})}\bigl(n_n({\bf q})
     \vert\quad.
\label{eq:A.3}
\end{equation}
Here $n_n({\bf q}) = \sum_{\bf k}\,f_{\bf k}({\bf q})$ is the electron
number density operator, and
$g({\bf q}) = \left(n_n({\bf q})\vert n_n({\bf q})\right)$ is the static
density susceptibility or wave vector dependent compressibility.
Using ${\cal P}$ twice, it is now easy to project onto the density in
Eq.\ (\ref{eq:A.1a}). We obtain
\begin{mathletters}
\label{eqs:A.4}
\begin{equation}
H_{int}^{(s)} \approx\frac{1}{2} {\sum_{\bf q}}^{\,\prime}\,\Gamma^{(s)}
         ({\bf q})\,n_n({\bf q})\,n_n(-{\bf q})\quad,
\label{eq:A.4a}
\end{equation}
where,
\begin{eqnarray}
\Gamma^{(s)}({\bf q}) = \frac{1}{g^2({\bf q})} \sum_{{\bf k},{\bf p}}
  \Gamma_{k,p}^{(s)}(q)\,\bigl(f_{\bf k}({\bf q})\vert n_n({\bf q})\bigr)
\nonumber\\
   \times\bigl(n_n({\bf q})\vert f_{\bf p}({\bf q})\bigr)\quad.
\label{eq:A.4b}
\end{eqnarray}
\end{mathletters}%
The phase space Kubo function, 
$g_{{\bf k}{\bf p}}({\bf q}) = \left(f_{\bf k}({\bf q})\vert f_{\bf p}
({\bf q})\right)$, for clean, free
electrons is proportional to 
$\delta_{{\bf k}{\bf p}} \delta ({\bf k}^2 - k_F^2)$, so in this case
Eq.\ (\ref{eq:A.4b}) results in pinning ${\bf k}$ and ${\bf p}$ to the
Fermi surface. In a disordered system, $g_{{\bf k}{\bf p}}({\bf q})$
has a width given by the inverse elastic mean--free path, and hence
$\Gamma^{(s)}({\bf q})$ is a weighted average over a region in the
vicinity of the Fermi surface, as mentioned in Sec.\ \ref{subsec:II.A}.
Finally, for physics that is controlled by soft modes and long--wavelength
processes we can replace $\Gamma^{(s)}({\bf q})$ by 
$\Gamma^{(s)}\equiv\Gamma^{(s)}({\bf q}=0)$. Switching back to a 
field--theoretic representation of the fermions, we obtain 
Eq.\ (\ref{eq:2.12a}).

\section{$\mathbf O(N)$ symmetric $\mathbf \phi^4$-theory}
\label{app:B}

In this appendix we perform an analysis of $\phi^4$-theory that is analogous
to that of the matrix field theory in Sec.\ \ref{subsec:III.B}. 
Much of this material can be found in the literature, e.g. in Zinn-Justin's 
book, Ref.\ \onlinecite{ZJ}. However, we
find it useful to include it here for pedagogical reasons. The remarkable
analogy between field theories for electrons and classical spin models
that was first noted by Wegner\cite{Wegner} substantially simplifies the
understanding of the former in terms of the technically much simpler
structure of the latter.

\subsection{Origin of the nonlinear sigma--model}
\label{app:B.1}

Let us consider an $O(N)$ symmetric $\phi^4$-theory with a magnetic field
$h$ in the $1$-direction. The action
\begin{mathletters}
\begin{eqnarray}
S[{\vec\phi}]&=&\int d{\bf x}\,\left[r\bigl({\vec\phi}({\bf x})\bigr)^2
               + c\bigl(\nabla{\vec\phi}({\bf x})\bigr)^2\right]
\nonumber\\
&&+ u \int d{\bf x}\,\left({\vec\phi}({\bf x})\cdot{\vec\phi}({\bf x})\right)^2
  - h\int d{\bf x}\,\phi_1({\bf x}) \quad,
\label{eq:B.1a}
\end{eqnarray}
determines the partition function
\begin{equation}
Z[h] = \int D[{\vec\phi}]\,e^{-S[{\vec\phi}]}\quad.
\label{eq:B.1b}
\end{equation}
\end{mathletters}%
In the low--temperature phase, where the $O(N)$ symmetry is spontaneously
broken, it is convenient to
decompose the vector field $\vec\phi$ into its modulus $\rho$ and a
unit vector field $\hat\phi$,
\begin{equation}
{\vec\phi}({\bf x}) = \rho({\bf x})\,{\hat\phi}({\bf x})\quad,\quad
                        {\hat\phi}^2({\bf x})\equiv 1\quad,
\label{eq:B.2}
\end{equation}
$\hat\phi$ parametrizes the unit sphere, and thus provides a representation
of the homogeneous space $O(N)/O(N-1)$.
In terms of $\rho$ and $\hat\phi$ the action reads,
\begin{mathletters}
\label{eqs:B.3}
\begin{eqnarray}
S[\rho,\hat\phi]&=&\int d{\bf x}\,\biggl[c^{(1)}\rho^2({\bf x})
      \left(\nabla{\hat\phi}({\bf x})
            \right)^2 + c^{(2)}\left(\nabla\rho({\bf x})\right)^2
\nonumber\\
&+&r\rho^2({\bf x})\biggr] + u\int d{\bf x}\,\rho^4({\bf x})
  - h\int d{\bf x}\,\rho({\bf x}){\hat\phi}_1({\bf x})\,,
\nonumber\\
\label{eq:B.3a}
\end{eqnarray}
and switching from the functional integration variables $\vec\phi$ to
$(\rho,\hat\phi)$ leads to a Jacobian or invariant measure
\begin{equation}
I[\rho] = \prod_{\bf x} \rho^{N-1}({\bf x})\quad.
\label{eq:B.3b}
\end{equation}
\end{mathletters}%
In Eq.\ (\ref{eq:B.3a}) the bare values of $c^{(1)}$ and $c^{(2)}$ are equal, 
and equal to $c$. Notice that the field $\hat\phi$ appears only in
conjunction with two gradient operators. $\hat\phi$ represents the $N-1$ soft
Goldstone modes of the problem, while $\rho$ represents the massive modes.
Now we parametrize $\hat\phi$,
\begin{mathletters}
\label{eqs:B.4}
\begin{equation}
{\hat\phi}({\bf x}) = \bigl(\sigma ({\bf x}),{\vec\pi}({\bf x})
                          \bigr)\quad,
\label{eq:B.4a}
\end{equation}
where
\begin{equation}
\sigma ({\bf x}) = \sqrt{1-{\vec\pi}^2({\bf x})}\quad.
\label{eq:B.4b}
\end{equation}
\end{mathletters}%
We split off the expectation value of the massive $\rho$-field,
$\rho({\bf x}) = M + \Delta\rho({\bf x})$, with 
$M = \langle\rho({\bf x})\rangle$, and expand in powers of $\vec\pi$ and
$\Delta\rho$. Rescaling the coupling constants with appropriate powers of
$M$, the action can be written
\begin{mathletters}
\label{eqs:B.5}
\begin{equation}
S[\rho,{\vec\pi}] = S_{NL\sigma M}[{\vec\pi}] + \Delta S[\rho,{\vec\pi}]\quad.
\label{eq:B.5a}
\end{equation}
Here
\begin{eqnarray}
S_{NL\sigma M}[{\vec\pi}] = \frac{1}{t}\int d{\bf x}\,\left[
             \left(\nabla{\vec\pi}({\bf x})\right)^2 
                            + \left(\nabla\sigma({\bf x})\right)^2\right]
\nonumber\\
- h\int d{\bf x}\,\sigma{(\bf x})\quad,
\label{eq:B.5b}
\end{eqnarray}
is the action of the $O(N)$ nonlinear sigma--model, and
\begin{eqnarray}
\Delta S[\rho,{\vec\pi}]&=&r \int d{\bf x}\,\left(\Delta\rho ({\bf x})\right)^2
            + c^{(2)} \int d{\bf x}\,\left(\nabla \Delta\rho ({\bf x})\right)^2
\nonumber\\
  &&+ u_3 \int d{\bf x}\,\left(\Delta\rho ({\bf x})\right)^3
     + u_4 \int d{\bf x}\,\left(\Delta\rho ({\bf x})\right)^4
\nonumber\\
  &&\qquad\qquad + O\left(\Delta\rho\,\sigma,\Delta\rho\left(\nabla{\hat\phi}
                                                \right)^2\right)\quad.
\label{eq:B.5c}
\end{eqnarray}
\end{mathletters}%
If we neglect all fluctuations of the $\rho$-field, then we are left with
the $O(N)$ symmetric nonlinear sigma--model in the usual parametrization.

\subsection{The low--temperature fixed point}
\label{app:B.2}

Now we define the scale dimensions of the fields $\pi_i$ and $\Delta\rho$ as
\begin{mathletters}
\label{eqs:B.6}
\begin{equation}
\left[\pi_i({\bf x})\right] = \frac{1}{2}\,(d-2+\eta')\quad,
\label{eq:B.6a}
\end{equation}
\begin{equation}
\left[\Delta\rho ({\bf x})\right] = \frac{1}{2}\,(d-2+\eta)\quad,
\label{eq:B.6b}
\end{equation}
\end{mathletters}%
and perform a momentum--shell RG procedure.\cite{WilsonKogut}
Here we have defined the scale dimension of a length $L$ to be $[L] = -1$,
and the above relations define the exponents $\eta$ and $\eta'$. For $\eta$, 
this definition coincides with that of the exponent usually denoted by this
symbol. The stable, low--temperature FP of the action describes the ordered 
phase. Physically, one expects short--ranged $\Delta\rho$--correlations, 
and diffusive $\pi$--correlations at this FP, which suggests
the choice\cite{etaFootnote2}
\begin{equation}
\eta = 2\quad,\quad \eta' = 0 \quad.
\label{eq:B.6'}
\end{equation}
Because of the positive scale dimension of $\vec\pi$, it is convenient to expand
the sigma--model action,
\begin{eqnarray}
S_{NL\sigma M}&=&\frac{1}{t} \int d{\bf x}\,\left(\nabla {\vec\pi}({\bf x})
                  \right)^2 + \frac{h}{2} \int d{\bf x}\,{\vec\pi}^2({\bf x})
\nonumber\\
&&+ v \int d{\bf x}\,\left({\vec\pi}({\bf x})\nabla{\vec\pi}({\bf x})\right)^2
   + O\left({\vec\pi}^4,\nabla^2{\vec\pi}^6\right)\quad.
\nonumber\\
\label{eq:B.7}
\end{eqnarray}
The FP action is given by the first terms in Eqs.\ (\ref{eq:B.7}) and
(\ref{eq:B.5c}), respectively. With respect to this FP,
the magnetic field is the only relevant operator, with a scale dimension
$[h] = 2$. $t$ and $r$ are marginal, and all other operators are
irrelevant with respect to this FP, so the FP is really stable.

The leading correction to scaling near $d=2$ is given by the operator $v$, 
with a scale dimension $[v] = -(d-2)$. This identifies $d_c^- = 2$ as the
lower critical dimension of the problem. $v$ also provides the leading
nonanalytic correction to scaling in {\em all} dimensions, since the
leading irrelevant operator related to the massive $\Delta\rho$-field
has a scale dimension $[c^{(2)}] = -2$, and therefore leads to analytic
corrections to scaling.
For $d>d_c^-$, the spontaneously
broken $O(N)$ symmetry leads to $N-1$ Goldstone modes, which are
represented by the correlations of the $n-1$ $\pi$-fields,
\begin{equation}
\langle\pi_i({\bf k})\,\pi_j({\bf -k})\rangle = \delta_{ij}\vert{\bf k}\vert
  ^{-2+\eta'} = \delta_{ij}/{\bf k}^2\quad.
\label{eq:B.8}
\end{equation}
The FP value of the magnetization, $m=\langle\phi_1({\bf x})\rangle$,
is equal to $M$, and the leading correction is given by the $\pi$-$\pi$
correlation function. By virtue of the scale dimensions of $\pi$ and $h$,
one finds for the leading field dependence of the magnetization
\begin{equation}
m(h) = {\rm const.} + h^{(d-2)/2}\quad,
\label{eq:B.9}
\end{equation}
which implies that the longitudinal susceptibility,
$\chi_L = \partial m/\partial h \sim h^{(d-4)/2}$, diverges in the
$h\rightarrow 0$ limit for all $d<4$.\cite{BrezinWallace} Alternatively,
the wave vector dependent zero--field susceptibility diverges in the
homogeneous limit like $\chi_L \sim \vert {\bf k}\vert^{d-4}$.
Notice that the leading nonanalytic corrections to scaling are given by
the Goldstone modes, and are thus contained in the nonlinear sigma--model.

The point of the above excercise, in the present context, is to demonstrate
how much more one gets out of power counting after performing a symmetry
analysis and separating the soft modes from the massive ones, as opposed
to doing the power counting on the action in the original $\vec\phi$
formulation.\cite{Ma} The analogy to the procedure in Sec.\ \ref{subsec:III.B}
is as follows: $\Delta\rho$ corresponds to the massive fields $\Delta P$ and
$\Delta\Lambda$, $\hat\phi$ corresponds to $\hat Q$ and $\cal S$ (the
matrix theory cannot be expressed in terms of $\hat Q$ only), and $\vec\pi$
corresponds to $q$. The external magnetic field $h$ plays the role of the
external frequency in the matrix theory, and the analogy with respect to
the Goldstone modes and the scaling behavior of the order parameter is
obvious.

\subsection{The critical fixed point}
\label{app:B.3}

Apart from the low--temperature FP discussed above, the nonlinear sigma--model, 
Eq.\ (\ref{eq:B.5b}), also contains a critical FP, as was noticed by 
Polyakov,\cite{Polyakov} and discussed by him and 
others.\cite{BrezinZJ,BrezinZJLeG,NelsonPelcovits}. 
This FP marks the instability of the
low--temperature or broken symmetry phase; the $O(N)$ symmetry is restored,
and hence the fields $\vec\pi$ and $\sigma$ have the same scale dimension.
$\vec\pi$ must then be dimensionless, which implies
\begin{mathletters}
\label{eqs:B.10}
\begin{equation}
\eta'=2-d\quad,
\label{eq:B.10a}
\end{equation}
in Eq.\ (\ref{eq:B.6a}). The exponent $\eta$ in Eq.\ (\ref{eq:B.6b}), on the
other hand, is the critical correlation function exponent that is related to
the order parameter exponent $\beta$, and the susceptibility exponent $\gamma$,
via
\begin{equation}
\eta = 2 - \gamma/\nu = 2 - d + 2\beta/\nu\quad.
\label{eq:B.10b}
\end{equation}
\end{mathletters}%

At criticality, the $\Delta\rho$ fluctuations become massless, and the region
of validity of the sigma--model shrinks to zero. Specifically, the $\Delta\rho$
fluctuations can be neglected only if one works at momenta $p$ that are larger
then the mass of the $\Delta\rho$-field, $m_{\rho} = \sqrt{r/c^{(2)}}$, see
Eq.\ (\ref{eq:B.5c}). Power counting, and Eq.\ (\ref{eq:B.6b}), shows that the
scale dimension of $m_{\rho}$ is $[m_{\rho}] = 1$, so upon approaching 
criticality it vanishes like 
$m_{\rho} \sim \xi^{-1} \sim \vert T-T_c\vert^{\nu}$. The criterion for the
validity of the sigma--model is therefore
\begin{equation}
p \ll \vert T-T_c\vert^{\nu}\quad.
\label{eq:B.11}
\end{equation}
In this region perturbation theory works, and the critical properties can
be explored by using the RG to sum the perturbative results.  In this way
one obtains a description of the Heisenberg critical behavior in the
vicinity of $d=2$ that complements the perturbative treatment of the
$\phi^4$-formulation of the problem, Eq.\ (\ref{eq:B.1a}), near $d=4$.
The critical FP of the matrix nonlinear sigma--model, 
Sec.\ \ref{subsubsec:III.B.3}, is again in many respects analogous to the
critical FP of the $O(N)$ model.

It has been suggested that terms
with more than two gradients, which appear as corrections to the sigma--model
upon explicitly integrating out the massive field, may lead to an instability
of the critical FP, even though these terms are irrelevant by power counting
at tree level.\cite{HigherGradients}. The status of this problem is currently
unclear, see Ref.\ \onlinecite{BrezinHikami} for a recent discussion.

\section{The invariant measure $I[P]$}
\label{app:C}

Here we derive explicitly the invariant measure $I[P]$, defined by
Eq.\ (\ref{eq:3.24}). Let us write Eq.\ (\ref{eq:3.23}) in block matrix form,
\begin{eqnarray}
\left( \begin{array}{cc} Q^{>>} & Q^{><}\\
                  Q^{<>} & Q^{<<}\\
       \end{array}\right)
= \left( \begin{array}{cc} {\cal S}^{>>} & {\cal S}^{><}\\
                           {\cal S}^{<>} & {\cal S}^{<<}\\
         \end{array}\right)
  \left( \begin{array}{cc} P^> & 0\\
                           0 & P^<\\
         \end{array}\right)
\nonumber\\
\times\left( \begin{array}{cc} \left( {\cal S}^{-1}\right)^{>>} &
                                          \left( {\cal S}^{-1}\right)^{><}\\
                               \left( {\cal S}^{-1}\right)^{<>} &
                                          \left( {\cal S}^{-1}\right)^{<<}\\
             \end{array}\right)\quad.
\label{eq:C.1}
\end{eqnarray}
Here each block is a $4Nn\times 4Nn$ matrix. $P$ is block diagonal by
construction, and the blocks of ${\cal S}$ are subject to constraints due to
the definition of the coset space. Now consider a variation $\delta Q$ of $Q$.
Since by symmetry the invariant measure is independent of the group
parameters, it suffices to consider an infinitesimal variation. To first order
in infinitesimal quantities, the variation of $\cal S$ is block
off--diagonal (see Eq.\ (\ref{eq:3.50})), and we have,
\begin{eqnarray}
&\mbox{}&\left( \begin{array}{cc} \delta Q^{>>} & \delta Q^{><}\\
                         \delta Q^{<>} & \delta Q^{<<}\\
       \end{array}\right) =
\nonumber\\
&\mbox{}&\left( \begin{array}{cc} \delta P^> & \delta {\cal S}^{><}\ P^< - P^>\
                                                       \delta {\cal S}^{><}\\
           \delta {\cal S}^{<>}\ P^> - P^<\ \delta {\cal S}^{<>} & \delta P^<\\
       \end{array}\right)\quad.
\nonumber\\
\label{eq:C.2}
\end{eqnarray}
By directly differentiating $Q$ with respect to $P$ we find from
Eq.\ (\ref{eq:C.2}),
\begin{equation}
I[P] = \det\left[\openone\otimes (P^<)^T - P^>\otimes\openone\right]\quad.
\label{eq:C.3}
\end{equation}
$P^>$ and $P^<$ are analytic continuations of anti--hermitian matrices, 
so they can be diagonalized by means of analytic continuations of
unitary $4Nn\times 4Nn$ matrices. Let their eigenvalues be
$\lambda_i^>$ and $\lambda_i^<$ ($i=1,\ldots 4Nn$). Putting the space
dependence back in, we finally obtain
\begin{equation}
I[P] = \left\vert\prod_{\bf x}\prod_{i,j=1}^{4Nn}\left(\lambda_{i}^{<}({\bf x})
                - \lambda_{j}^{>}({\bf x})\right)\right\vert\quad.
\label{eq:C.4}
\end{equation}

%
%
\end{document}